\documentclass{article}%

\usepackage[T1]{fontenc}%
\usepackage{lmodern}%
\usepackage{textcomp}%
\usepackage{lastpage}%
\usepackage{float}%
\usepackage[dvipsnames]{xcolor}%
\usepackage[margin=3cm]{geometry}
\usepackage{graphicx}%
\usepackage[backref=page,breaklinks=true,hidelinks]{hyperref}
\hypersetup{
    colorlinks,
    linkcolor={red!50!black},
    citecolor={blue!50!black},
    urlcolor={blue!80!black}
}
\renewcommand*{\backref}[1]{}
\renewcommand*{\backrefalt}[4]{%
    \ifcase #1 {\footnotesize(Not cited.)}%
    \or        {\footnotesize(Cited on page~#2.)}%
    \else      {\footnotesize(Cited on pages~#2.)}%
    \fi}
\usepackage{caption}%
\usepackage{natbib}%
\usepackage{multirow}
\usepackage{amsmath}%
\usepackage{amssymb}%
\usepackage{mathtools}%
\usepackage{booktabs}%
\usepackage{epstopdf}%
\usepackage{adjustbox}%
\usepackage{array}%
\usepackage{lscape}%
\usepackage{cleveref}
\usepackage{titling}%
\usepackage{amsthm}
\usepackage{xargs}
\usepackage{xcolor}
\usepackage{mathtools}
\usepackage{subcaption}
\usepackage{placeins}
\usepackage[normalem]{ulem}
\usepackage{setspace}
\usepackage{mdframed}
\usepackage{bbm}

\graphicspath{{figures_replication/}}

\renewcommand{\P}{\mathop{}\!\textnormal{P}}
\newcommand{\E}{\mathop{}\!\textnormal{E}}

\usepackage[color=yellow!20]{todonotes} %
\usepackage{xfrac}
\graphicspath{{figures/}}
  
\title{
\vspace{-3em}
Machine Learning Who to Nudge: \\
Causal vs Predictive Targeting in a Field Experiment \\on Student Financial Aid Renewal
}%
\author{Susan Athey \and Niall Keleher \and Jann Spiess
}%
\date{
May 31, 2024
}

\begin{document}%
\renewcommand{\sectionautorefname}{Section}
    
    \begin{titlepage}

    \maketitle%

    \begin{abstract}
        In many settings, interventions may be more effective for some individuals than others, so that targeting interventions may be beneficial. We analyze the value of targeting in the context of a large-scale field experiment with over 53,000 college students, where the goal was to use ``nudges'' to encourage students to renew their financial-aid applications before a non-binding deadline. 
        We begin with baseline approaches to targeting.
        First, we target based on a causal forest that estimates heterogeneous treatment effects and then assigns students to treatment according to those estimated to have the highest treatment effects.
        Next, we evaluate two alternative targeting policies, one targeting students with low predicted probability of renewing financial aid in the absence of the treatment, the other targeting those with high probability. The predicted baseline outcome is not the ideal criterion for targeting, nor is it a priori clear whether to prioritize low, high, or intermediate predicted probability. Nonetheless, targeting on low baseline outcomes is common in practice, for example because the relationship between individual characteristics and treatment effects is often difficult or impossible to estimate with historical data. We propose hybrid approaches that incorporate the strengths of both predictive approaches (accurate estimation) and causal approaches (correct criterion); we show that targeting \emph{intermediate} baseline outcomes is most effective in our specific application, while targeting based on low baseline outcomes is detrimental. In one year of the experiment, nudging all students improved early filing by an average of 6.4 percentage points over a baseline average of 37\% filing, and we estimate that targeting half of the students using our preferred policy attains around 75\% of this benefit.
    \end{abstract}

    \vspace{2em}

    \footnotesize
    Susan Athey, Stanford University, Stanford, CA, USA, \href{mailto:athey@susanathey.com}{\texttt{athey@susanathey.com}};
    Niall Keleher, Innovations for Poverty Action, New Haven, CT, USA, \href{mailto:nkeleher@poverty-action.org}{\texttt{nkeleher@poverty-action.org}};
    Jann Spiess, Stanford University, Stanford, CA, USA, \href{mailto:jspiess@stanford.edu}{\texttt{jspiess@stanford.edu}}.
    We thankfully acknowledge support from the Alfred P. Sloan Foundation and Schmidt Futures through the ``Computational Applications for Behavioral Science'' project with ideas42. Matthew Schaelling, Janelle Nelson, and Keshav Agrawal provided exceptional research assistance. For their support, we also thank Octavio Medina, Rebecca Nissan, Rachel Rosenberg, and Josh Wright of ideas42, Vitor Hadad and Henrike Steimer during their time as postdocs at the Golub Capital Social Impact Lab at Stanford GSB, and the Office of Institutional Research and Assessment of the City University of New York.
    For helpful comments and discussions, we are grateful to Mohsen Bayati, Michal Koles\'{a}r, and
    seminar audiences at Stanford GSB, ETH Zurich,
    the Machine Learning in Economics Summer Institute,
    the INFORMS Annual Meeting,
    Chicago Booth,
    LMU Munich,
    J-PAL,
    and
    UCSC.
    \end{titlepage}
    
\section{Introduction}
\label{sec:introduction}%

    A growing number of randomized experiments set out to measure the effectiveness of behaviorally-informed nudges. Typically, these experiments are designed and analyzed to assess whether a nudge works well on average. In this article, we utilize causal machine learning to move beyond average effects towards optimal targeting of nudges. In a large-scale experiment that randomized behaviorally-informed reminders to increase student financial-aid renewal applications, we estimate not just whether the nudge worked on average but also whether it worked for some students better than others. We then ask how such heterogeneous effect estimates can improve the effectiveness of policy interventions.
    
    Our application considers data from a field experiment among over 53,000 college students. The experiment aimed to measure the causal effect of behavioral nudges on timely applications for financial aid. Across two randomized controlled trials run in 2017 and 2018 by ideas42 and the City University of New York, enrolled students were randomly assigned to receive behaviorally informed text and email reminders about renewing their federal financial aid. The average treatment effect of the behavioral nudges was noteworthy. Students who received nudges were on average $6.4\pm0.6$%
    \footnote{Here, the number after the $\pm$ sign denotes a standard error estimate.}
    (2017) and $12.1\pm0.7$ (2018) percentage points more likely to submit their Free Application for Federal Student Aid (FAFSA) forms by the priority deadline \citep[as previously reported in][]{nissan2020}.
  
    Our goal is to estimate for whom the nudges are most effective, and thus whom to target if there is a limited budget of time or attention that creates opportunity cost associated with the nudges. For example, alternative messages might be used or developed for those where the nudges are less effective.
    The problem of whom to target arises in many settings, ranging from precision medicine to prioritization of salespeople to allocation of advertising spend. A common approach in practice is based on predicting which individuals are most at risk of undesirable outcomes, such as customer churn or a poor medical outcome. Predictive approaches are attractive because they can be applied with observational, historical data without the need to run an experiment. In the context of nudges to file financial aid forms, we could imagine forming a hypothesis about which type of student would be most influenced by a nudge and building a model to predict that outcome using data about student behavior in the absence of the treatment. If we hypothesized that students who were otherwise at risk of not filing would be most influenced by the nudge, we would target the students with the lowest predicted probability of filing in the absence of the treatment. 
    
    However, it is an empirical question as to what type of student is most likely to be influenced and what factors drive heterogeneity in treatment effects. In this article, we use causal machine learning to estimate how treatment effects vary with individual characteristics; that is, we estimate conditional average treatment effects (CATEs).
    Based on information available before the nudges are sent, we can predict differences in the response to the nudge.
    We identify subgroups across which differences in treatment effects are statistically significant, and the magnitudes of differences can be as large as a factor of two.
    We estimate the difference between those with below-median predicted effect and those with above-median predicted effect to be around three to five percentage points.
    Once it becomes available, enrollment status is highly predictive of treatment effects. Students who were unenrolled at the time of the behavioral nudge campaign had smaller treatment effects. However, we also uncover additional variation in CATEs among the enrolled students.

    We then estimate the benefits of assigning students with the highest estimated CATEs to receive the nudge (counterfactually, under a budget constraint, which may be financial or based on opportunity cost). We find that non-parametric estimates of the CATE are noisy, so we explore several hybrid models that incorporate predicted baseline outcomes and nonparametric estimates of CATEs as covariates in a parametric model. We find that both non-parametric and hybrid models do significantly better than a random policy,
    with hybrid models substantially outperforming the non-parametric model.
    A policy that targets students \textit{least} likely to file does very poorly, while targeting students \textit{most} likely to file performs well. Based on our experiment, the best approach is to prioritize students with \textit{intermediate} predicted baseline outcomes first and those with low predicted baseline outcomes last.
    
    Our results provide an example where naive risk-based targeting from a machine-learning prediction of outcomes (particularly, one that targeted those who might have seemed to need the nudges most) performs substantially worse than targeting based on estimated treatment effects.
    On the other hand, if we had correctly guessed that it was effective to target those who were already most likely to file, we would have achieved acceptable performance.
    Our findings highlight the value of augmenting machine-learning algorithms -- which provide powerful prediction tools -- with careful causal inference to tackle policy problems. 
    
    Our application also speaks to the value of modeling treatment effects carefully when treatment effect variation is moderate and noisy.
    A simple causal model based on baseline predictions substantially outperforms the fully non-parametric causal machine learning model.
    This semi-parametric model estimates a simple logistic regression on top of a random-forest prediction of baseline filing probabilities, which can be interpreted as a summary statistic of an individual's characteristics and is straightforward to implement as a prediction within the control group only.
    Treatment effects are assumed to be constant in log-odds (and thus U-shaped in baseline outcomes), and we do not find evidence that modeling the treatment effects in more complex ways substantially improves the performance of our targeting policy. 
    This approach is related to the insight in \cite{athey2021semiparametric} that it is often much easier to produce quality estimates of baseline responses while estimating treatment effects directly may be noisy and, therefore, benefit from parametric models.
    While the simple semi-parametric logistic regression model captures treatment effects well in our example, it has the disadvantage that it may perform poorly if the relationship between baseline outcomes and treatment effects is not a good fit with the logistic functional form, or if there is substantial variation in treatment effects that is not captured by the baseline.
    We, therefore, also consider more flexible models, including a non-parametric CATE model as a function of baseline predictions and a hybrid model that incorporates both predicted baseline outcomes and non-parametric CATE estimates.
    
    Our results suggest four general conclusions. First, they clarify the importance of integrating causal inference and randomized trials into machine learning to analyze and improve policy rather than relying on predictive tools based on non-experimental baseline data alone.
    Second, our analysis uncovers important challenges in applying machine learning to improve the analysis and targeting of nudges. The environment we study has a fairly low signal-to-noise ratio, and we find that treatment effect estimates are unlikely to be well-calibrated. Thus, describing heterogeneity requires careful out-of-sample evaluation to avoid small-sample biases.
    Third, the effects of treatments that are only small interventions, as is often the case for nudges, may not always accrue for those with low baseline outcomes.
    While the relationship between baseline outcomes and treatment effect may vary from case to case, in our specific experiment, the effects are stronger for those with intermediate and high baseline outcomes who just need to be ``nudged'' over the finish line.
    Finally, we demonstrate the value of combining non-parametric predictive tools with simple econometric models for causal estimation and targeting.

    We build upon a growing literature that combines causal estimation with prediction techniques from machine learning \citep{Mullainathan2017-qw,Athey2019-zc} and discusses the application of machine learning to policy problems \citep{Kleinberg2015-ix}. Methods for estimating heterogeneous treatment effects using machine learning have been proposed in, among others, \cite{imai2013estimating, Athey2016-zi, Wager2018-pe, Athey2019-grf, kunzel2019metalearners, nie2021learning}. \cite{Chernozhukov2019-ti} discusses the analysis of heterogeneous treatment effects with arbitrary machine-learning estimators and suggests diagnostic tools.
    Diverse literature across econometrics, statistics, computer science, operations, and marketing have been studying the optimal targeting of policies, including \cite{zhao2012estimating,swaminathan2015batch,kitagawa2018should}.
    \cite{yadlowsky2021evaluating} proposes metrics for assessing the benefits of targeting similar to those applied in this article. \cite{athey2021policy} and \cite{zhou2023offline} analyze efficient estimation of targeted policies with constraints on the policy class. \citet{Hitsch2018-bw, Knaus597, yang2020targeting}, among others, carry out empirical studies analyzing targeted policies derived using machine learning methods. 
    \cite{zhang2022coarse} proposes a two-step procedure involving treatment-effect estimation and discretization to decide which treatments to offer to whom.

    Until recently, there was little empirical evidence about the relative performance of predictive and causal targeting. An early article to do so is \cite{Ascarza2018-md}, which shows that targeting a customer retention program based on predicted churn performs considerably worse than targeting based on estimated treatment effects. Subsequent studies in marketing have similar findings \citep[e.g.][]{devriendt2021you}.
    Like us, \cite{fernandez2022causal} compares targeting by negative and positive baseline to causal targeting, and shows that the former can be beneficial when causal targeting is noisy.
    In development economics,
    \cite{haushofer2022targeting} considers targeting by ``impact'' (treatment effect) vs ``deprivation'' (baseline).
    In medicine, \cite{inoue2023machine} analyzes the tradeoff between ``high-risk'' and ``high-benefit'' approaches when targeting treatments for high blood pressure.

    We also connect to a literature on behaviorally-informed nudges \citep{Sunstein2008-pr} and their empirical validation. In the context of student financial aid, \cite{Castleman2016-tb} shows that a simple text-based intervention that encouraged FAFSA submission increased sophomore year retention by 14\%. ideas42 research at Arizona State University showed that behaviorally informed student reminder emails increased priority deadline FAFSA renewal by 11 percentage points, from 29\% to 40\% \citep{Ideas422016-ue}.

\section{Experiment and Data}
\label{sec:experiment_data}%

\begin{figure}
        \centering
        \includegraphics[width=\textwidth]{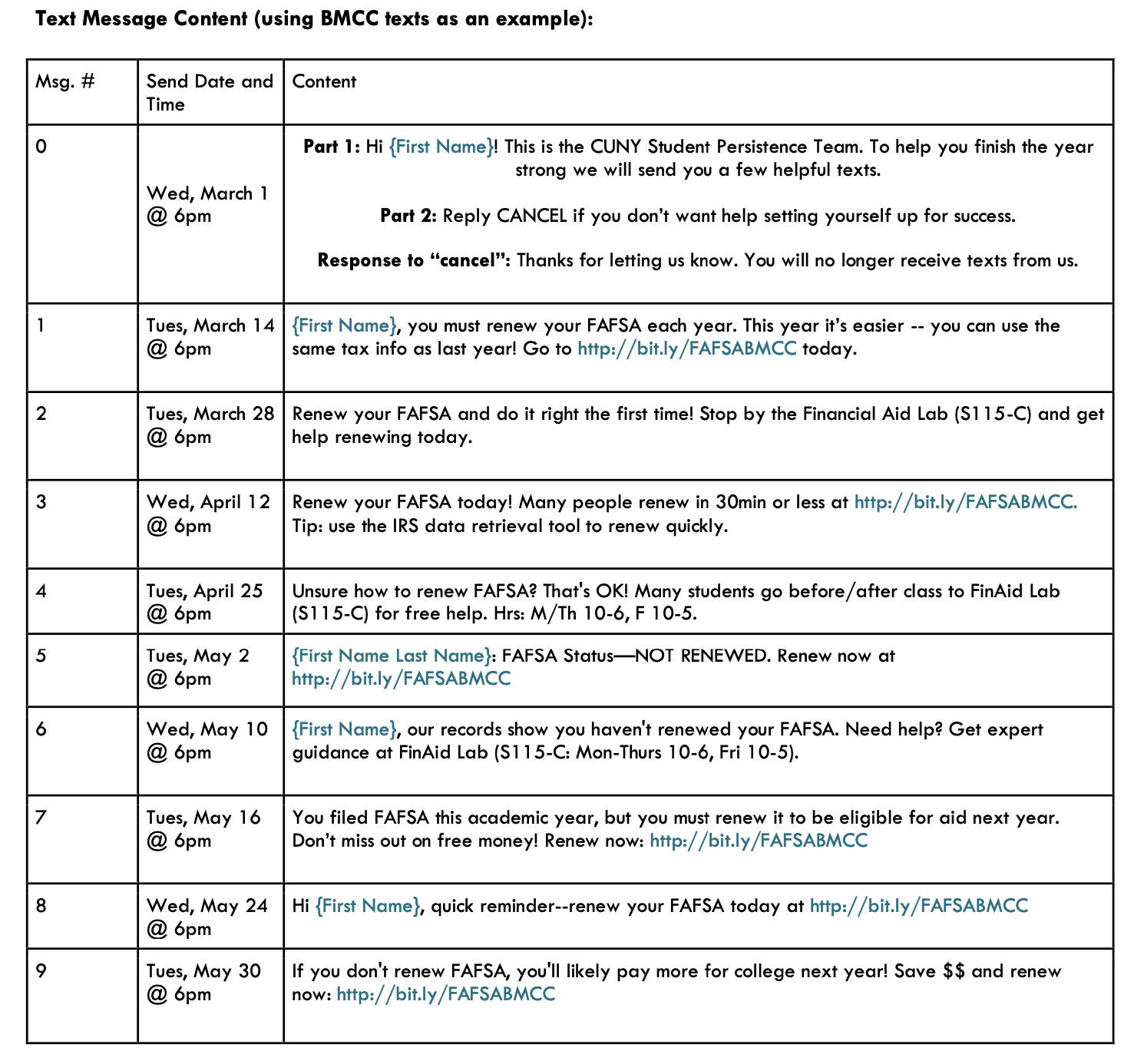}
        \caption{Example reminder text messages sent to students in the treatment group.}
        \label{fig:textmessages}
    \end{figure}

    This article analyzes data from a multi-year experiment conducted in New York City. Students were randomly assigned to receive behaviorally informed text and email reminders to renew their federal financial aid. The field experiment, run in 2017 and 2018 by ideas42 and the City University of New York (CUNY), aimed at increasing applications for Free Application for Federal Student Aid (FAFSA) financial support by the June 30 priority deadline. Students randomly assigned to the control group received only business-as-usual emails from the college. Students assigned to the treatment groups also received supplementary behavioral emails and text messages. These emails and text messages were designed to trigger loss aversion, plan-making, and commitment.\footnote{During 2017, the experiment had one treatment arm. The two treatment arms in the 2018 experiment differed in whether they used one-way texts or two-way texts that prompted students to respond. For this article, we pool the two treatment arms in the 2018 study.}
    \autoref{fig:textmessages} shows example text messages sent to students in the treatment group.
    
    The experiment involved matriculated students from CUNY community colleges. Eligible students were those who had not yet renewed FAFSA in February of the study year. The 2017 study sample includes 25,167 students from three community colleges, of which 50\% were randomly assigned to treatment. The 2018 sample includes 40,638 students from five community colleges, which were included in the intervention in two batches: an early batch of 30,627, of which 45\% were assigned to each of the two treatment arms and 10\% to control, and a late batch of 10,011 with a larger control group of 25\% and roughly equal treatment groups. We pool the late and early cohorts from 2018 for a total combined fraction of 86\% treated across the two treatment arms.
    Throughout our analysis of 2018 data, we adjust estimates for the varying propensity scores between early and late cohorts by inverse probability-weighted estimators.
    
    Our data include baseline demographic, academic, and administrative information about the community college students in the experiment. On average, students were 24 years old, with a considerable standard deviation of almost seven years. Our sample includes more women than men, with 57\% women in the 2017 experiment as well as 56\% (early schools) and 53\% (late schools) women in 2018. A plurality of students was Hispanic (52\% in 2017, 45\% in 2018), followed by Black non-Hispanic students who made up around a third of the student body in this study. Almost 20\% of students were enrolled part-time. Overall, we do not observe large imbalances between treatment and control groups; for nine baseline characteristics we tested across the 2017 and 2018 cohorts, only one variable, GPA for late 2018 schools, is significantly different between treatment and control at the 5\% level. Estimated propensity scores are concentrated around their batch-wise mean and balanced between the respective treatment and control groups. Details are available in Tables \ref{tbl:balance2017} and \ref{tbl:balance2018} in the appendix.
    
    Across the two randomized controlled trials, those who received the treatment interventions were on average $6.4\pm0.6$ (2017) and $12.1\pm0.7$ (2018) percentage points more likely to submit their FAFSA forms by the priority deadline, increasing early filing rates from 37\% to 43\% and 38\% to 50\%, respectively. These estimates, which are based on simple averages between treatment and control units within batches, are robust with respect to two alternatives, augmented inverse propensity weighted (AIPW) estimators that leverage covariate information to reduce noise (see \autoref{tbl:ate} in the appendix for details). The first of these estimators assumes constant propensity scores within batches, while the second corrects for possible imbalances by estimating the propensity score. Both are based on random forest estimation of the outcome model and propensity score.

\section{Estimating Treatment-Effect Heterogeneity}
\label{sec:heterogeneity}%
    
    Above, we noted a sizable average effect of behaviorally-informed reminders in this study of increasing FAFSA filing rates by the priority deadline by $6.4\pm0.6$ (2017) and $12.1\pm0.7$ (2018) percentage points. In this section, we use machine-learning tools to estimate treatment effects as a function of available individual covariates.
    We repeat the analysis for each study year (2017 and 2018) as well as for two sets of explanatory covariates -- first, those available at the time of randomization before the spring semester, and second, all information available halfway through the semester (which adds enrollment information and additional academic records) just before reminders were sent out.
    We report tests and diagnostics based on analysis of the full sample of 25,167 students in 2017 and  40,638 in 2018.

    Our goal in this section is to estimate conditional average treatment effects (CATEs), that is, the expected effect of an intervention adjusting for a person's observed covariates.
    Before describing how we estimate these treatment effects, we formally define the object of interest.
    We denote by $Y \in \{1,0\}$ the random variable that expresses whether a student has filed by the priority deadline ($Y{=}1$) or not ($Y=0$), and $T \in \{1,0\}$ for whether the student is in the treatment group ($T{=}1$) or not ($T{=}0$).
    We use standard potential-outcomes notation and write $Y(1)$ for the filing status a student would have had if they had been assigned to treatment, and $Y(0)$ for the filing status had they been assigned to control.
    The treatment of that student is then $Y(1) - Y(0)$, and we are interested in how this treatment effect varies with some baseline student characteristics $X$.
    Specifically, we aim to estimate the conditional average treatment effect
    \begin{align*}
      \tau(x) = \E[Y(1) - Y(0) | X{=}x]
    \end{align*}
    of students with characteristics $X{=}x$.
    When estimating $\tau(x)$, we face the challenge that for every student we only observe one of the potential outcomes $Y(1)$ and $Y(0)$, namely the realized filing decision $Y = Y(T)$ for their actual treatment status $T$.
    However, since treatment has been randomized, the realization of $Y(1)$, $Y(0)$ and $X$ are independent of $T$, so we can identify treatment effects from $\tau(x) = \E[Y| T{=}1, X{=}x] - \E[Y| T{=}0, X{=}x]$.
    In words, while we cannot compare outcomes within student, we can compare outcomes across similar students who have been treated or assigned to control, which yields the same conditional average effect when treatment is assigned randomly.
    
    We employ the causal forest algorithm, an instance of generalized random forests \citep{Athey2019-grf} that is specifically adapted to solve the causal-inference problem of estimating CATEs ($\tau(x)$) in settings like ours.
    The specific causal-forest implementation we employ estimates CATEs as follows.%
    \footnote{
        All of the main causal-forest estimates reported in our article are obtained via the \texttt{causal\_forest} function from the \texttt{grf} package \citep{grf} in R.  
    }
    First, the algorithm draws 3000 bootstrap samples from the training data.
    For each of the bootstrap samples, it forms a causal tree \citep{Athey2016-zi} by recursively partitioning the covariate space
    so that estimated average treatment effects vary as much as possible between subsets, up until the size of the subsets crosses a minimum threshold.
    Second, the causal forest aggregates all these 3000 causal trees into weights $\hat{w}_i(x)$.
    For each training observation $(Y_i,T_i,X_i)$, the weight $\hat{w}_i(x)$ indicates how often a target vector $x$ of covariates shares the same cell as the training observation $X_i$ across the partitions from different trees.
    Third, to estimate the CATE $\tau(x)$ for a particular vector $x$ of covariates, the causal forest returns a weighted average treatment effect
    \begin{align*}
      \hat{\tau}(x) = \frac{\sum_{T_i=1} \hat{w}_i(x) Y_i}{\sum_{T_i=1} \hat{w}_i(x)} - \frac{\sum_{T_i=0} \hat{w}_i(x) Y_i}{\sum_{T_i=0} \hat{w}_i(x)}
    \end{align*}
    of nearby observations $(Y_i,T_i,X_i)$.
    Moreover, when aggregating trees into weights $\hat{w}_i(x)$, we use a careful rule for sample splitting \citep[``honesty,''][]{Wager2018-pe} that ensures that $Y_i$ is not used in the construction of $\hat{w}_i(x)$, which guarantees that the resulting estimates are consistent and asymptotically normal. 

    \begin{figure}
      \centering
      \begin{subfigure}[t]{0.5\textwidth}
        \centering
        \includegraphics[width=.7\textwidth]{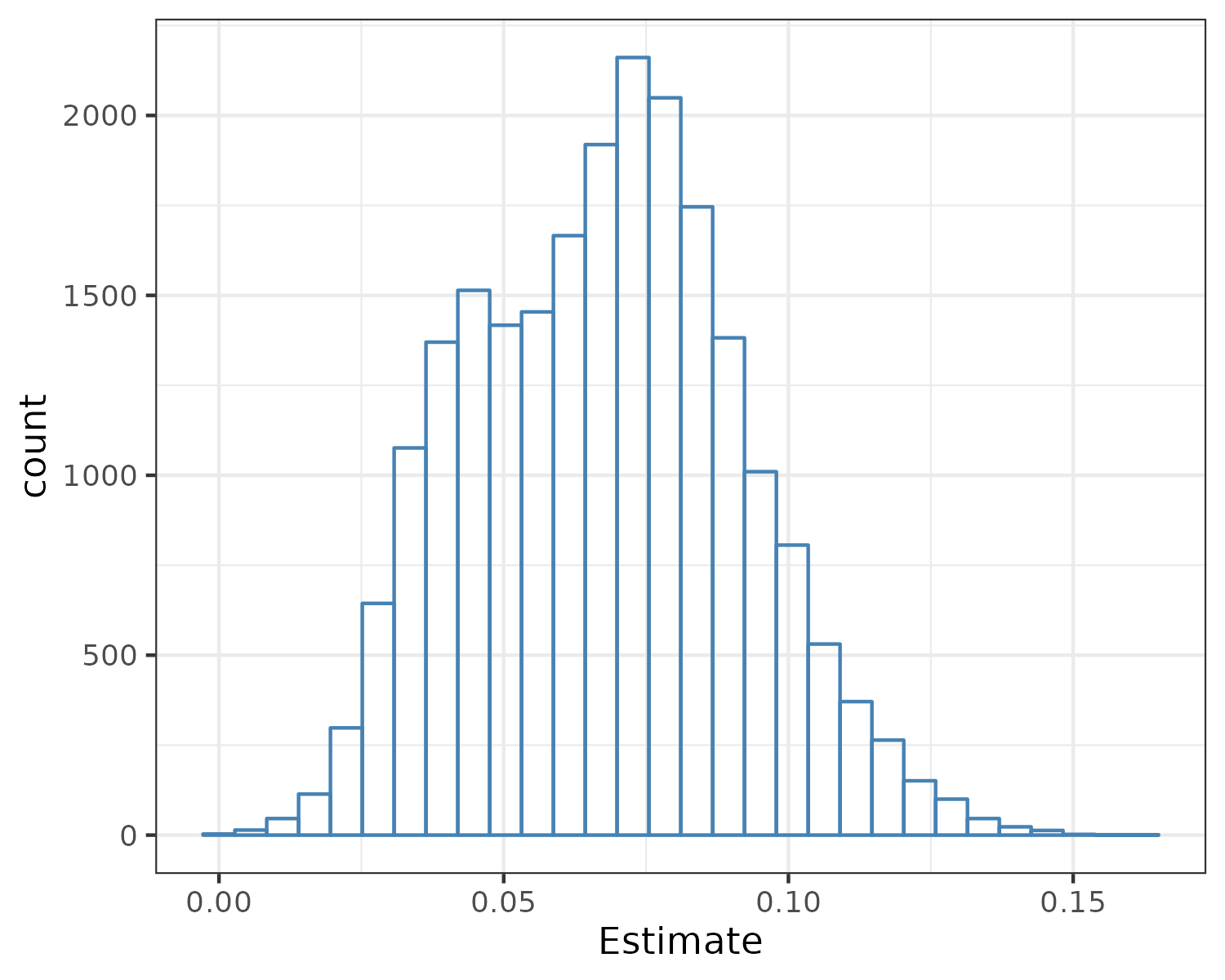}
        \caption{2017, early covariates}
      \end{subfigure}%
      ~
      \begin{subfigure}[t]{0.5\textwidth}
          \centering
          \includegraphics[width=.7\textwidth]{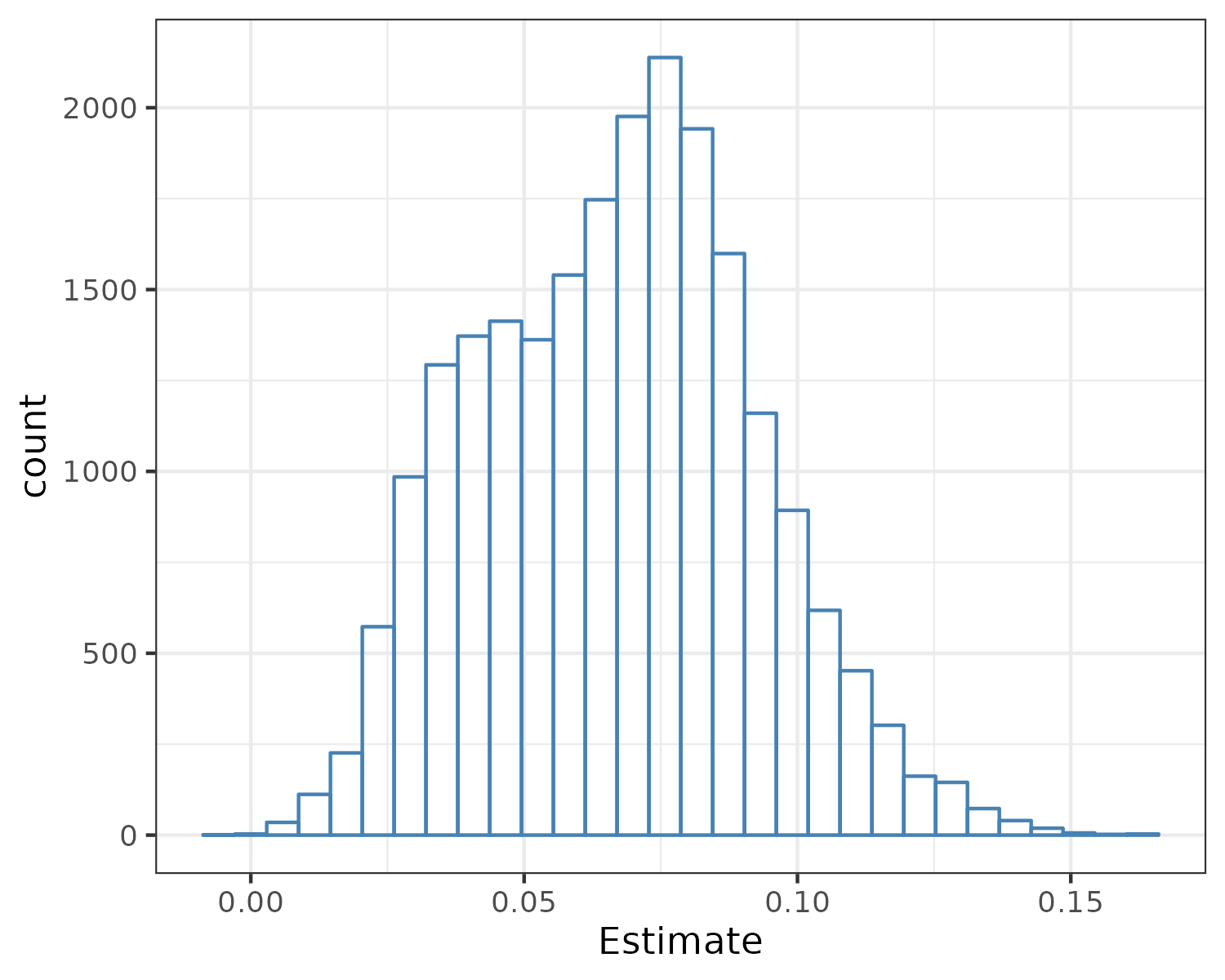}
          \caption{2017, late covariates}
      \end{subfigure}
      \bigskip
      \\
      \begin{subfigure}[t]{0.5\textwidth}
        \centering
        \includegraphics[width=.7\textwidth]{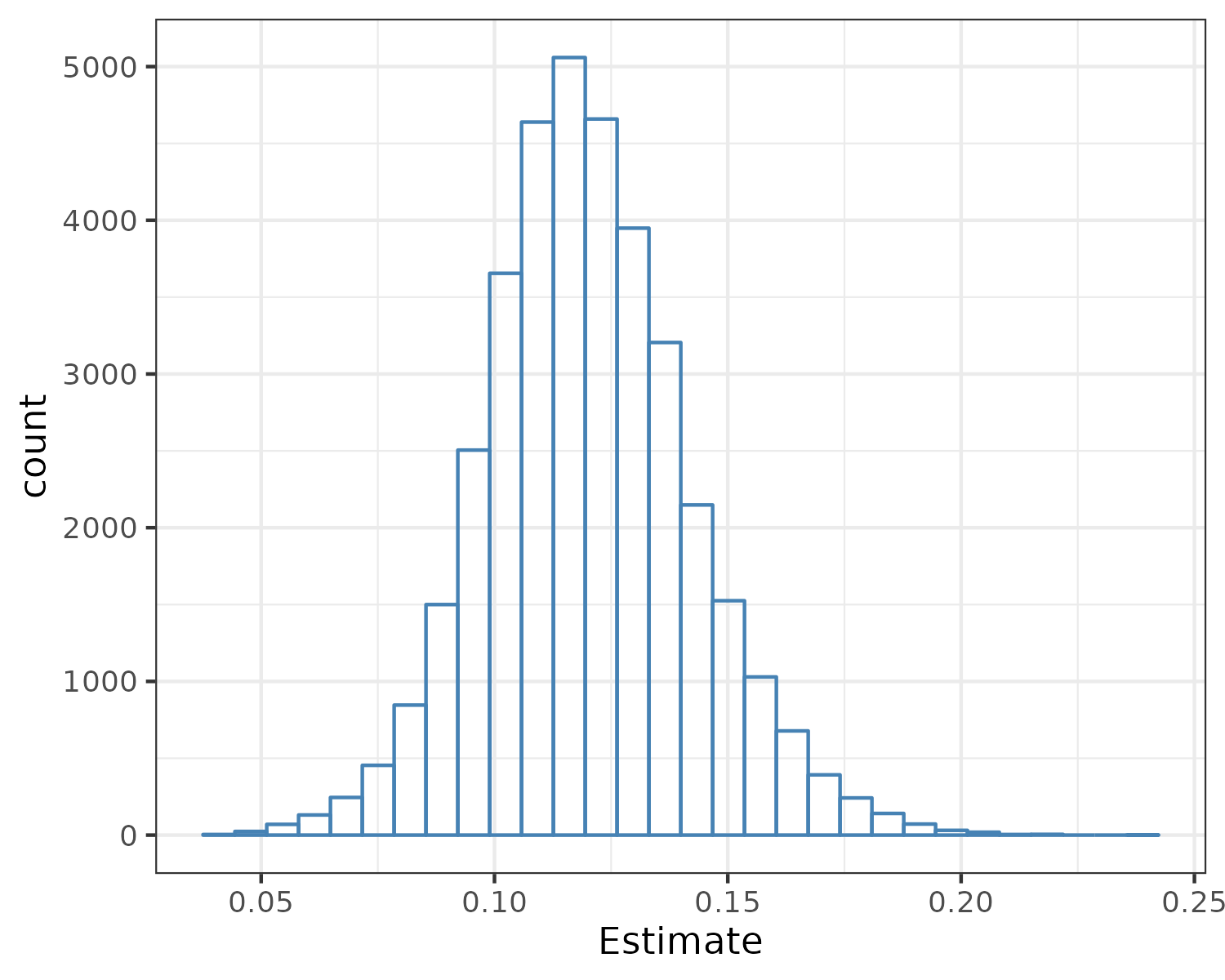}
        \caption{2018, early covariates}
      \end{subfigure}%
      ~
      \begin{subfigure}[t]{0.5\textwidth}
          \centering
          \includegraphics[width=.7\textwidth]{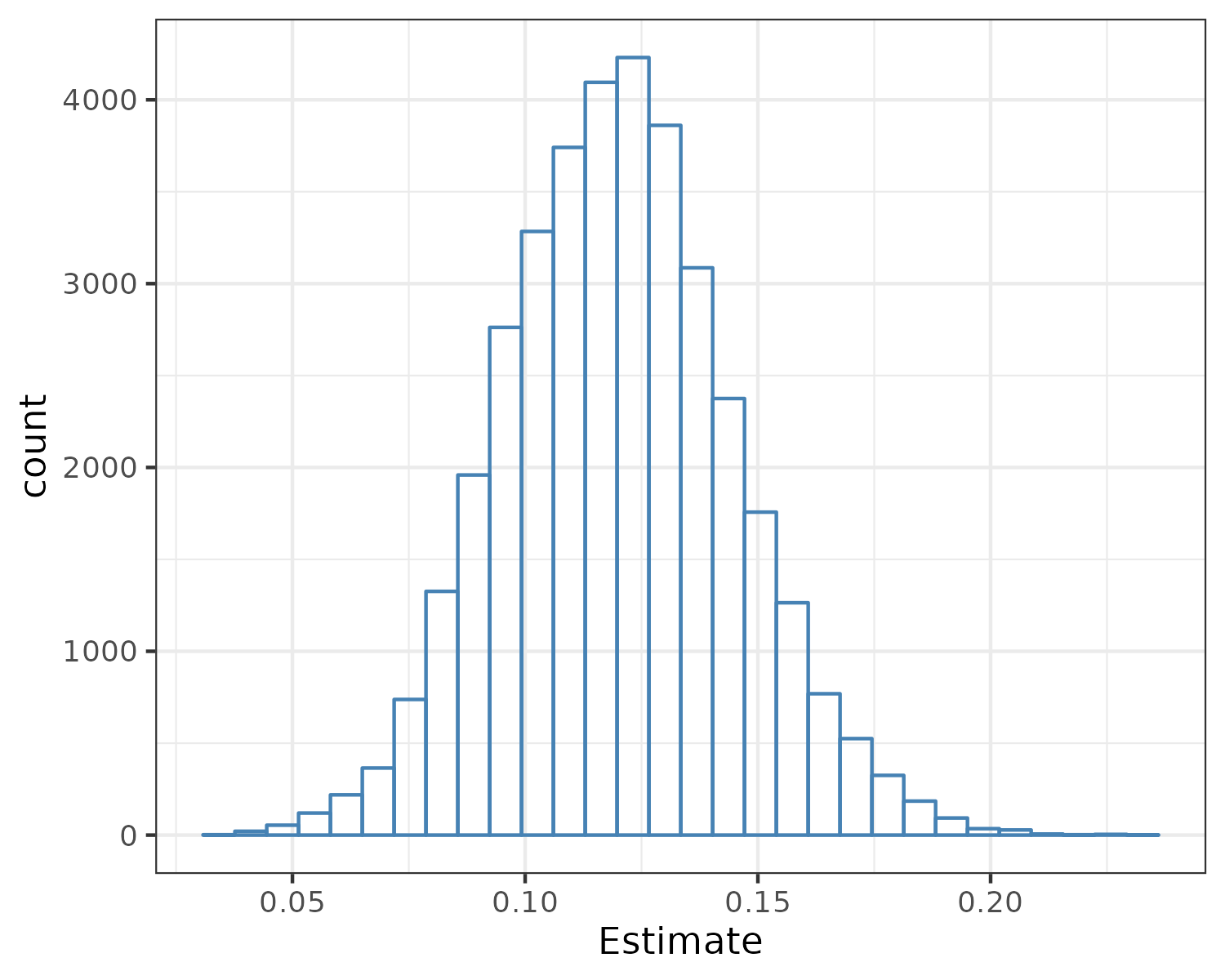}
          \caption{2018, late covariates}
      \end{subfigure}      
      
      \caption{Histograms of estimated treatment effects across years and set of covariates, using honest estimates from the causal forest method.}
      \label{fig:histograms}
    \end{figure}

    \autoref{fig:histograms} shows the distribution of estimated treatment effects across the two study years we consider (2017, top, and 2018, bottom), differentiating between covariates available before the beginning of the semester (``early covariates'', left) and those only available later, but still before the interventions (``late covariates'', right).
    The histogram shows treatment effects ranging from around 0 to 15 percentage points in the 2017 data, and from around 5 to 20 percentage points for 2018. Heterogeneity seems to be similar between early and late covariates for 2017.
    In 2018, treatment effect estimates using the late covariates are somewhat more spread out than those using information from before the start of the semester only.
    We do not find any evidence for negative treatment effects, so reminders are unlikely to have dissuaded any students from filing by the priority deadline.

    Despite the theoretical guarantees for the honest treatment-effect estimates, in practice, the signal-to-noise ratio is often such that a very large sample is required for the theory to be a useful guide.  For more realistic sample sizes, estimates $\hat{\tau}(x)$ of CATEs $\tau(x)$ are often miscalibrated. For a particular covariate vector $x$, the estimates $\hat{\tau}(x)$ may be biased towards the overall average treatment effect $\tau = \E[Y(1) - Y(0)]$, as there will not, in general, be enough observations with similar covariate vectors to a particular target.  On the other hand, in a setting with very little true heterogeneity relative to the noise, sampling variation will still induce a distribution of estimated treatment effects over different covariate vectors, potentially overstating heterogeneity. We do not rely on the theoretical guarantees about the estimator in this article; rather, we use our estimates of CATEs as an input to other analyses, such as constructing policies, and evaluate those policies using model-free methods.

    In order to assess the importance of treatment effect heterogeneity, we present additional tests based on \cite{Chernozhukov2019-ti}.
    For these tests, all of the treatment effect estimates are based on ten-fold cross-fitting in order to avoid biases.
    Specifically, we randomly divide the sample into ten folds, and for an observation $i$ in a given fold $k$ estimate their CATE $\tau(X_i)$ by $\hat{\tau}_{-k}(X_i)$ using a causal forest $\hat{\tau}_{-k}$ fitted only on the other folds.
    We first provide a calibration-based analysis of the estimates of heterogeneous treatment effects.
    Across both years and both sets of covariates, those available earlier and those later,
    we report results in \autoref{tbl:calibration} of a calibration regression of actual outcomes $Y_i$ on treatment-effect estimates $\hat{\tau}(X_i)$ interacted with normalized treatment $T_i - p$, where $p$ is the overall probability of being treated.%
    \footnote{In order to avoid biases when heterogeneity is limited, we run the calibration regression separately by fold and aggregate the resulting coefficient and standard error estimates.}
    If treatment effects were perfectly calibrated, we would expect the slope estimate to be close to one, while it would be close to zero if treatment effect estimates are purely spurious.
    Our estimates in \autoref{tbl:calibration} are significantly larger than zero (at the 5\% significance level), providing evidence that our estimates indeed do capture treatment-effect heterogeneity.
    For robustness, we present results based on two separate approaches to estimating heterogeneous treatment effects, namely using the fact that we know that the propensity score is constant across groups of schools (``known'') or estimating the propensity score from the data to be robust against possible issues with randomization (``AIPW'').
    Although we find evidence of heterogeneity across these specifications, the model is not perfectly calibrated (the slope coefficients are substantially less than one), and as such, we cannot assume that the magnitudes of our CATE estimates $\hat{\tau}(x)$ are unbiased for $\tau(x)$.
        
    \begin{table}[h]
    \centering
    \begin{tabular}{l l l r r r}
        \toprule
        Year & Covariates & Propensity score & Slope estimate & SE & $p$-value \\
        \midrule
        2017 & early & known & 0.76171 & 0.23041 & 0.00047 \\
         &  & AIPW & 0.74504 & 0.22908 & 0.00057 \\
         \cmidrule(lr){2-6}
         & late & known & 0.78549 & 0.21804 & 0.00016 \\
         &  & AIPW & 0.62928 & 0.21639 & 0.00182 \\
         \midrule
        2018 & early & known & 0.55267 & 0.29159 & 0.02902 \\
         &  & AIPW & 0.65692 & 0.28802 & 0.01128 \\
         \cmidrule(lr){2-6}
         & late & known & 0.75689 & 0.25451 & 0.00147 \\
         &  & AIPW & 0.67651 & 0.25451 & 0.00393 \\
        \bottomrule
    \end{tabular}

    \caption{
      Slope coefficient estimates for the calibration regression of actual outcomes on treatment-effect estimates interacted with normalized treatment following \citet{Chernozhukov2019-ti}.}
    \label{tbl:calibration}

\end{table}
        
    Even if CATE estimates are miscalibrated and thus cannot be taken at face value, they may still be reliable for assessing which units have higher treatment effects than others. While it is impossible to assess the accuracy of a treatment effect estimate for a single observation, since we can only observe the outcome $Y$ for an individual with one of the two possible treatment assignments $T \in \{1,0\}$, we can construct an unbiased estimate of the average treatment effect $\E[Y(1) - Y(0)|G]$ for a sufficiently large group of individuals, where the group $G = g(X)$ is defined by covariates $X$.
    Following the ``Sorted Group Average Treatment Effects (GATES)'' methodology of \cite{chernozhukov2018generic}, we proceed by creating such groups based on our out-of-fold CATE estimates. Specifically, within each fold $k$, we then divide the covariate space into four groups $G \in \{1,2,3, 4\}$ based on the quartile of estimated treatment effects $\hat{\tau}_{-k}(x)$. Finally, we estimate average treatment effects $\E[Y_i(1) - Y_i(0)|G=g]$ for each of the four groups $g$ by the average difference between treated and control outcomes within that group, combining groups across all folds.  These estimates are unbiased estimates of the average treatment effect for the groups (recalling groups are defined by the covariates), since the outcomes of the units in fold $k$ were not used in any part of the process of assigning units to groups.
    The resulting estimates are model-free, unbiased estimates of average treatment effects in each group.
    
\begin{figure}
      \centering
      \begin{subfigure}[t]{0.5\textwidth}
        \centering
        \includegraphics[width=\textwidth]{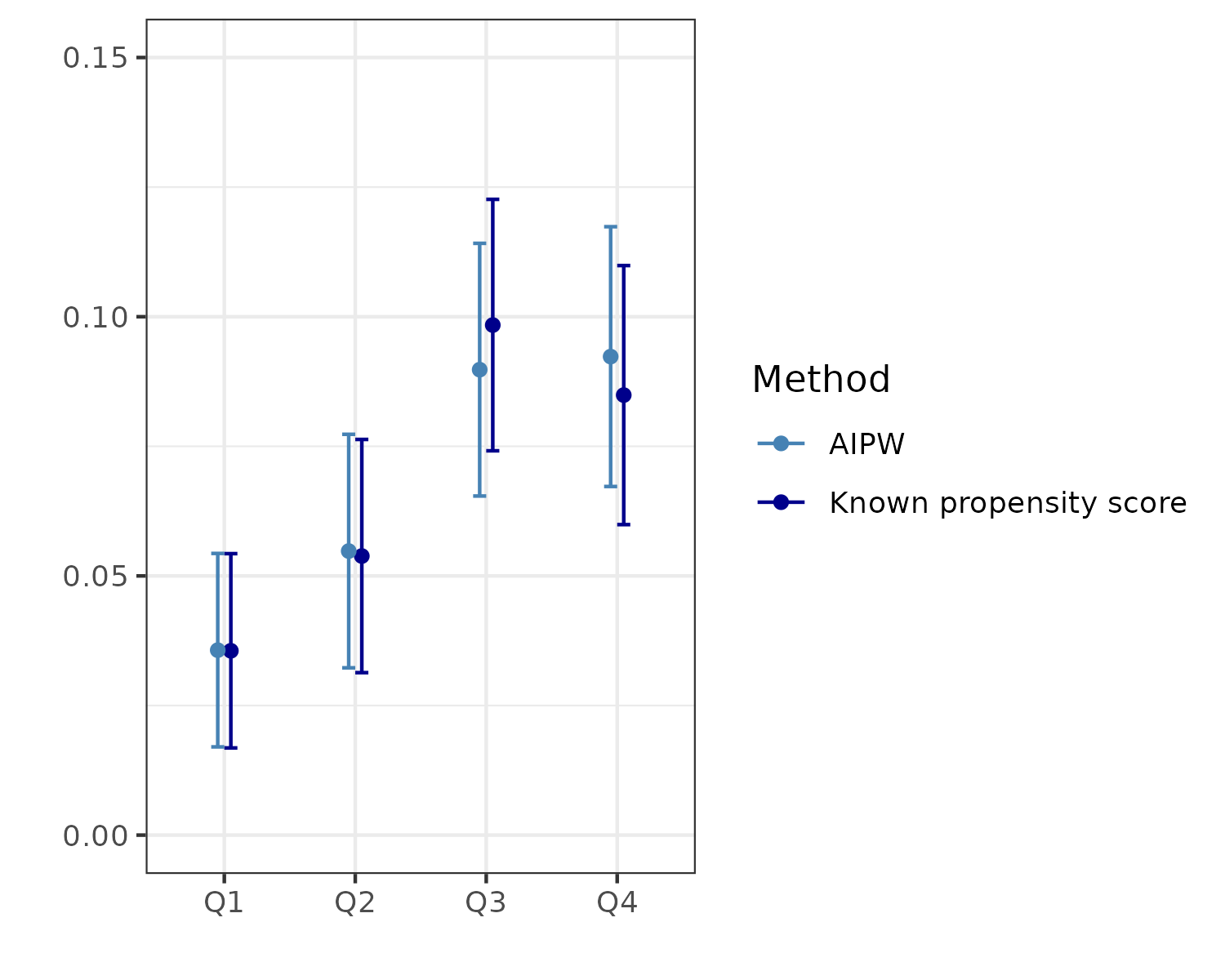}
        \caption{2017, early covariates}
      \end{subfigure}%
      ~
      \begin{subfigure}[t]{0.5\textwidth}
          \centering
          \includegraphics[width=\textwidth]{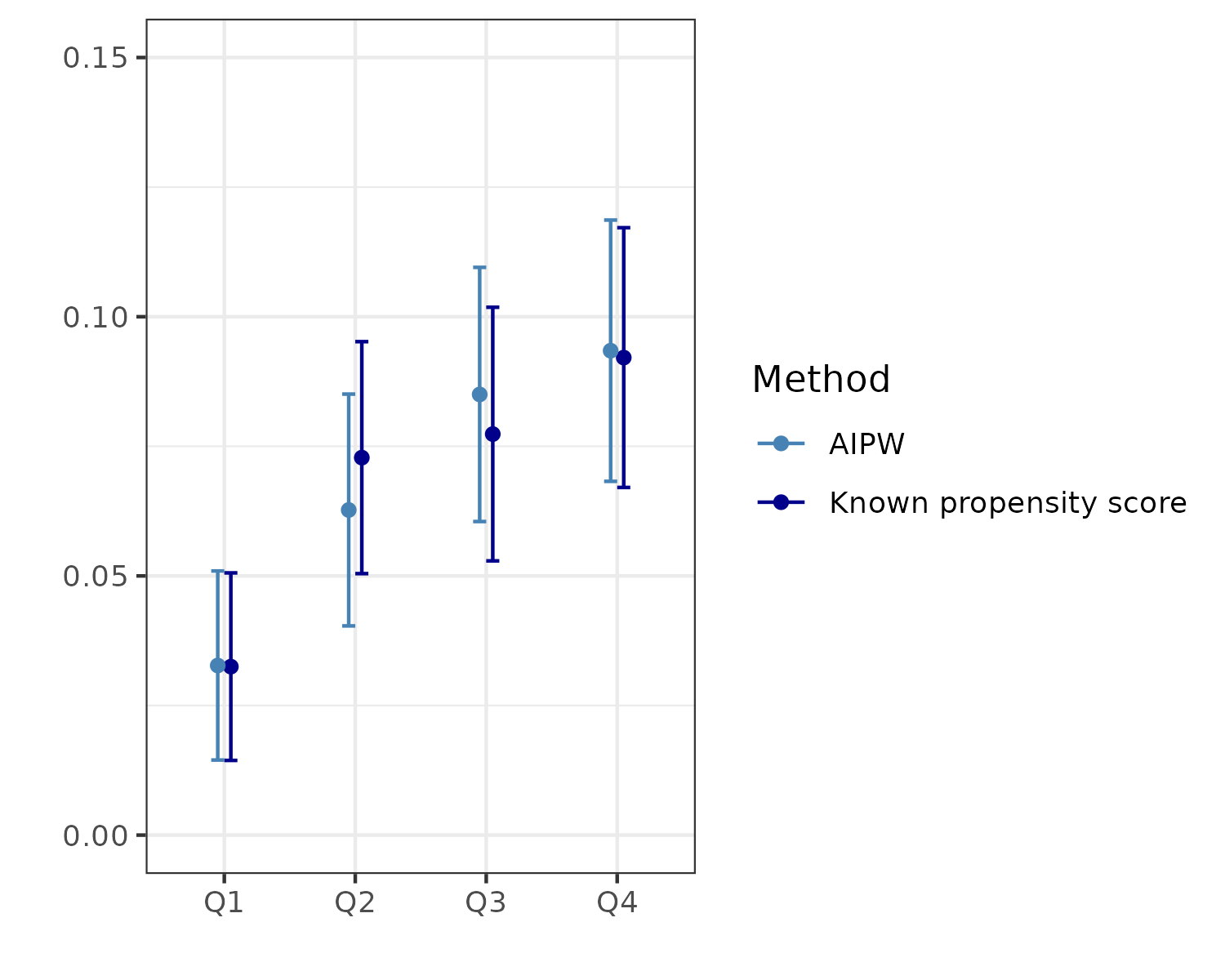}
          \caption{2017, late covariates}
      \end{subfigure}
      \bigskip
      \\
      \begin{subfigure}[t]{0.5\textwidth}
        \centering
        \includegraphics[width=\textwidth]{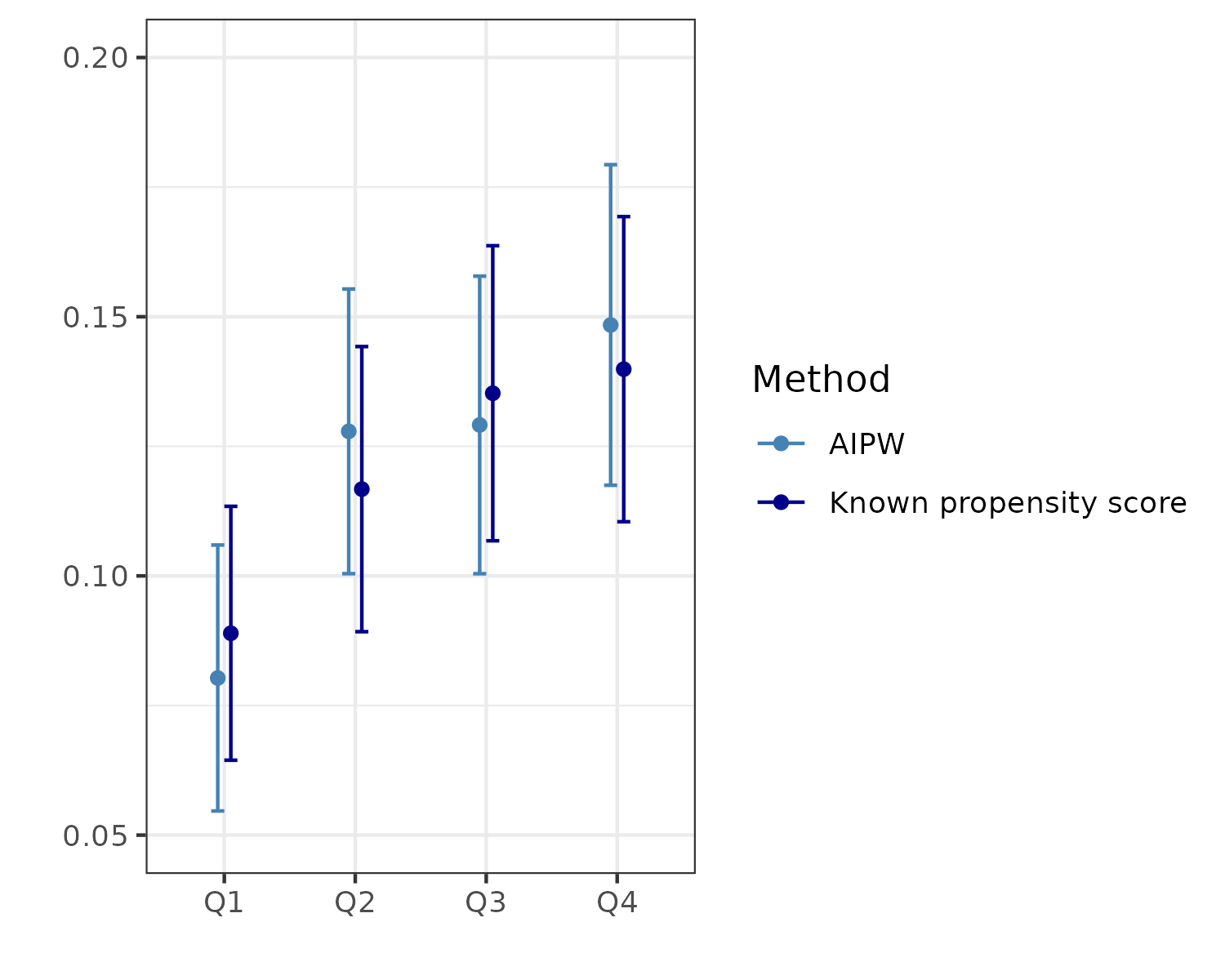}
        \caption{2018, early covariates}
      \end{subfigure}%
      ~
      \begin{subfigure}[t]{0.5\textwidth}
          \centering
          \includegraphics[width=\textwidth]{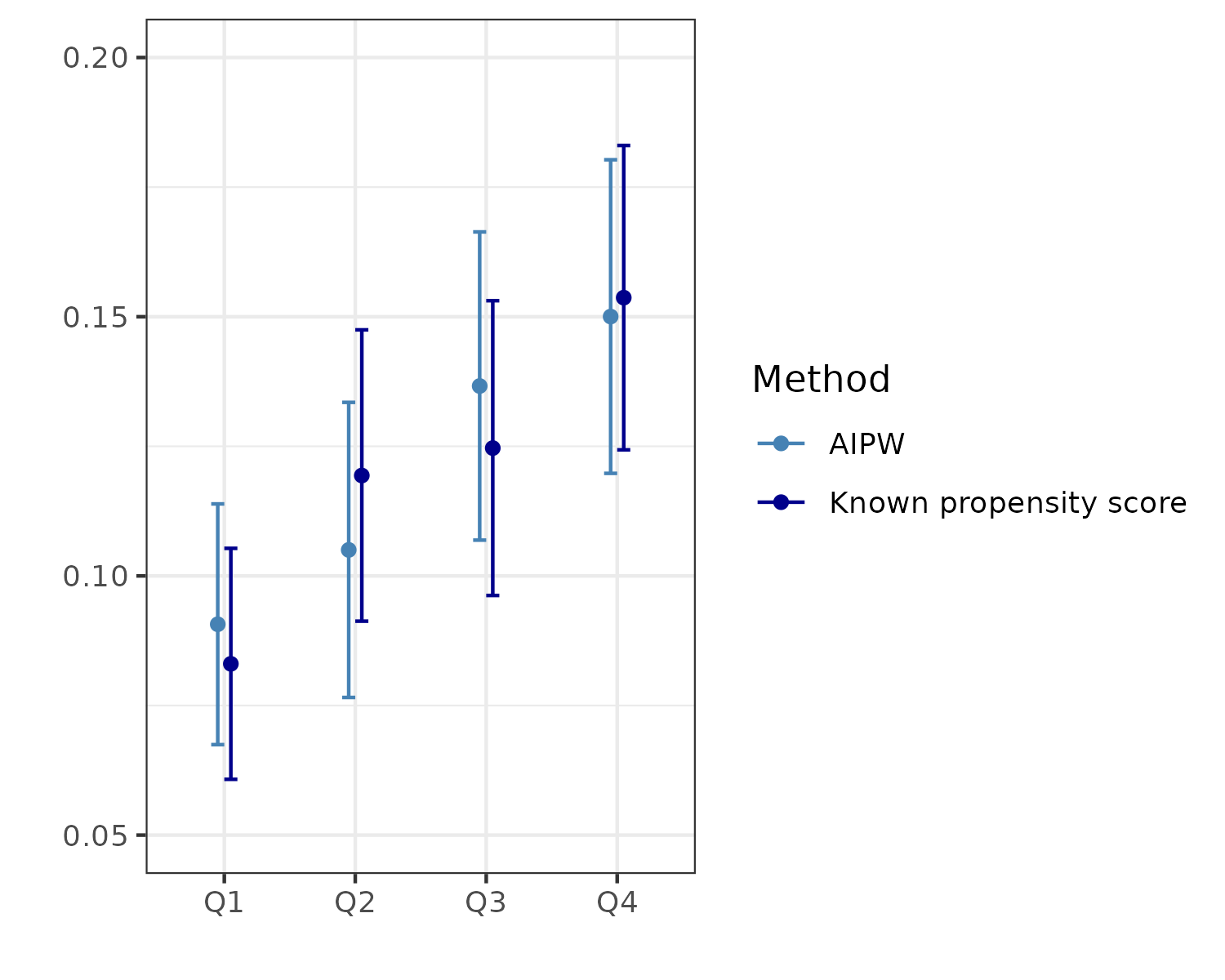}
          \caption{2018, late covariates}
      \end{subfigure}      
      
      \caption{Average treatment effects by quartiles of estimated treatment effects. The $x$-axis divides the sample into quartiles of predicted cross-fitted treatment effects using ten folds. The $y$-axis plots groups and estimates based on an augmented inverse-propensity weighted estimator using an estimated (``AIPW'') or the known propensity score (``Known propensity score''), along with a 95\% confidence interval.}
      \label{fig:quartiles}
    \end{figure}

    Our resulting estimates of treatment effects are noisy, and the heterogeneity across groups is moderate. \autoref{fig:quartiles} plots unbiased estimates of the group-wise average treatment effect across quartiles of estimated treatment effects for the 2017 (top) and 2018 (bottom) cohort, differentiating between covariates available before the beginning of the semester (left) and those only available later, but still before the interventions (right).
    Both graphs document that treatment effects $\tau(x)$ vary across the distribution, although the forest-based estimates $\hat{\tau}(x)$ differ from unbiased estimates and even produce non-monotonic rankings in one instance.
    Across both years and irrespective of the set of covariates, we find statistically significant heterogeneity in treatment effects.
    For 2017, the average treatment effect in the two bottom quartiles (early covariates) or in the bottom quartile (late covariates) is significantly below that of the remaining sample, with a difference of around five percentage points separating the respective group averages.
    For 2018, the average outcome in the bottom quartile is around 7 percentage points lower than the average among the rest.
    Effects for the set of covariates available earlier are not considerably noisier.
    The differences across quartile AIPW estimates are reported in \autoref{tbl:quartiles} in the appendix along with standard error estimates.
    
    We next inspect which variables drive treatment effect heterogeneity.
    To do so, we report a simple variable-importance measure based on the causal forest used for estimating heterogeneous treatment effects.
    This variable-importance measure counts how often each variable is used in a split within the causal forest, with higher weight given to splits higher up in each tree.
    It is thus a way of assessing the contribution of a variable to estimating heterogeneous treatment effects.
    The top variables identified in this way across both years and early and late data include age, GPA, and how many credits a student attempted and actually earned.
    While suggestive of drivers of heterogeneity, this measure tells us neither the direction nor the degree to which treatment effects are affected.

    Measuring the importance of a variable in terms of the number of times it is used for splitting may understate the importance of variables with only a few levels.
    Some binary variables, for example, may have a large impact on treatment effects, but some partitions may split on them only once.
    We, therefore, also inspect which \emph{binary} variables the causal forest splits consistently on. Through this analysis, we identify enrollment as another potential driver of heterogeneity.
    There is no enrollment restriction on who can apply for FAFSA, and students could unenroll in a given year yet remain eligible to apply for FAFSA for the subsequent academic year.
    However, students who drop out midyear may not only be harder for administrators to track but they may also be less likely to re-enroll for the subsequent academic year. For the randomized experiment, unenrolled students were still eligible for behavioral nudges.
    Using the later set of covariates that become available during the semester, we can identify students who were enrolled at the start of the academic year yet dropped out by the time of the behavioral nudge treatment.
  
\begin{figure}
      \centering
      \begin{subfigure}[t]{0.5\textwidth}
        \centering
        \includegraphics[width=\textwidth]{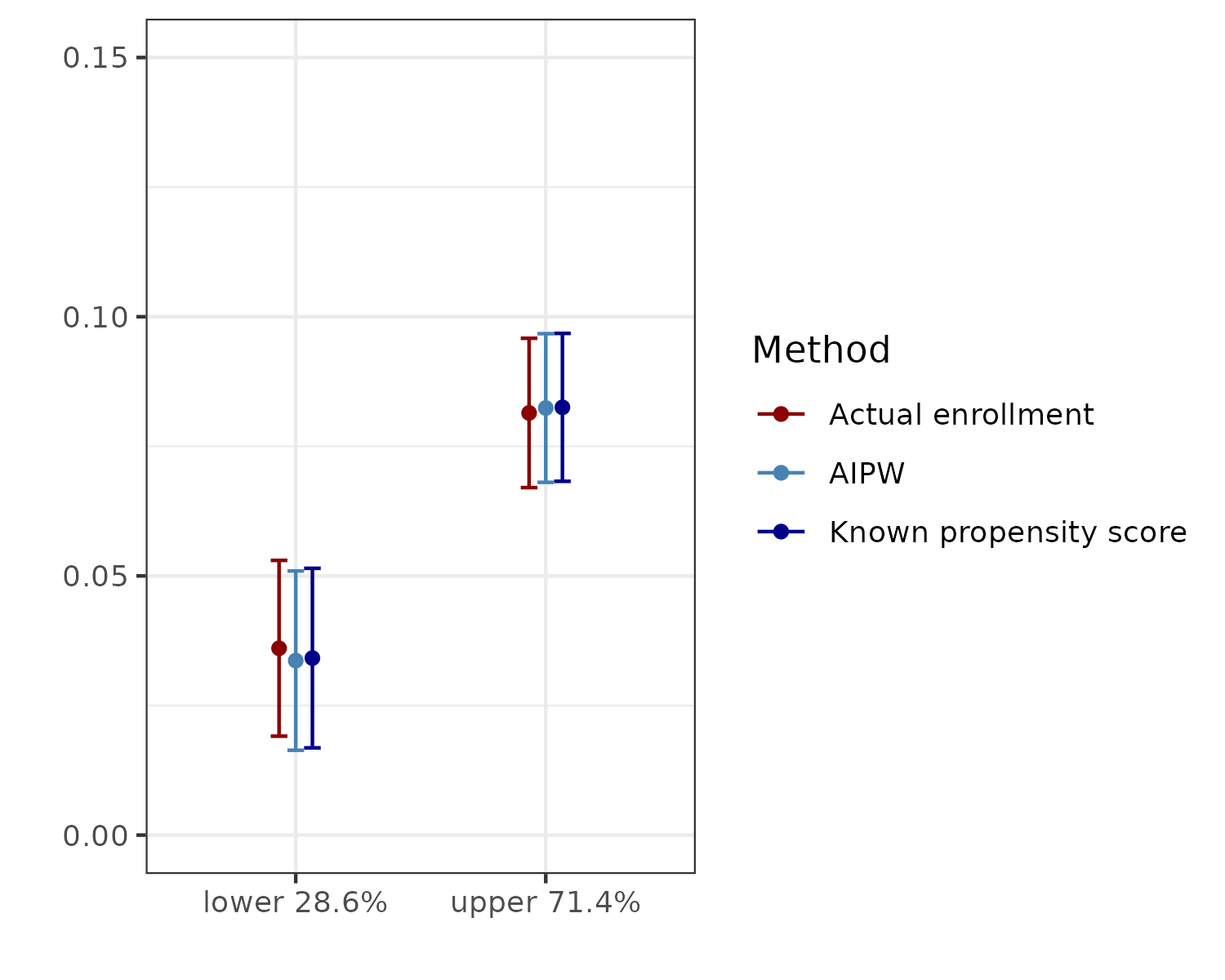}
        \caption{2017}
      \end{subfigure}%
      ~ 
      \begin{subfigure}[t]{0.5\textwidth}
          \centering
          \includegraphics[width=\textwidth]{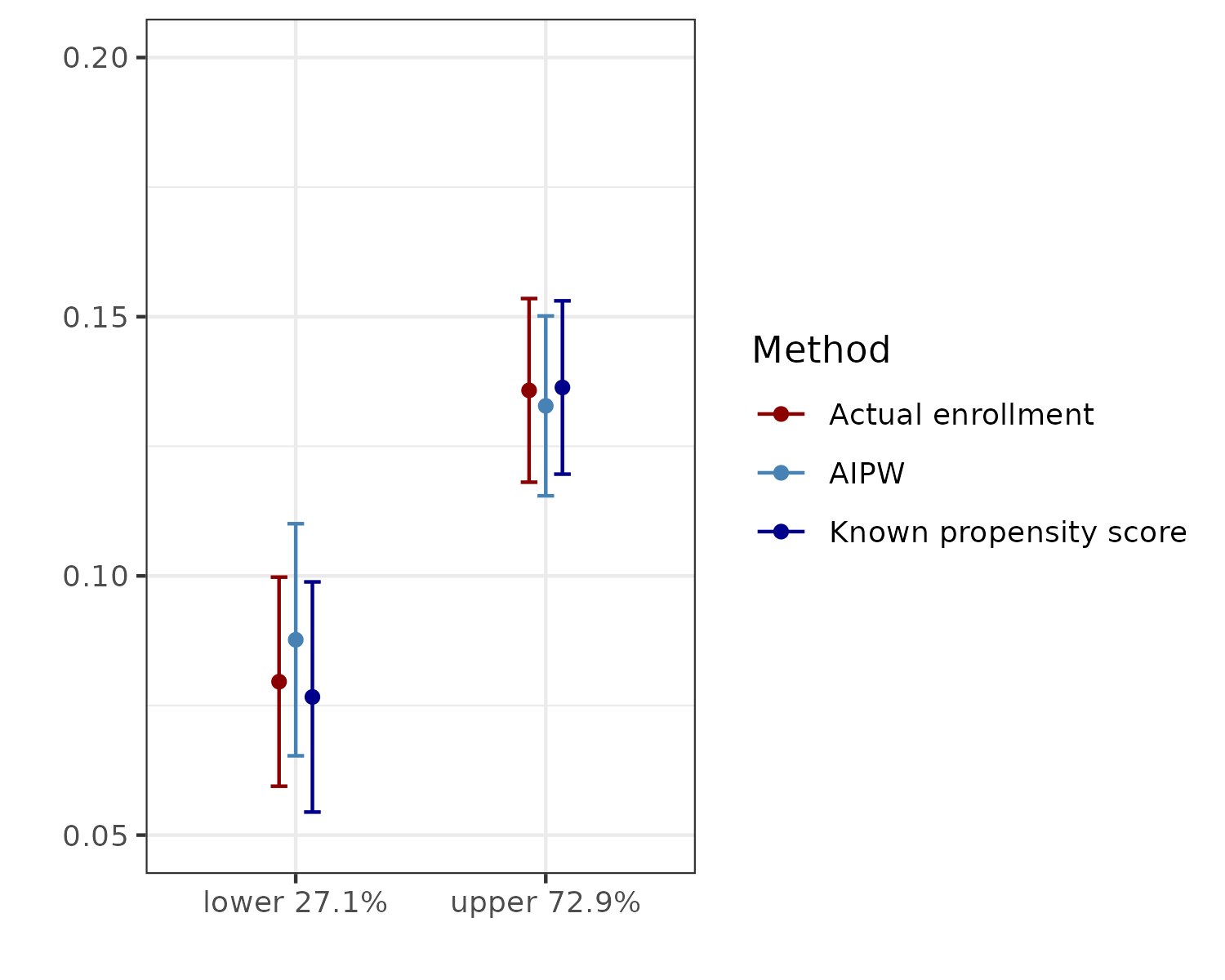}
          \caption{2018}
      \end{subfigure}
      \caption{Average treatment effects by enrollment status (red) and by whether predicted cross-fitted treatment effects are below or above the quantile corresponding to the proportion of enrolled students (blue), using all data up to the start of the intervention. The $y$-axis plots a simple difference-in-averages for actual enrollment as well as two estimators based on augmented inverse-propensity weighted estimators using the estimated (``AIPW'') or the known propensity score (``Known propensity score'') constant and with estimated propensity scores, along with 95\% confidence intervals.}
      \label{fig:enrolled_vs_hte}
    \end{figure}

    We find that enrollment status, once it becomes available, is indeed highly predictive of treatment effects. 
    Indeed, if we calculate average treatment effects across enrollment status, we find that it partitions treatment effects just as well as if we had formed two groups of students based on our treatment-effect estimates (\autoref{fig:enrolled_vs_hte}), holding the size of the two groups fixed.
    However, there seems to be additional heterogeneity in treatment effects even among those enrolled.
    Comparing effects across the top three quartiles on the right side of \autoref{fig:quartiles} to the average effect of enrolled students in \autoref{fig:enrolled_vs_hte} suggests that the highest quartile of estimated effects has a higher treatment effect than enrolled students overall, although the difference is noisy.
            
    We note that the results on enrollment as an important treatment-effect moderator are intuitive but not mechanical. Indeed, reminders affect the filing of both students who are enrolled and who are not enrolled at the time enrollment is measured for the spring semester. Since students may drop out and re-enroll, providing reminders for unenrolled students remains effective. Rather, we consider it interesting that there seems to be only minor additional heterogeneity, making enrollment a powerful proxy for the effectiveness of reminders already by itself.
    
\section{Causal vs Predictive Targeting}
\label{sec:targeting}
    
    In the previous section, we found that there is heterogeneity in students' responses to the reminder.
    We now ask how we can use this predictable variation in the response to reminders to improve their targeting.
    We start this section by leveraging the heterogeneous-treatment effect estimates $\hat{\tau}(x)$ we constructed above to target reminders.
    We then compare the performance of this policy to a policy purely based on predictions of the probability of filing at baseline.
    Throughout, we focus on the 2017 cohort of students, where we have a large control group available. 

  One finding of our analysis of heterogeneous treatment effects was that a good part of the heterogeneity in conditional average treatment effects $\tau(x)$ can be predicted by enrollment once we know whether a student has dropped out. Hence, one natural way of targeting reminders that we investigate below is to send them only to enrolled students.
  Notwithstanding this finding, there are two reasons why targeting based on individual covariates is still meaningful:
  First, we ask whether we could have used predictions of heterogeneous treatment effects to select students to send behaviorally-informed reminders before the beginning of the semester, when enrollment information was not yet available.
  Second, even among those students who are known to have enrolled, there is still some heterogeneity in the response to reminders.

    \subsection{Targeting based on non-parametric treatment effect estimates}

    We start by evaluating the value of the estimates $\hat{\tau}(x)$ of heterogeneous treatment effects from \autoref{sec:heterogeneity} for targeting reminders better.
    That is, rather than estimating the quality of our predictions $\hat{\tau}(x)$ in terms of how well they estimate true treatment effects $\tau(x)$,
    we quantify which proportion of total gain from the reminders we could have realized from targeting only a selected fraction of students.
    Specifically, we consider assignment policies
    \begin{align}\label{eq:assignt}
      \hat{\pi}^{\text{causal}}_t(x) = \mathbbm{1}(\hat{\tau}(x) \geq t)
    \end{align}
    that assigning all students to treatment whose estimated treatment effect is above some threshold $t$.
    
\begin{figure}
      \centering
      \begin{subfigure}[t]{0.75\textwidth}
        \centering
        \includegraphics[width=\textwidth]{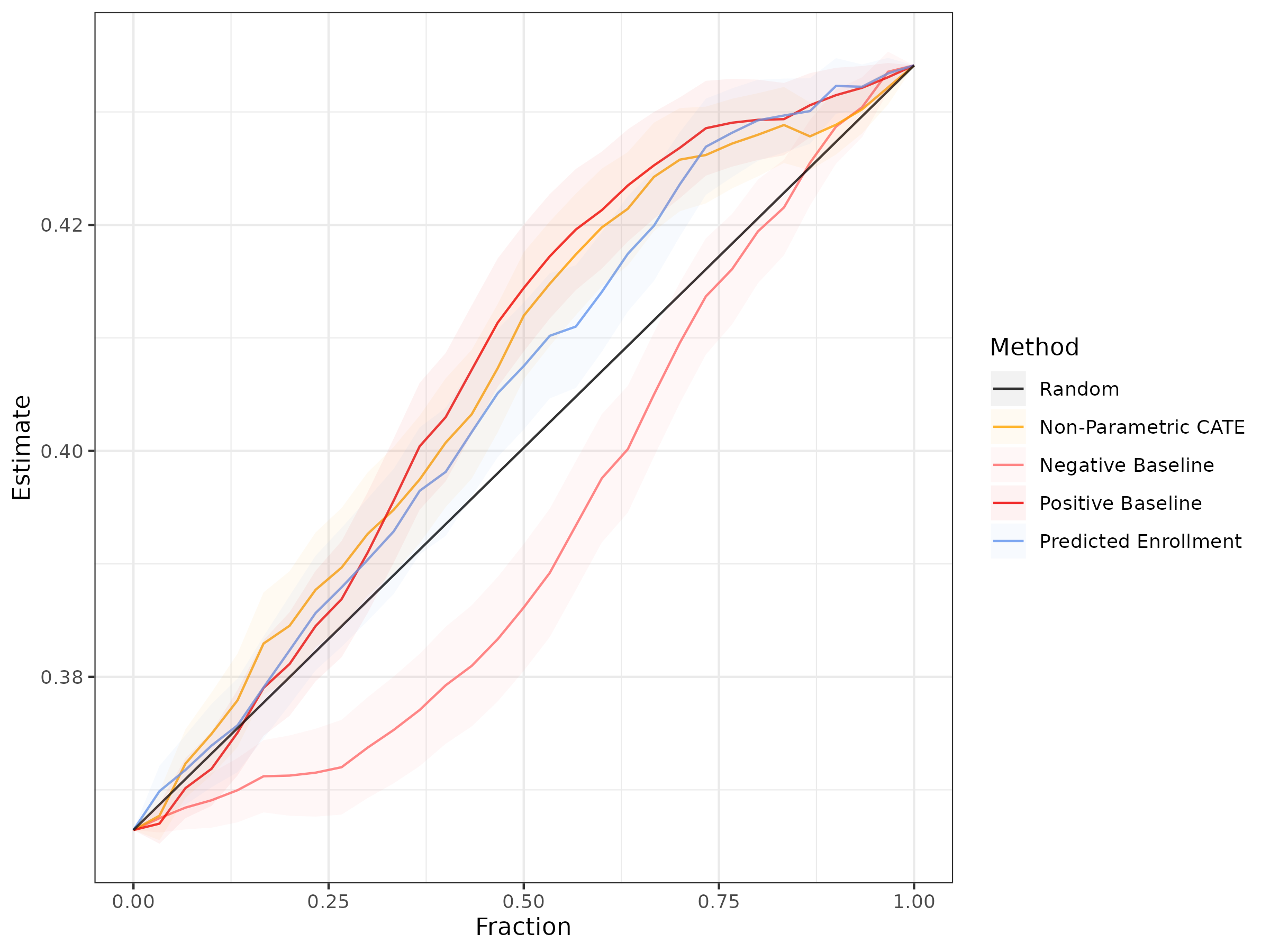}
        \caption{2017, early covariates}
      \end{subfigure}%

      \begin{subfigure}[t]{0.75\textwidth}
          \centering
          \includegraphics[width=\textwidth]{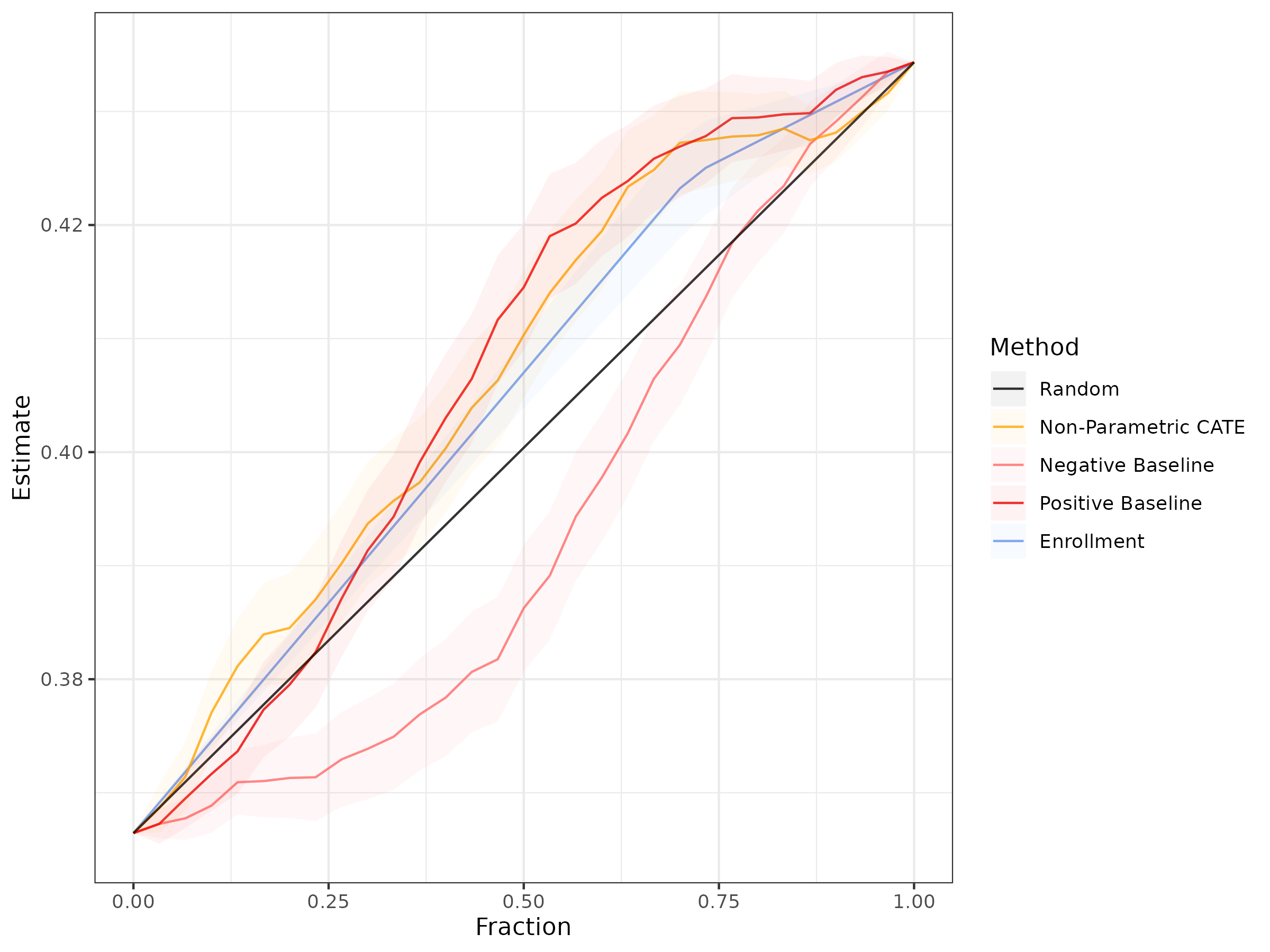}
          \caption{2017, late covariates}
      \end{subfigure}
      \caption{Total estimated FAFSA renewal rate ($y$-axis) by targeting a given fraction ($x$-axis) of students according to different cross-fitted predictions in the 2017 data, including targeting by estimated treatment effects using the causal forest (``Non-Parametric CATE''),
      by a random-forest prediction of outcomes absent treatment (``Negative Baseline'' for low baseline treated first, and ``Positive Baseline'' the reverse),
        and by predicted or actual enrollment (``Predicted Enrollment'' and ``Enrollment'').
      Shown are model-free unbiased augmented inverse propensity weighted estimates with 95\% confidence intervals that represent the point-wise uncertainty of the difference in renewal rate relative to the random policy (``Random'') that assigns the same fraction to treatment, with details provided in \autoref{apx:inference}.
      }
      \label{fig:policy_V1}
    \end{figure}

    The results of our assignment exercise are summarized in \autoref{fig:policy_V1}, separately for leveraging early covariates (those available before the beginning of the semester) and late covariates (those available by the time the reminders are sent) for the 2017 cohort of the FAFSA experiment.
    We plot (under the counterfactual policy \eqref{eq:assignt}) the estimated fraction of students who file by the priority deadline against the fraction $t$ of students who are assigned to treatment.%
    \footnote{
        Such policy ROC curves (also called ``uplift curve'', ``profit curve'', or ``cost curve'') have also been used to represent benefits of targeting at varying costs e.g. in \citet{Rzepakowski2012-pi,Zhao2013-la,Sun2021-ed,yadlowsky2021evaluating,Hitsch2018-bw}.
    }
    Our main benchmark is a random assignment policy that chooses a random group of students to receive reminders, leading to a linear relationship between the fraction of students assigned to treatment and the estimated fraction of students who file (black line in \autoref{fig:policy_V1}).
    The value of using heterogeneous treatment effects estimates $\hat{\tau}(x)$ from the causal forest is represented by the ``CATE'' line, which shows a persistent gain over the random assignment policy.%
    \footnote{
        In the main text, we discuss estimates of the value of targeting based on CATE estimates obtained similarly to \autoref{sec:heterogeneity} but with tuning parameters chosen to optimize targeting performance.
        In \autoref{sec:heterogeneity}, the minimum size of partitions in the causal forest is chosen to work well for achieving low mean-squared estimation error, while the minimum size of partitions in this section is chosen to achieve a high value when used for targeting.
        In the appendix, we provide additional robustness checks.
        Specifically, \autoref{fig:robustness_tuning} evaluates the targeting performance of the causal forest across different tuning parameter values, specifically varying the minimum size of the partitions and the number of trees.
        In addition, \autoref{fig:robustness_reestimation} changes how the same policies are estimated, by separating the estimation of the CATE used for targeting from the estimation of nuisance components used in evaluation.
        Our results appear largely insensitive to these changes, and the chosen tuning parameters seem to perform well.
    }

    In order to estimate the numbers in \autoref{fig:policy_V1}, we leverage the insight that the outcome under alternative assignment policies can be estimated from existing trial data by exploiting randomized assignment.%
    \footnote{Such counterfactual estimation goes back to the  \cite{horvitz1952generalization} estimator and has been leveraged in our context of evaluating policies for example by \cite{athey2021policy} and \cite{Hitsch2018-bw}.}
    This permits the estimation of $\E[Y(\pi(X))] = \E[\pi(X) \: Y(1) + (1-\pi(X)) \: Y(0)]$ for an assignment policy that maps characteristics $x$ to an assignment $\pi(x) \in \{1,0\}$.
    Indeed, $\E[Y(\pi(X))]$ is identified by
    \[
      \E[Y(\pi(X))] = \E\left[
        \pi(X) \: \frac{T}{p(X)} \: Y 
        + (1-\pi(X)) \: \frac{1-T}{1-p(X)}  \: Y
      \right],
    \]
    where $p(x) = \P(T=1|X=x)$ is the (known) propensity score.
    For each fraction, $q$, we estimate the average outcome $U^{\text{causal}}(q) = \E[Y(\hat{\pi}^{\text{causal}}_{t(q)}(X))]$ we could have achieved using this policy, where the threshold $t(q)$ is chosen such that a fraction $q \in [0,1]$ of individuals receives the treatment.
    For every fraction $q$, \autoref{fig:policy_V1} plots the resulting estimate of $U^{\text{causal}}(q)$ on the $y$-axis, allowing us to evaluate and compare policies based on the total increase in FAFSA renewal relative to the number of students who are sent reminders.

    When we estimate the value $U^{\text{causal}}(q)$ of assigning a fraction of $q$ students to treatment, we use sample splitting to obtain an unbiased and model-free estimate.
    Specifically, we estimate all quantities in \autoref{fig:policy_V1} by ten-fold cross-fitting.
    For each fold $k$, we obtain an estimated CATE function $\hat{\tau}_{-k}$ from data of the \emph{other} folds only.
    We then rank all observations within fold $k$ by their out-of-sample treatment effect estimates $\hat{\tau}_{-k}(X_i)$, and choose the top $q$ fraction to be treated.
    We use the realized outcomes $Y_i$ and treatment statuses $T_i$ to estimate the value of the policy in this fold.
    To do so, we employ a doubly-robust augmented inverse propensity weighted estimator using fixed (known) propensity scores, with details provided in \autoref{apx:inference}.
    Finally, we average the estimates across all folds to obtain the estimates reported in the figures.
    As a result, all estimates can be interpreted as unbiased out-of-sample evaluations of the benefit one would achieve if one used the same policies on new data coming from the same distribution.%
    \footnote{To be more precise, since the policies can vary slightly from fold to fold, we estimate the \emph{average} benefit of using one of the ten policies estimated in this way.}
    In particular, since we never use the same observations to construct our assignment policy and to evaluate it, there is no bias from using the existing RCT data for constructing policies and for evaluating them.

    In economic terms, our estimates suggest a moderate gain from targeting using treatment-effect estimates from the causal forest.
    For example, we could have increased FAFSA renewal at the priority deadline from 36.5\% to 41.2\% (realizing about 68\% of the gain, 95\% confidence interval $[60\%,77\%]$) by targeting those 50\% of students with the highest predicted effect based using early covariates, with basically the same performance for late covariates.
    In statistical terms, the point-wise 95\% confidence bands in \autoref{fig:policy_V1} document a significant gain over random assignment for a substantial range of fractions of assigned students.
    However, since these confidence bands are estimated point-wise, we complement our analysis with a joint test based on the RATE framework of \cite{yadlowsky2021evaluating}.
    \autoref{fig:rate} in the appendix shows average treatment effects by targeting fraction, and \autoref{tbl:rate} provides corresponding estimates of the area under this curve. An omnibus test of average gain over random assignment based on these estimates (which is also a test of the null hypothesis of no heterogeneity) rejects at the 5\% level for early and late covariates in the 2017 sample, and also for late covariates in the 2018 data.

    We benchmark the performance of targeting based on causal-forest estimates of heterogeneous treatment effects against targeting based on predicted and realized enrollment.
    In \autoref{sec:heterogeneity}, we argued that a good fraction of heterogeneity in treatment effects can be attributed to enrollment.
    Here, we therefore consider what happens if we prioritize reminders to go to those students who have the highest predicted probability of being enrolled.
    The ``Predicted Enrollment'' method in Panel (a) of \autoref{fig:policy_V1} evaluates that policy and shows that it does not perform as well as the causal policy, suggesting additional variation in treatment effects beyond predicted enrollment, although differences are noisy.
    In Panel (b), we instead target based on realized enrollment, which is available as part of the late covariates.
    That policy has a similar performance to predicted enrollment.

    \subsection{Targeting based on baseline predictions}
    
    We compare targeting by causal treatment effects to a purely predictive targeting rule.
    A natural approach to targeting would have been to prioritize reminders for those students who, absent treatment, would have been least likely to file for FAFSA renewal. 
    Such a policy is intuitive because students with low baseline probability of filing have the largest potential to gain from the nudge.
    It is also practical because it only requires information from the control group, which could be learned before the introduction of the intervention.
    A similar comparison to predictive approaches is performed by \cite{Ascarza2018-md} in the context of churn, and by \cite{Hitsch2018-bw} for targeting based on the predicted potential outcome under treatment in a catalog-mailing application.
    \cite{fernandez2022causal} provides a theoretical and empirical comparison of targeting based on predicted outcomes vs treatment effects.

    In order to implement a targeting policy by baseline predictions, we estimate the probability $f(x) = \E[Y(0)|X{=}x]$ that a student would have filed by the priority deadline absent the behaviorally-informed reminders, and give treatment first to those with the lowest predicted probability. That is, we let
    \begin{align*}
      \hat{\pi}^{\text{predictive}}_b(x) = \mathbbm{1}(\hat{f}(x) \leq b)
    \end{align*}
    where $\hat{f}(x)$ is an estimate of $f(x)$ and $b$ is a threshold.
    We can implement this policy efficiently using any off-the-shelf machine-learning predictor that predicts filing by the deadline from available variables in the absence of an experiment, since $f(x) = \E[Y|X{=}x, T{=}0]$.
    
    We provide an evaluation of this predictive policy using random-forest predictions $\hat{f}(x)$ in \autoref{fig:policy_V1} as the ``Negative Baseline'' method.
    This specific policy
    performs significantly worse than the policy based on the causal estimation of treatment effects. Indeed, we estimate that the outcome $\E[Y(\hat{\pi}^{\text{predictive}}_b(X))]$ of the prediction-based approach is considerably worse than assigning the same number of people randomly, across choices of the threshold $b$:
    If we chose students with a below-median probability of filing at baseline, we would only increase total filing from 36.5\% to around 38.5\%, for a gain of less than a third of the total.

    At the same time, we could have ranked students by who would have been \textit{most} likely to file for renewal by the deadline absent of the treatment (``Positive Baseline'' in \autoref{fig:policy_V1}).
    This policy even outperforms the non-parametric CATE estimate in our example for large fractions of targeted students.
    Here, a likely driver of this observation is that students who are unlikely to file for FAFSA at baseline are also unlikely to be convinced to do so by the reminder.
    Rather, the relatively weak behavioral nudge seems to work best for students already close to filing.
    A baseline policy that targets unlikely filers first, therefore, achieves exactly the opposite of the desired effect.

    While both the positive and negative baseline policies may have been plausible ex-ante, we learned their properties only through the ex-post evaluation from the experiment.
    The causal approach has the advantage of directly estimating a policy that we can expect to work well based solely on the empirical relationships of covariates to treatment effects, rendering guessing a policy that may work well (or testing a large number of them explicitly) unnecessary.
    At the same time, the example suggests gains from leveraging baseline predictions.
    Below, we analyze how we can let the data decide how to use baseline predictions to get the best of both worlds.

    \section{Improving Targeting by Combining Predictive and Causal Modeling}
    \label{sec:combined}

    Our above results demonstrate the importance of modeling an intervention's causal effect to better target it rather than relying blindly on ad hoc predictive targeting rules.
    At the same time, these results show a strong relationship between baseline predictions and causal effects.
    In this section, we ask whether we can profitably combine predictive information and causal modeling to achieve better targeting.
    
    This section explores three related approaches to integrate non-parametric predictions into the modeling causal effects.
    We first consider parametric models that take baseline predictions as an input, but otherwise impose simple parametric forms.
    We then explore simple models that employ non-parametric baseline predictions and CATE estimates.
    Finally, we mention non-parametric models that use intermediate baseline and CATE estimates as input features.

    \subsection{Model-based targeting from baseline predictions}
    \label{subsec:modelbased}

    \newcommand{\logit}{\textnormal{logit}}

    We start with a simple model that combines non-parametric baseline predictions with a logistic regression model of treatment effects.
    The main idea behind this approach is to use the full flexibility of the easy-to-predict baseline, while constraining treatment effects so that they are easier to estimate.
    Specifically, we model filing status as
    \begin{align*}
      \P(Y{=}1|X{=}x, T{=}t) = \frac{1}{1+\exp(-\alpha(x) -\beta \: t)},
    \end{align*}
    or equivalently,
    \begin{align*}
        \logit \: \P(Y{=}1|X{=}x, T{=}t) = \log(\P(Y{=}1|X{=}x, T{=}t)/(1-\P(Y{=}1|X{=}x, T{=}t))) = \alpha(x) +\beta \: t.
    \end{align*}
    Here, $\alpha(x)$ can be any non-parametric function and $\beta$ represents a fixed treatment effect in log-odds.
    This model could be derived from a binary choice model where $\alpha(x)$ is the perceived expected net utility gain from filing, which is unrestricted and can vary arbitrarily across students.
    In this model, $\beta$ can be interpreted as the effect of the reminder in terms of reducing hassle cost or increasing the perceived value of renewing financial aid. Even though the mean utility is additive in the baseline and the treatment, the nonlinear form of the logit model implies that treatment effects vary with the baseline in a particular way. Specifically, treatment effects are largest at intermediate values of the mean outcome. That is, when the probability of filing is close to zero or one, the incremental impact of the treatment is low. This relationship is discussed more fully below and illustrated in \autoref{fig:catemodel}.
    
    In order to estimate our simple model of FAFSA filing,
    we note that the probability of filing at baseline fulfills $\logit \: \P(Y(0)=1|X{=}x) = \alpha(x)$. Hence, a simple way we can estimate the model based on a baseline prediction $\hat{f}(x)$ of the probability $f(x) = \P(Y(0){=}1|X{=}x)$ of filing is to let $\tilde{f}(x) = \logit(\hat{f}(x))$ and to estimate the logistic regression
    \begin{align}
    \label{eqn:logit}
      \logit \: \widehat{\P}(Y{=}1|X{=}x, T{=}t)
      = \hat{\alpha} + \hat{\alpha}_{\tilde{f}} \: \tilde{f}(x) + \hat{\beta} \: t.
    \end{align}
    We implement this procedure with the random-forest estimates $\hat{f}(x)$ of baseline filing from the previous section.%
    \footnote{We use honest out-of-bag estimates when estimating the model parameters to avoid over-fitting \citep{Wager2018-pe}. This means that we employ a specific sample-splitting scheme that ensures that all treatment effect estimates do not use data from the observation they are applied to, even those in the training data itself.}
    We then estimate the implied difference in predictions $\tau(x) = \P(Y{=}1|X{=}x, T{=}1) - \P(Y{=}1|X{=}x, T{=}0)$ by $\hat{\tau}^*(x) = \widehat{\P}(Y{=}1|X{=}x, T{=}1) - \widehat{\P}(Y{=}1|X{=}x, T{=}0)$, and prioritize students with high estimated treatment effects, that is,
    \begin{align*}
      \hat{\pi}^{\text{logit}}_t(x) = \mathbbm{1}(\hat{\tau}^*(x) \geq t).
    \end{align*}
    As before, we evaluate all policies using sample splitting so that all results can be interpreted as out-of-sample estimates.
    In particular, the logit coefficients themselves are also estimated on the training folds only.

     \begin{figure}
      \centering
      \begin{subfigure}[t]{0.75\textwidth}
        \centering
        \includegraphics[width=\textwidth]{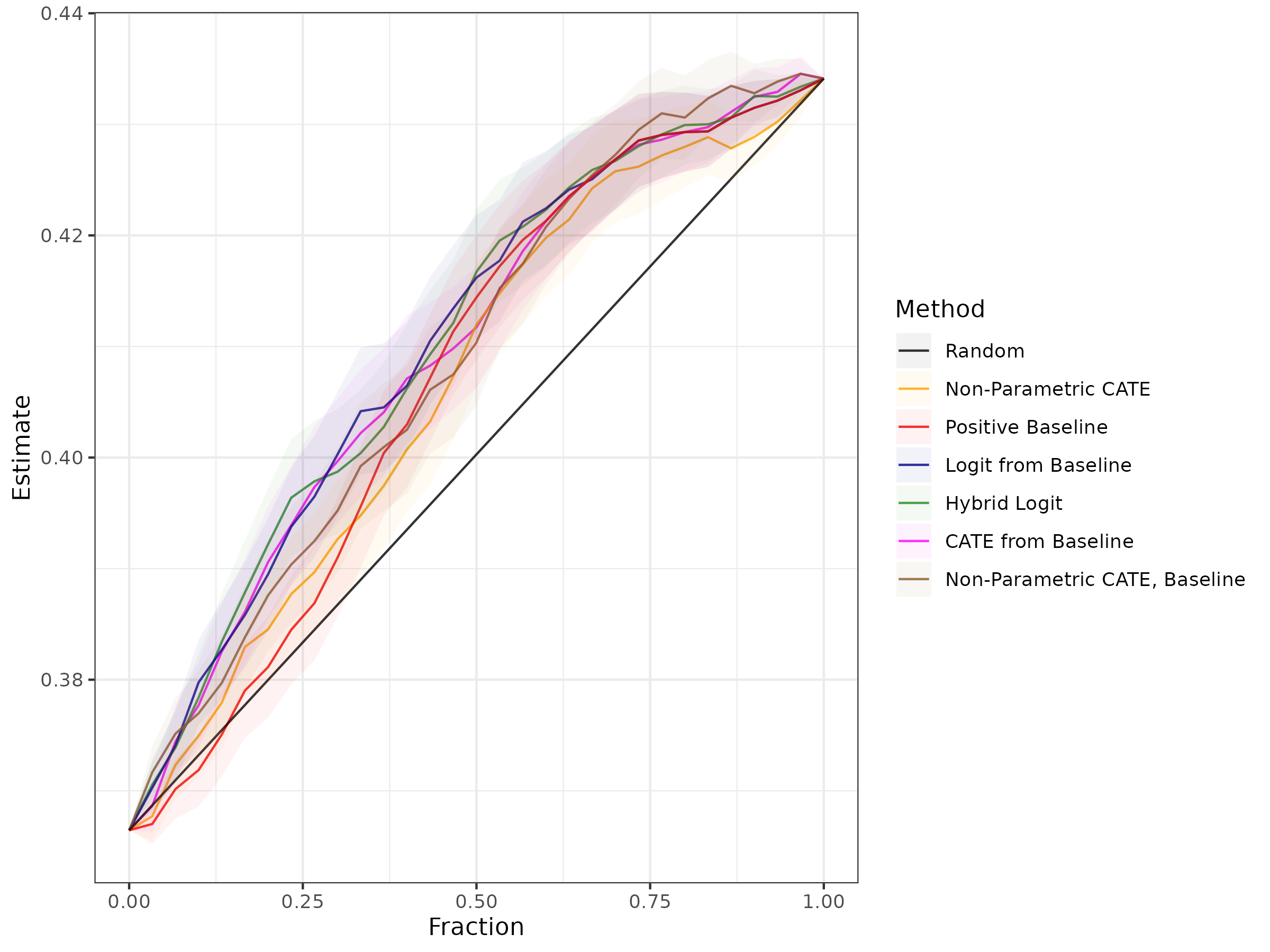}
        \caption{2017, early covariates}
      \end{subfigure}%

      \begin{subfigure}[t]{0.75\textwidth}
          \centering
          \includegraphics[width=\textwidth]{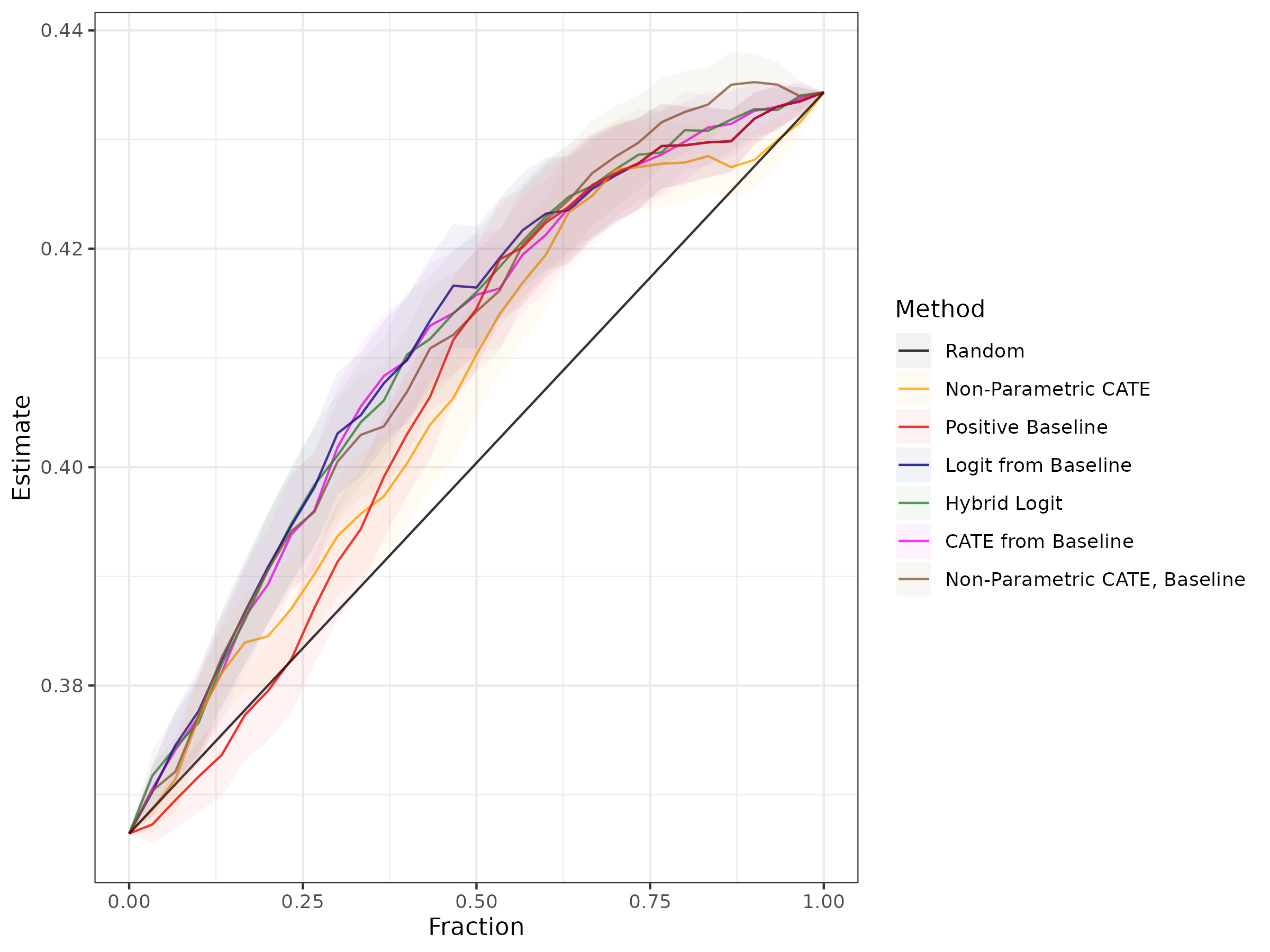}
          \caption{2017, late covariates}
      \end{subfigure}
      \caption{Total estimated FAFSA renewal rate ($y$-axis) by targeting a given fraction ($x$-axis) of students according to different cross-fitted predictions in the 2017 data as in \autoref{fig:policy_V1}, with additional targeting rules based on
      logistic regression with a random-forest prediction of the baseline response as covariate (``Logit from Baseline''),
      a hybrid logit model that uses transformed baseline and treatment-effect estimates as features (``Hybrid Logit''),
      a non-parametric estimate of treatment effects from predicted baseline only (``CATE from Baseline''),
      and a non-parametric estimate of treatment effects from predicted baseline and estimated treatment effects (``Non-parametric CATE, Baseline'').
      }
      \label{fig:policy_V2}
    \end{figure}

    The targeting policy derived from our simple logit model is presented as the ``Logit from Baseline'' method in \autoref{fig:policy_V2}.
    Across both early and late covariates, this simple policy outperforms the non-parametric policy based on the causal forest.
    With this semiparametric policy, we can achieve around 75\% of the gain from targeting only half of the students (95\% confidence interval $[67\%,84\%]$).
    The policy also improves substantially over targeting based on predicted or realized enrollment, showing that there is predictable heterogeneity in treatment effects beyond enrollment.
    It also outperforms the policy that targets those students with \textit{highest} baseline probability of filing first (``Positive Baseline'').

    \subsection{Targeting based on a hybrid model that adapts to heterogeneity}
    \label{subsec:hybrid}

    While the simple logistic regression model from \eqref{eqn:logit} performs well in our example, it may perform poorly when constant treatment effects in log-odds do not approximate the relationship of treatment effects to the baseline well or when there is additional variation in treatment effects that is not captured by variation in the baseline.
    We therefore consider the hybrid logistic regression
    \begin{align}
    \label{eqn:hybrid}
      \logit \: \widehat{\P}(Y{=}1|X{=}x, T{=}t)
      = \hat{\alpha} + \hat{\alpha}_{\tilde{f}} \: \tilde{f}(x) + \hat{\alpha}_{\tilde{g}} \: \tilde{g}(x) + (\hat{\beta} + \hat{\beta}_{\tilde{f}} \: \tilde{f}(x) + \hat{\beta}_{\tilde{g}} \: \tilde{g}(x)) \: t
    \end{align}
    that includes the transformed baseline $\tilde{f}(x) = \logit(\hat{f}(x))$
    as well as a logit transformation $\tilde{g}(x)$ of the non-parametric CATE estimates $\hat{\tau}(x)$ from \autoref{sec:heterogeneity}, where we construct $\tilde{g}(x)$ by
    \begin{align*}
        \tilde{g}(x) = \logit(\hat{f}(x) + \hat{\tau}(x)) - \logit(\hat{f}(x)).
    \end{align*}
    Here, we assume that the estimates $\hat{f}(x) + \hat{\tau}(x)$ of the probability of $Y(1){=}1$ as well as the estimates $\hat{f}(x)$ of the probability of $Y(0){=}1$ are all within the unit interval, which is the case in our data.
    All of the estimates $\hat{f}(X_i)$ and $\hat{\tau}(X_i)$ in the data are fitted using sample splitting so that the outcome of an observation is never used to estimate its baseline or treatment effects, and all policies are evaluated out-of-sample and separately from the estimation of the model parameters.
    The performance of this hybrid logit model is documented in \autoref{fig:policy_V2}, where it performs on par with or slightly worse than the simpler logit model that uses baseline information only while outperforming targeting based on the fully non-parametric estimate throughout.
    
    This specific way of parametrizing treatment effect estimates and baseline predictions has the advantage that it allows the model to capture both the simple model of constant log-odds treatment effects from \eqref{eqn:logit} (by setting $\alpha_{\tilde{g}} = \beta_{\tilde{g}} = \beta_{\tilde{f}} = 0$) as well as to recover the nonparametric treatment effect estimate $\hat{\tau}(x)$ itself by setting $\hat{\alpha}_{\tilde{f}} = \hat{\beta}_{\tilde{g}} = 1$
    and
    $\hat{\alpha} = \hat{\alpha}_{\tilde{g}} = \hat{\beta} = \hat{\beta}_{\tilde{f}} = 0$
    in which case
    \begin{align*}
        \widehat{\P}(Y{=}1|X{=}x, T{=}1) - \widehat{\P}(Y{=}1|X{=}x, T{=}0)
        =
        \frac{1}{1 + \exp(-\tilde{f}(x)-\tilde{g}(x))} - \frac{1}{1 + \exp(-\tilde{f}(x))}
        =
        \hat{\tau}(x).
    \end{align*}
    By fitting the model, we can let the data decide which of these models works best.
    This procedure can be thought of as data-driven regularization using cross-validation.
    Since all treatment effect estimates and baseline predictions are fitted out of sample,
    the model parameters adapt to how well $\hat{f}$ and $\hat{\tau}$ capture baseline variation and treatment effects, respectively.
    For example, if treatment effects are imprecise, then the optimal regression coefficients associated with $\tilde{g}$ are small and put little weight on estimated treatment effects.
    If, on the other hand, $\hat{\tau}$ captures true CATEs well, then $\hat{\beta}_{\tilde{g}}$ may be close to one.
    Crucially, we expect this regression to still perform well even if treatment effect estimates are noisy since it can effectively learn to shrink more noisy estimates.

    \subsection{Comparison of non-parametric, simple, and hybrid targeting strategies}

    The above results show that a simple causal model on top of baseline predictions substantially outperforms a fully non-parametric causal model, while a hybrid model that combines baseline predictions and CATEs also performs well.
    When would we expect this pattern to hold?
    In general, if sample sizes are large, treatment effects vary considerably, and covariates are not too high-dimensional, then we would expect a direct non-parametric estimation of causal effects to yield good targeting rules since these are the conditions under which tools like the causal forest are able to estimate true CATEs well.
    However, if treatment effects become difficult to estimate because sample sizes are small or their variation is moderate, then semi-parametric and hybrid approaches may be more appropriate.
    Effectively, these models improve efficiency by imposing simple functional forms as a form of regularization.
    If baseline outcomes can be estimated precisely and predict treatment effects well, then relating the two by a simple model can prove an effective restriction that reduces variance, as in our example.
    On the other hand, if the non-parametric CATE estimate provides helpful complementary information for targeting, then including it in a simple model may reduce bias at a limited cost in variance.

\begin{figure}
      \centering
          \includegraphics[width=\textwidth]{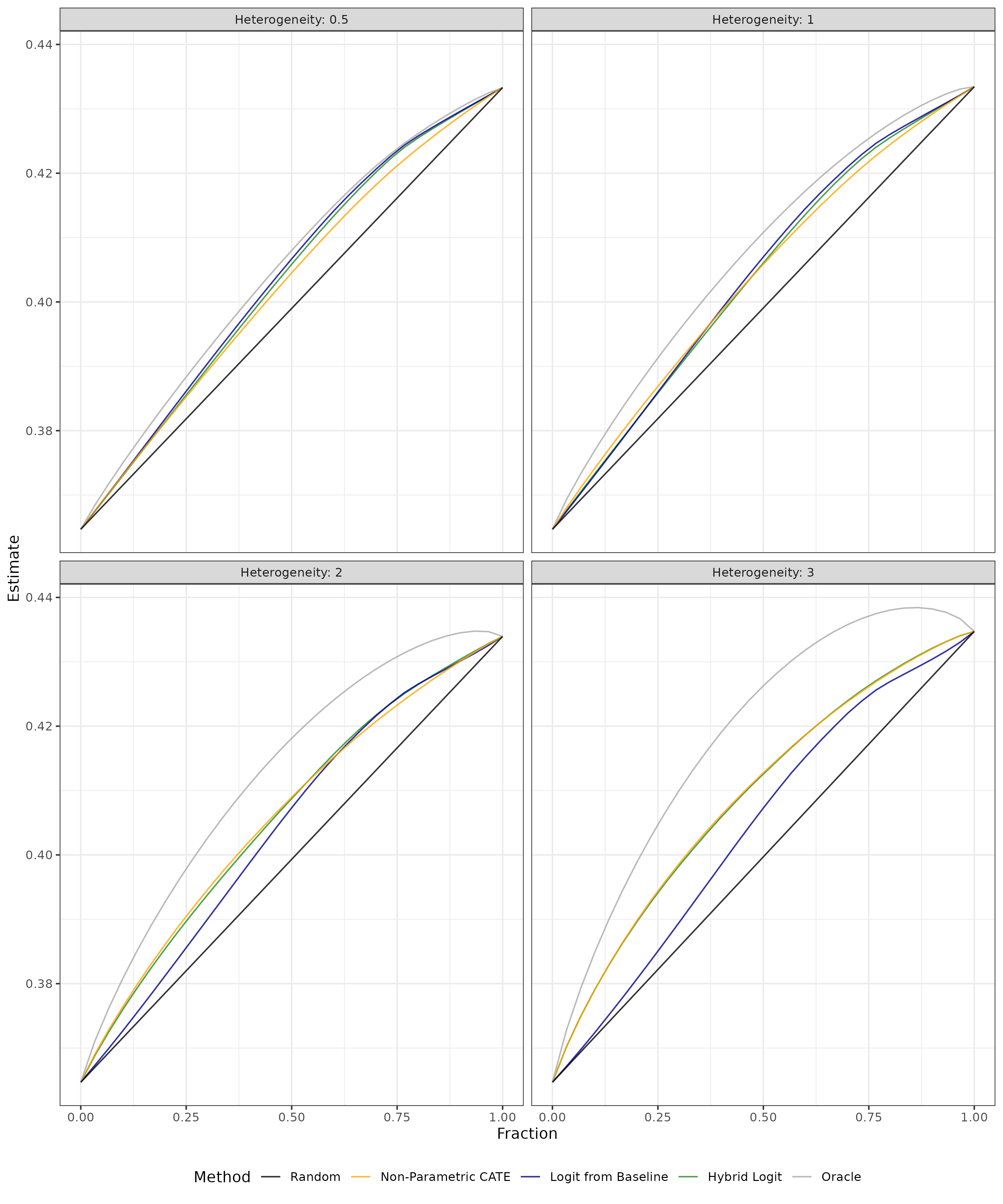}
      
      \caption{Performance of selected models from \autoref{fig:policy_V2} in a simulation based on the 2017 data with early covariates and varying strengths of heterogeneity. Shown are averages over 20 random draws of treatment assignment, using the same sample of 22,192 students and ten-fold cross-fitting scheme as in \autoref{fig:policy_V2} and comparing based on known simulated treatment effects.
      The ``Oracle'' method ranks students based on these known effects.}
      
      \label{fig:simulation}
    \end{figure}

    We now compare these approaches in a small simulation study, which further documents the ability of the hybrid model from \autoref{subsec:hybrid} to adapt to the level of heterogeneity.
    We present the main results in \autoref{fig:simulation}.
    Here, we take the non-parametric baseline and treatment-effect estimates $\hat{f}(x)$ and $\hat{\tau}(x)$,%
    \footnote{We re-calculate non-parametric honest estimates $\hat{f}(x)$ and $\hat{\tau}(x)$ using the random forest and causal forest, respectively, for which we use more complex trees than for the results reported in \autoref{sec:heterogeneity} to obtain enough heterogeneity.}
    re-randomize treatment assignment, and re-draw outcomes by the model
    \begin{align*}
        \logit \: \P(Y{=}1|X{=}x,T{=}t)
        =
        \tilde{f}(x) + \lambda \:  (\tilde{g}(x) - \E[\tilde{g}(X)]) \: t + \E[\tilde{g}(X)] \: t.
    \end{align*}
    $\lambda$ expresses the degree of treatment effect heterogeneity.
    For $\lambda = 1$, this model simply simulates based on the non-parametric estimates $\hat{f}(x)$ and $\hat{\tau}(x)$.
    For $\lambda = 0$ this model would assume that treatment effects are constant in log-odds, while $\lambda > 1$ corresponds to additional variation.

    The simulation results highlight the benefit of the hybrid model: for low heterogeneity ($\lambda = \sfrac{1}{2}$ and $\lambda = 1$, top row in \autoref{fig:simulation}), the simple logit regression from baseline from \eqref{eqn:logit} performs well, while the hybrid model comes close to its performance.
    For high variation ($\lambda = 3$, bottom right panel), the fully non-parametric estimate of heterogeneous treatment effects performs best, with the hybrid model achieving the same performance.
    In the intermediate regime ($\lambda = 2$, bottom left panel), the non-parametric model performs best for small treatment fractions, while the simple logit performs well for high fractions. 
    Here, the hybrid model represents a compromise between both, coming close to the non-parametric model, while outperforming the simple logistic model.
    Overall, the hybrid targeting model thus adapts to the level of signal and provides comparatively good performance throughout.

    The simulation presented in \autoref{fig:simulation} only considers one determinant of the relative performance of different modeling approaches, namely, how much heterogeneity there is in treatment effects. But relative performance and the value of regularization from imposing simpler models also vary with other important factors.
    These include the sample size, the complexity of covariates, the variation in the baseline, the noise in the outcome variable, and how much of the treatment-effect variation can be explained by the baseline alone.
    As an illustration of these additional factors, we repeat the simulation exercise with the same models as above, but reducing the sample size to only half of the original sample, from 22,192 to 11,096.
    The results, reported in \autoref{fig:simulation_halfsample} in the appendix, suggest that the non-parametric causal estimates perform comparably worse in small sample sizes when heterogeneity is weak, while the simple logit model from \autoref{subsec:modelbased} performs comparably better.
    The direct application of the causal forest only dominates for very high heterogeneity.
    The simple logit model does well relative to the more complicated hybrid model even for moderately high heterogeneity, showing the additional value of stronger regularization for smaller sample sizes.

    \subsection{Non-parametric targeting with baseline predictions as a feature}

    The parametric models of \autoref{eqn:logit} make a functional-form assumption about the relationship of treatment effects to baseline predictions.
    Even if treatment effects can be expressed well in term of this baseline probability of filing by the priority deadline, specific functional forms can be restrictive.
    As an alternative, we now model treatment effects as a non-parametric function of the predicted baseline.
    The performance of a policy that estimates treatment effects using a strongly regularized causal forest from predicted baseline probabilities is plotted as the ``CATE from Baseline'' method in \autoref{fig:policy_V2}, where it is shown to obtain performance slightly worse than and on par with the simple logit, respectively, and substantially outperforms the direct non-parametric estimate of treatment effects.
    In general, we would expect this method to do well whenever treatment effects mainly vary with the baseline and a simple relationship of treatment effects to baseline may not be sufficient to express them, but treatment effects themselves may be hard to estimate.

\begin{figure}
      \centering

      \begin{subfigure}[t]{0.7\textwidth}
        \centering
          \includegraphics[width=\textwidth]{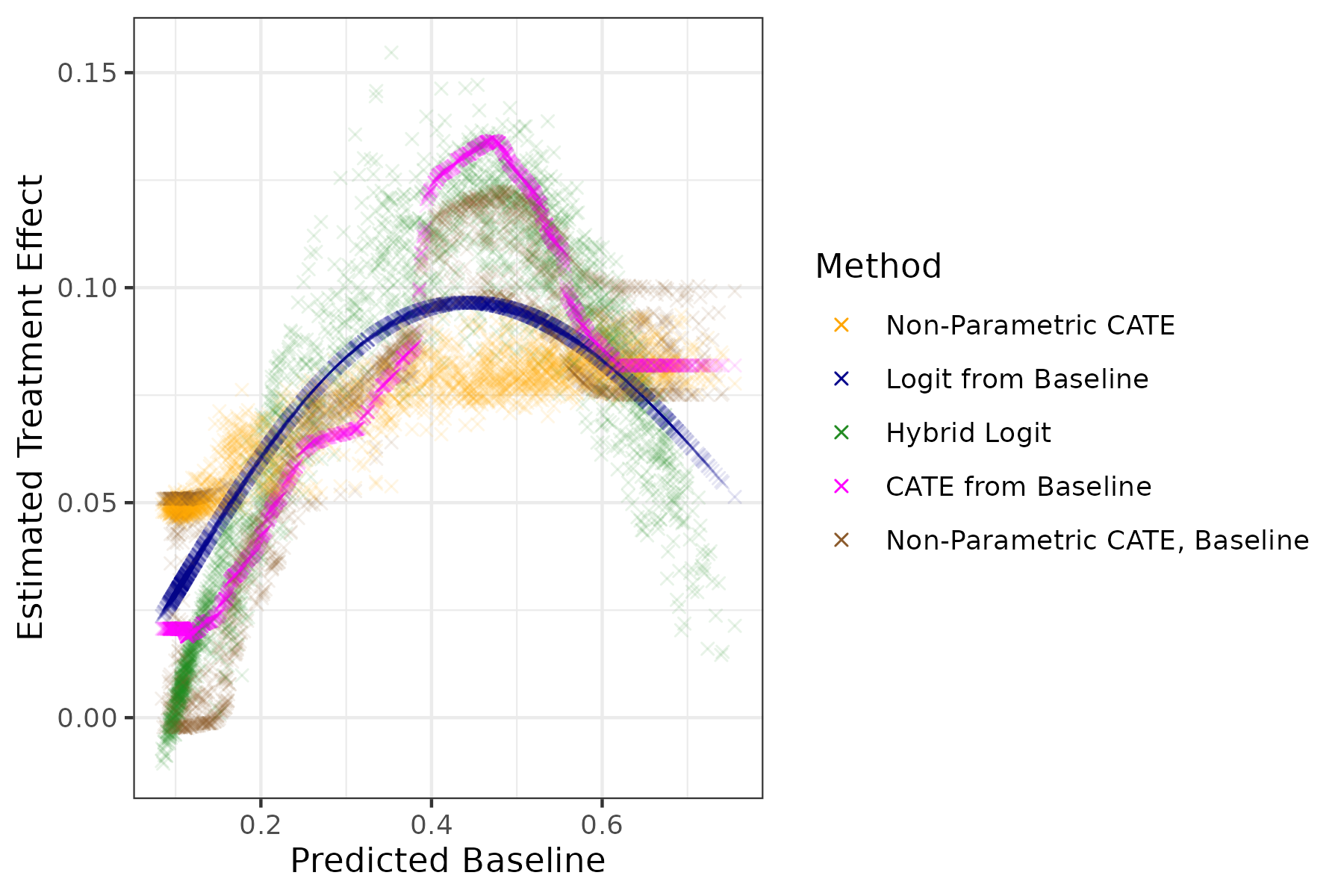}

        \caption{2017, early covariates}
      \end{subfigure}%

      \begin{subfigure}[t]{0.7\textwidth}
          \centering
        \includegraphics[width=\textwidth]{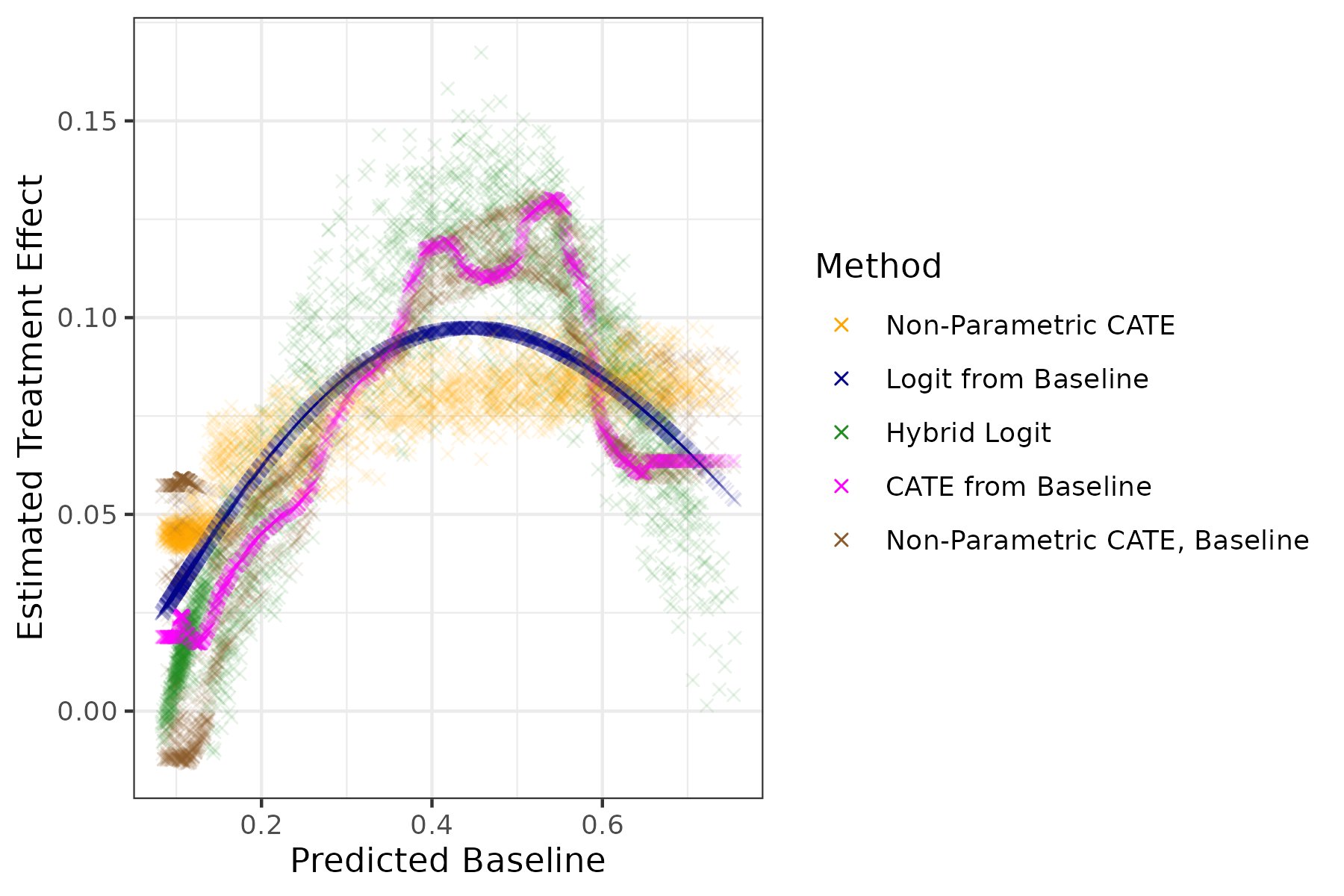}
          \caption{2017, late covariates}
      \end{subfigure}   
      
      \caption{Heterogeneous treatment effect estimates by estimated baseline (``Non-Parametric CATE''), along with treatment effects re-estimated 
      by a simple logistic regression with a constant coefficient on treatment (``Logit from Baseline''),
      by the hybrid logit model from \autoref{subsec:hybrid},
      from baseline using the causal forest (``CATE from Baseline''),
      and from baseline and non-parametric CATE using the causal forest (``Non-parametric CATE, Baseline''), shown here for a single fold.}
      
      \label{fig:catemodel}
    \end{figure}

    To visualize this strategy,
    \autoref{fig:catemodel} plots estimated treatment effects against baseline predictions of filing for the 2017 data with early covariates, presenting data from a single fold.
    The non-parametric treatment-effect estimates themselves only vary moderately, while the models that explicitly leverage baseline predictions are able to capture additional variation.
    Among them, the hybrid logit model has additional spread in treatment effect estimates for the same baseline level, but also shows a clear overall curved relationship similar to that of the simple logit.

    Modeling treatment effects directly as a function of the predicted baseline is only one way to use baseline predictions. In principle, we could obtain additional targeting models by combining baseline predictions with additional features, such as predicted enrollment, quartiles of estimated treatment effects, or some of the original covariates.
    \autoref{fig:policy_V3} in the appendix presents such additional specifications.
    However, the simple logit model introduced above generally performs at least comparably to those more complicated alternatives.
    We also note that it improves over a simple parametric logit model based on a few selected covariates.

    A specific model that can adapt to treatment effect heterogeneity could be obtained by estimating treatment effects non-parametrically from baseline predictions and non-parametric CATE estimates.
    Specifically, we could estimate treatment effects from a heavily causal forest that takes as input out-of-bag baseline estimates $\hat{f}(x)$ as well as non-parametric CATE estimates $\hat{\tau}(x)$ obtained as in \autoref{sec:heterogeneity}.
    Like the semi-parametric logistic regression model from \autoref{subsec:hybrid}, this approach is in principle able to recover the non-parametric CATE estimates when they capture heterogeneity well, while also allowing for modeling heterogeneous treatment effects as a function of the baseline.
    This policy is labeled ``Non-Parametric CATE, Baseline'' in \autoref{fig:policy_V2} and \autoref{fig:catemodel}. It performs better than targeting by the non-parametric CATE alone, but not quite as well as the logistic-regression model.
    
    We interpret our results in this section in terms of the importance of regularization, choice of features, and functional form.
    When treatment effect variation is small and their estimation is noisy,
    then leveraging first-order variation in the baseline and modeling its relationship to treatment effects in simpler ways provides a form of regularization that preserves variation in treatment effect estimates and can improve performance.
    This finding extends the semiparametric approach of \cite{athey2021semiparametric} from the estimation of average to the estimation of heterogeneous effects.
    The performance of simple semi-parametric models depends crucially on their functional form.
    In our example, a standard logit model with constant treatment effect coefficient and non-parametric baseline predictions appears to capture the structure of treatment effects remarkably well.
    At the same time, the baseline itself is straightforward to estimate in our data, and staying fully non-parametric for the baseline appears to provide better performance than imposing a model on it as well.

\section{Conclusion}

    In this article, we compared and combined predictive and causal approaches to targeting interventions.
    Our analysis provides an example of the value of integrating careful experimentation, causal inference, and predictive modeling. By itself, predictive machine learning without experimental evaluation could have led to a bad policy, while combining predictive modeling and causal targeting based on a coherent analysis of heterogeneous treatment effects can ultimately lead to sizeable gains.
        
    Our analysis also points to the challenges of evaluating existing experiments with machine learning. While sample and effect sizes seem large for estimating average treatment effects, an intervention designed to work well on average in an experiment powered for estimating averages makes the precise estimation of heterogeneous treatment effects statistically and technically challenging. Future experiments could also rely on an integration of targeting into their design.
    
    Our results come with important caveats that limit statistical power, generalizability, and policy applicability. Since this experiment was not designed for heterogeneous treatment effect analysis, treatment arms were chosen to work well on average rather than for specific subgroups. Overall treatment effects are moderate since they come from relatively modest nudges, and the experiment was powered to detect average effects rather than effects on many subgroups. Finally, sending behaviorally informed reminders is cheap and does not appear to have any negative treatment effects in the experiment, so while these results can help target reminders to those for whom they will work best, this would mainly avoid inundating students with reminder texts and emails for who the effect would be small. In this sense, the value of targeting would remain limited.
    
    We believe that overcoming these shortcomings in future studies requires designing experiments \emph{ex ante} to estimate heterogeneous treatment effects in the first place. This includes designing individual treatment arms that are likely to affect different people differently so that differentiated treatments can be matched to appropriate individuals and situations. It also involves updating power analyses to the higher sample size demands for estimating heterogeneous treatment effects rather than average effects alone. Finally, policies based on heterogeneous treatment effects will be particularly important when treatment delivery is costly or when we need to make choices between treatments for which none dominates others across individuals.

    We close this article by discussing questions that our results raise about fairness and equity in targeting.
    Our analysis suggests that nudge-type interventions may be most effective for those who are already more likely to engage in the desired behavior at baseline, and that not targeting those with low expected outcomes may therefore be efficient. However, there may be reasons why a planner may attach higher importance to improving outcomes of individuals with a low baseline.
    While we leave a more careful treatment to future research, we note that our approach can help capture such welfare considerations in two ways:
    First, we could directly optimize for an assignment that puts higher weight on the outcomes of individuals with, say, a low expected baseline.
    Second, we could use our estimates of causal effects to understand better for which group the current intervention is \textit{not} effective and thereby inform the design of better interventions.
    
\FloatBarrier

    \bibliographystyle{apalikefull3}%
    \bibliography{references}%

\clearpage

    \appendix

    \section{Additional Tables and Figures}

    \label{apx:additional}

\begin{table}[!htb]
    \centering
    \small
    \begin{tabular}{lrrrr}
            \toprule  
             &        Control        &        Treatment        & $p$-value &   $N$ \\ 
         &     $N=12,658$     &     $N=12,480$     &         &       \\ 
            \midrule
            COLLEGE\_IN\_INTERVENTION\_SPR: &                 &                 &  0.69   & 25138\\ 
        $\qquad$0 & 3953 (31.229\%)  & 3928 (31.474\%)  &         &      \\ 
        $\qquad$1 & 8705 (68.771\%)  & 8552 (68.526\%)  &         &      \\ 
        AGE & 23.680 (6.635)  & 23.556 (6.518)  &  0.19   & 19153\\ 
        GENDER: &                 &                 &  0.51   & 25138\\ 
        $\qquad$Men & 5396 (42.629\%)  & 5372 (43.045\%)  &         &      \\ 
        $\qquad$Women & 7262 (57.371\%)  & 7108 (56.955\%)  &         &      \\ 
        ETHNICITY: &                 &                 &  0.89   & 25138\\ 
        $\qquad$American Indian or Native Alaskan &   39 (0.308\%)   &   33 (0.264\%)   &         &      \\ 
        $\qquad$Asian or Pacific Islander &  1072 (8.469\%)  &  1091 (8.742\%)  &         &      \\ 
        $\qquad$Black, Non-Hispanic & 4107 (32.446\%)  & 4067 (32.588\%)  &         &      \\ 
        $\qquad$Hispanic, Other & 6558 (51.809\%)  & 6422 (51.458\%)  &         &      \\ 
        $\qquad$White, Non-Hispanic &  882 (6.968\%)   &  867 (6.947\%)   &         &      \\ 
        TRANSFER: &                 &                 &  0.32   & 25138\\ 
        $\qquad$0 & 9104 (71.923\%)  & 8874 (71.106\%)  &         &      \\ 
        $\qquad$1 &  591 (4.669\%)   &  584 (4.679\%)   &         &      \\ 
        $\qquad$'Missing' & 2963 (23.408\%)  & 3022 (24.215\%)  &         &      \\ 
        FT\_PT\_STATUS: &                 &                 &  0.39   & 25138\\ 
        $\qquad$FULL-TIME & 6610 (52.220\%)  & 6415 (51.402\%)  &         &      \\ 
        $\qquad$PART-TIME & 2441 (19.284\%)  & 2473 (19.816\%)  &         &      \\ 
        $\qquad$'Missing' & 3607 (28.496\%)  & 3592 (28.782\%)  &         &      \\ 
        GPA\_CUMU\_BF &  2.475 (0.952)  &  2.472 (0.947)  &  0.84   & 14633\\ 
        CRD\_CUMU\_ATMPT\_BF & 18.706 (16.565) & 18.785 (16.524) &  0.75   & 17939\\ 
        CRD\_CUMU\_EARN\_BF & 17.475 (15.661) & 17.387 (15.505) &  0.70   & 17939 \\ 
            \bottomrule
            \end{tabular}
        \vspace{1em}
        \caption{Balance table for the 2017 FAFSA experiment.}
        \label{tbl:balance2017}
\end{table}

\FloatBarrier

\begin{table}[!htb]
    \centering

    \begin{subfigure}[t]{\textwidth}
        \centering
        \scriptsize
        \begin{tabular}{lrrrrr}\\
            \toprule  
             &        Control        &        Treatment 1 & Treatment 2        & $p$-value &   $N$ \\ 
             &     $N=3226$      &     $N=13698$     &     $N=13610$     &         &       \\ 
             \midrule
            AGE & 23.412 (6.566)  & 23.625 (6.764)  & 23.425 (6.601)  &  0.07   & 23164\\ 
            GENDER: &                 &                 &                 &  0.40   & 30534\\ 
            $\qquad$Men & 1464 (45.381\%)  & 6062 (44.255\%)  & 5998 (44.071\%)  &         &      \\ 
            $\qquad$Women & 1762 (54.619\%)  & 7636 (55.745\%)  & 7612 (55.929\%)  &         &      \\ 
            ETHNICITY: &                 &                 &                 &  0.83   & 30534\\ 
            $\qquad$American Indian or Native Alaskan &   13 (0.403\%)   &   69 (0.504\%)   &   68 (0.500\%)   &         &      \\ 
            $\qquad$Asian or Pacific Islander &  503 (15.592\%)  & 2065 (15.075\%)  & 2000 (14.695\%)  &         &      \\ 
            $\qquad$Black, Non-Hispanic &  971 (30.099\%)  & 4112 (30.019\%)  & 4019 (29.530\%)  &         &      \\ 
            $\qquad$Hispanic, Other & 1419 (43.986\%)  & 6092 (44.474\%)  & 6147 (45.165\%)  &         &      \\ 
            $\qquad$White, Non-Hispanic &  320 (9.919\%)   &  1360 (9.928\%)  & 1376 (10.110\%)  &         &      \\ 
            TRANSFER: &                 &                 &                 &  0.59   & 30534\\ 
            $\qquad$0 & 2319 (71.885\%)  & 9809 (71.609\%)  & 9776 (71.830\%)  &         &      \\ 
            $\qquad$1 &  132 (4.092\%)   &  593 (4.329\%)   &  535 (3.931\%)   &         &      \\ 
            $\qquad$'Missing' &  775 (24.024\%)  & 3296 (24.062\%)  & 3299 (24.240\%)  &         &      \\ 
            FT\_PT\_STATUS: &                 &                 &                 &  0.67   & 30534\\ 
            $\qquad$FULL-TIME & 1706 (52.883\%)  & 7350 (53.657\%)  & 7342 (53.946\%)  &         &      \\ 
            $\qquad$PART-TIME &  643 (19.932\%)  & 2692 (19.653\%)  & 2601 (19.111\%)  &         &      \\ 
            $\qquad$'Missing' &  877 (27.185\%)  & 3656 (26.690\%)  & 3667 (26.943\%)  &         &      \\ 
            GPA\_CUMU\_BF &  2.538 (0.934)  &  2.508 (0.942)  &  2.500 (0.948)  &  0.28   & 19020\\ 
            CRD\_CUMU\_ATMPT\_BF & 21.635 (17.409) & 21.285 (16.829) & 21.275 (16.995) &  0.63   & 22334\\ 
            CRD\_CUMU\_EARN\_BF & 19.330 (15.676) & 19.001 (15.169) & 18.968 (15.317) &  0.58   & 22334 \\ 
            \bottomrule
            \end{tabular}
            \caption*{(i) Early schools}
    \end{subfigure}%

    \begin{subfigure}[t]{\textwidth}
        \centering
        \scriptsize
        \begin{tabular}{lrrrrr}\\
            \toprule  
            &        Control        &        Treatment 1 & Treatment 2        & $p$-value &   $N$ \\ 
            &     $N=2497$      &     $N=3802$      &     $N=3699$      &         &      \\ 
            \midrule
            AGE & 23.742 (6.448)  & 23.906 (6.724)  & 24.076 (6.803)  &  0.23   & 7866\\ 
GENDER: &                 &                 &                 &  0.50   & 9998\\ 
$\qquad$Men & 1200 (48.058\%)  & 1770 (46.554\%)  & 1745 (47.175\%)  &         &     \\ 
$\qquad$Women & 1297 (51.942\%)  & 2032 (53.446\%)  & 1954 (52.825\%)  &         &     \\ 
ETHNICITY: &                 &                 &                 &  0.54   & 9998\\ 
$\qquad$American Indian or Native Alaskan &   9 (0.360\%)    &   10 (0.263\%)   &   6 (0.162\%)    &         &     \\ 
$\qquad$Asian or Pacific Islander &  195 (7.809\%)   &  313 (8.233\%)   &  278 (7.516\%)   &         &     \\ 
$\qquad$Black, Non-Hispanic &  814 (32.599\%)  & 1241 (32.641\%)  & 1219 (32.955\%)  &         &     \\ 
$\qquad$Hispanic, Other & 1100 (44.053\%)  & 1717 (45.160\%)  & 1646 (44.499\%)  &         &     \\ 
$\qquad$White, Non-Hispanic &  379 (15.178\%)  &  521 (13.703\%)  &  550 (14.869\%)  &         &     \\ 
TRANSFER: &                 &                 &                 &  0.36   & 9998\\ 
$\qquad$0 & 1799 (72.046\%)  & 2753 (72.409\%)  & 2692 (72.776\%)  &         &     \\ 
$\qquad$1 &  151 (6.047\%)   &  258 (6.786\%)   &  213 (5.758\%)   &         &     \\ 
$\qquad$'Missing' &  547 (21.906\%)  &  791 (20.805\%)  &  794 (21.465\%)  &         &     \\ 
FT\_PT\_STATUS: &                 &                 &                 &  0.18   & 9998\\ 
$\qquad$FULL-TIME & 1309 (52.423\%)  & 2109 (55.471\%)  & 2019 (54.582\%)  &         &     \\ 
$\qquad$PART-TIME &  497 (19.904\%)  &  689 (18.122\%)  &  682 (18.437\%)  &         &     \\ 
$\qquad$'Missing' &  691 (27.673\%)  & 1004 (26.407\%)  &  998 (26.980\%)  &         &     \\ 
GPA\_CUMU\_BF &  2.408 (0.935)  &  2.459 (0.935)  &  2.494 (0.936)  &  0.02   & 6223\\ 
CRD\_CUMU\_ATMPT\_BF & 22.097 (17.048) & 22.089 (17.149) & 22.415 (17.162) &  0.74   & 7305\\ 
CRD\_CUMU\_EARN\_BF & 19.888 (15.367) & 19.900 (15.465) & 20.184 (15.339) &  0.74   & 7305 \\ 
            \bottomrule
            \end{tabular}
            \caption*{(ii) Late schools}
    \end{subfigure}%

    \caption{Balance tables for the 2018 FAFSA experiment.}
    \label{tbl:balance2018}
\end{table}

\FloatBarrier

\begin{table}[!htb]
    \centering
    \begin{tabular}{lllrr}
    \toprule
    Year & School timeline & Method & ATE & SE \\
    \midrule
    2017 &  & Mean difference & 0.0641 & 0.0061 \\
     & & Constant propensity & 0.0687 & 0.0053\\
     & & AIPW & 0.0686 & 0.0053 \\
    \midrule
    2018 & all & Mean difference & 0.1209 & 0.0074 \\
     & & Constant propensity & 0.1182 & 0.0065 \\
     & & AIPW & 0.1182 & 0.0065 \\
    \cmidrule(lr){2-5}
     & early & Mean difference & 0.1213 & 0.0091 \\
     & & Constant propensity & 0.1198 & 0.0079 \\
     & & AIPW & 0.1200 & 0.0080 \\
    \cmidrule(lr){2-5}
     & late & Mean difference & 0.1201 & 0.0109 \\
     & & Constant propensity & 0.1134 & 0.0099 \\
     & & AIPW & 0.1132 & 0.0100 \\
    \bottomrule
    \end{tabular}
    \caption{Overall average treatment effects, estimated by simple (propensity-adjusted) differences in averages (``Mean difference'') as well as by an augmented inverse propensity score estimator based on random forests with constant (``Constant propensity'') and flexible propensity score (``AIPW''), respectively.}
    \label{tbl:ate}
\end{table}

\FloatBarrier

\begin{table}
\newcommand{\significant}{\fontseries{b}\selectfont}
\centering
\begin{tabular}{llrrrr}
\toprule
    Year  & Covariates &  & Q1            & Q2                   & Q3                   \\
\midrule
    2017  & early      & Q2         & 0.018 (0.015) &                      &                      \\
          &            & Q3         & \significant 0.063 (0.015) & \significant 0.045 (0.017)        &                      \\
          &            & Q4         & \significant 0.049 (0.016) & \significant 0.031 (0.017)        & $-0.014$ (0.017)       \\
    \cmidrule(lr){2-6}
      & late       & Q2         & \significant 0.040 (0.014)  &                      &                      \\
          &            & Q3         & \significant 0.045 (0.015) & 0.005 (0.017)        &                      \\
          &            & Q4         & \significant 0.060 (0.015)  & 0.019 (0.017)        & 0.015 (0.018)        \\
        \midrule
    2018  & early      & Q2         & 0.028 (0.018) &                      &                      \\
          &            & Q3         & \significant 0.046 (0.019) & 0.018 (0.020)         &                      \\
          &            & Q4         & \significant 0.051 (0.019) & 0.023 (0.020)         & 0.005 (0.020)         \\
        \cmidrule(lr){2-6}
      & late       & Q2         & \significant 0.036 (0.018) &                      &                      \\
          &            & Q3         & \significant 0.042 (0.018) & 0.005 (0.020)         &                      \\
          &            & Q4         & \significant 0.071 (0.018) & \significant 0.034 (0.020)         & 0.029 (0.020)         \\
\bottomrule
\end{tabular}
\caption{Pairwise difference of quartile treatment effects from \autoref{fig:quartiles} based on AIPW estimates, with standard error estimate in parentheses. {\significant Bold estimates} denote statistically significant increases at the 5\% level based on a one-sided test.}
\label{tbl:quartiles}
\end{table}

\FloatBarrier

    \begin{figure}[hbtp]
      \centering
      \begin{subfigure}[t]{0.5\textwidth}
        \centering
        \includegraphics[width=\textwidth]{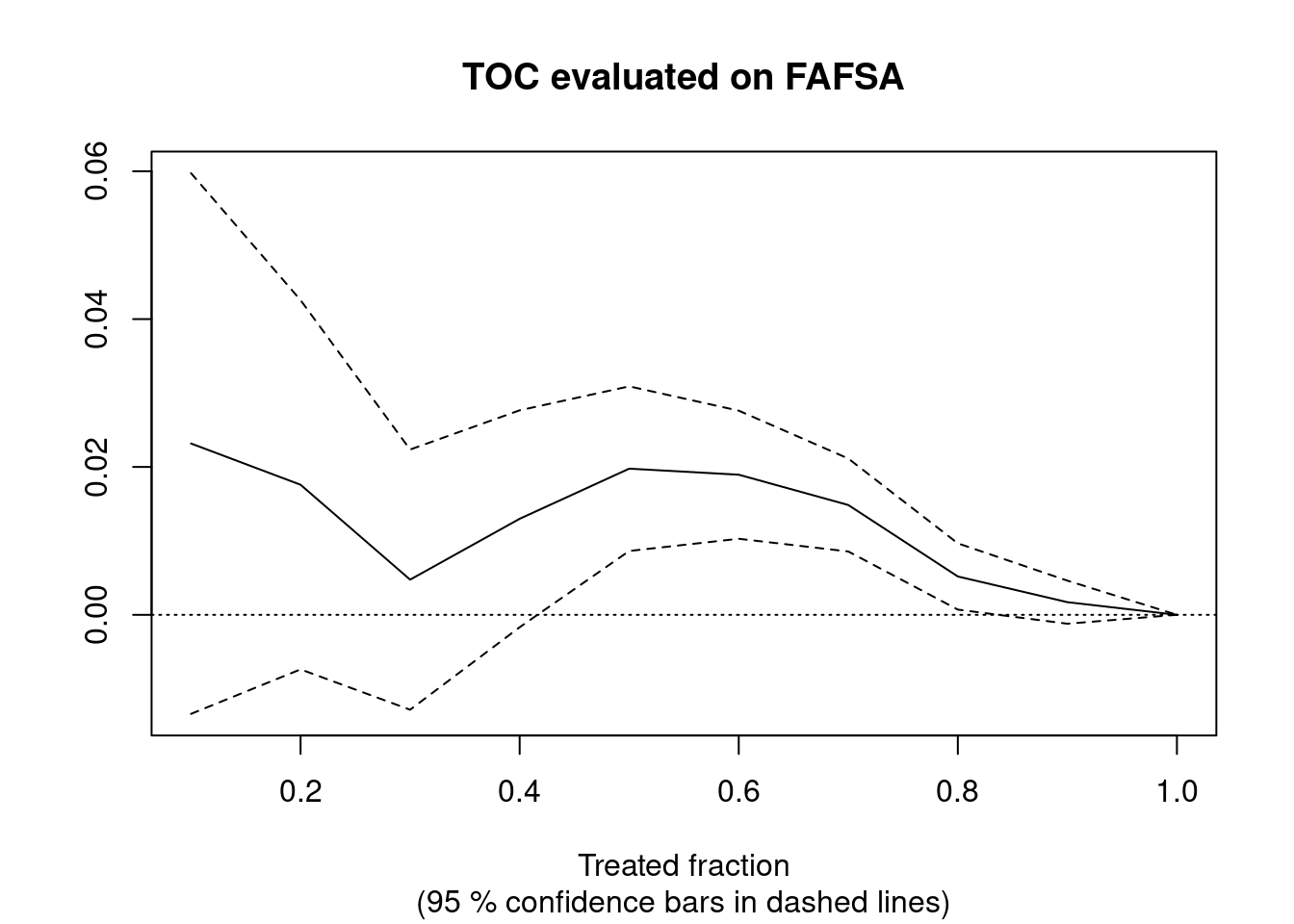}
        \caption{2017, early covariates}
      \end{subfigure}%
      ~
      \begin{subfigure}[t]{0.5\textwidth}
          \centering
          \includegraphics[width=\textwidth]{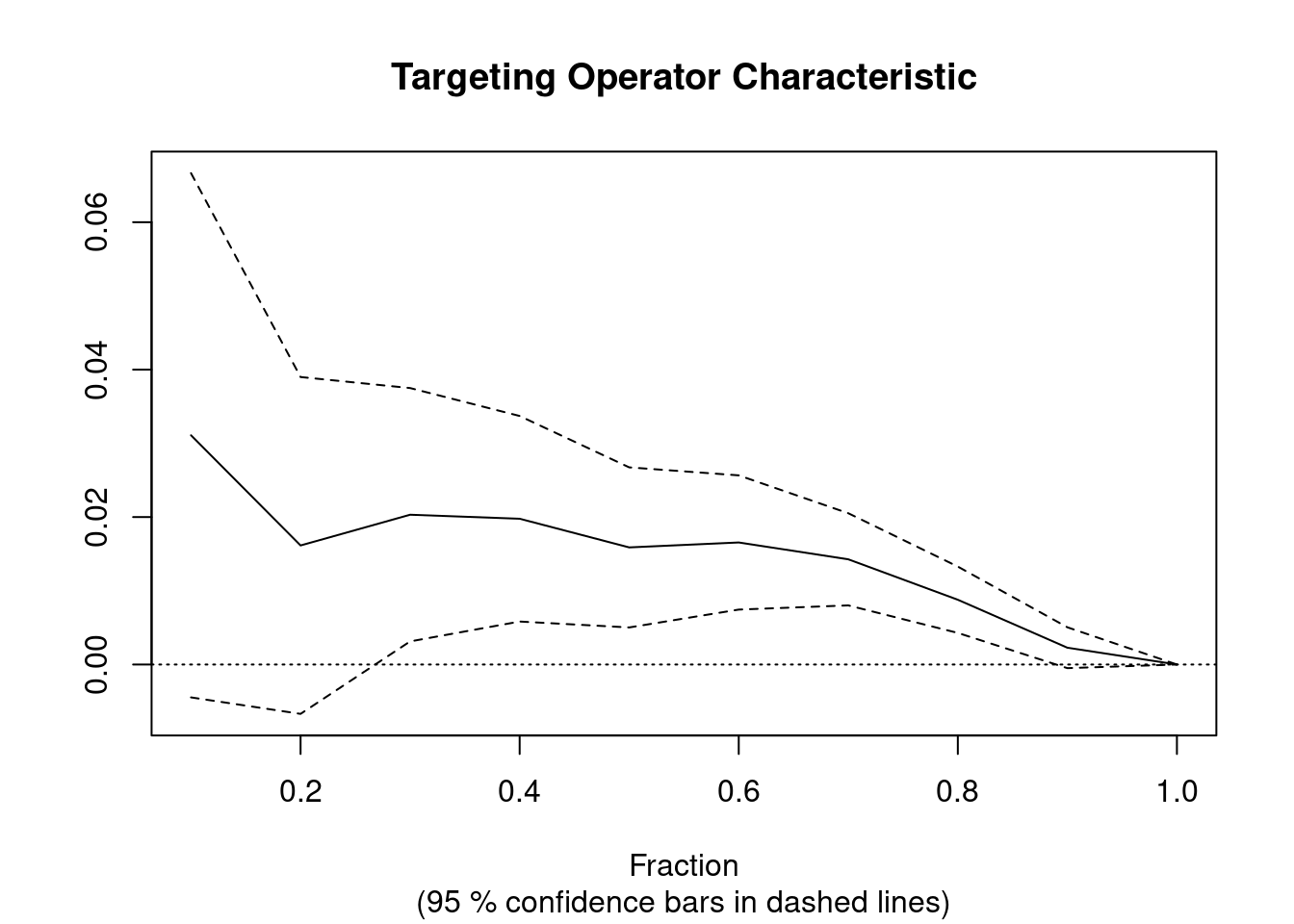}
          \caption{2017, late covariates}
      \end{subfigure}
              \bigskip
      \\
            \begin{subfigure}[t]{0.5\textwidth}
        \centering
        \includegraphics[width=\textwidth]{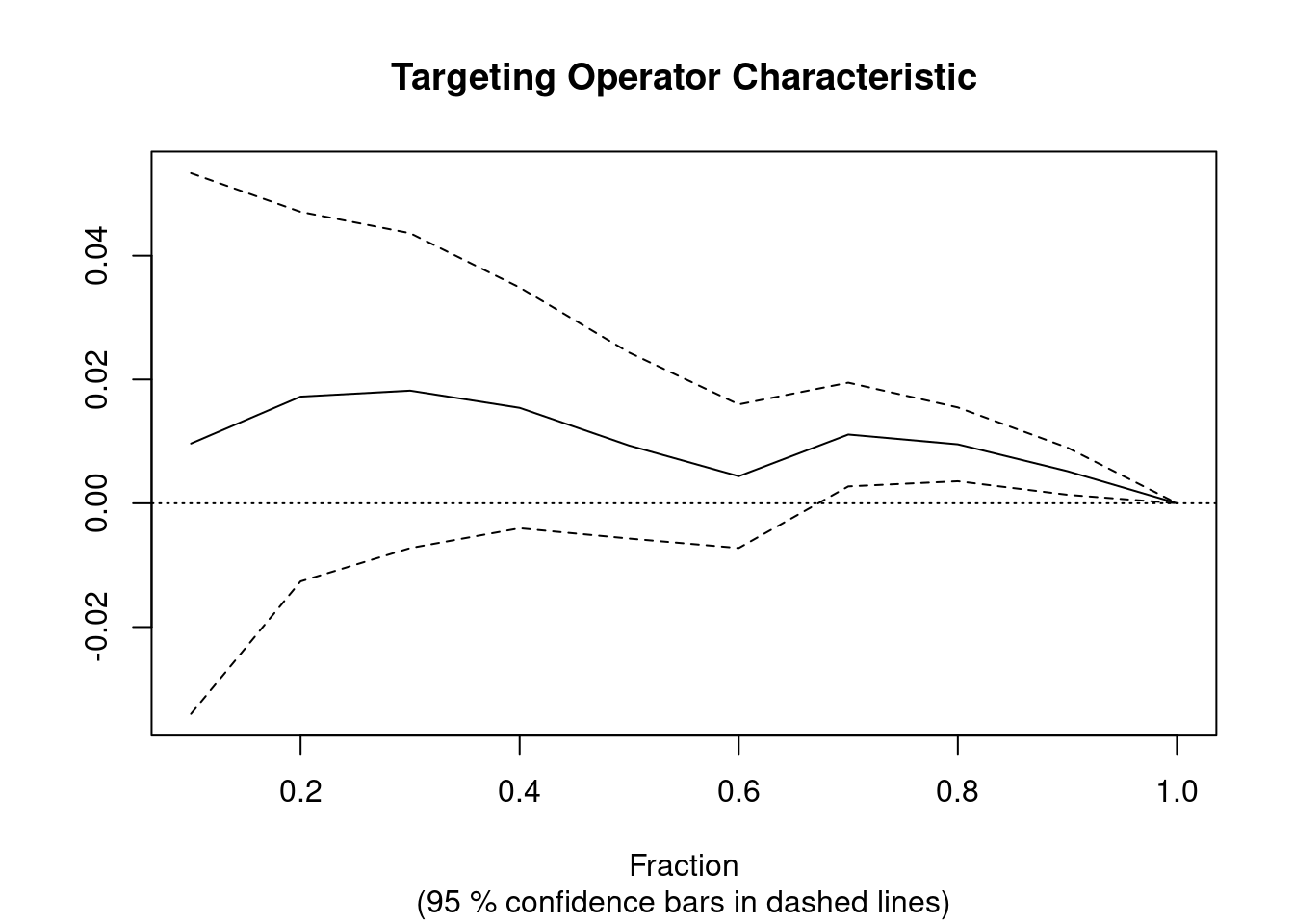}
        \caption{2018, early covariates}
      \end{subfigure}%
      ~
      \begin{subfigure}[t]{0.5\textwidth}
          \centering
          \includegraphics[width=\textwidth]{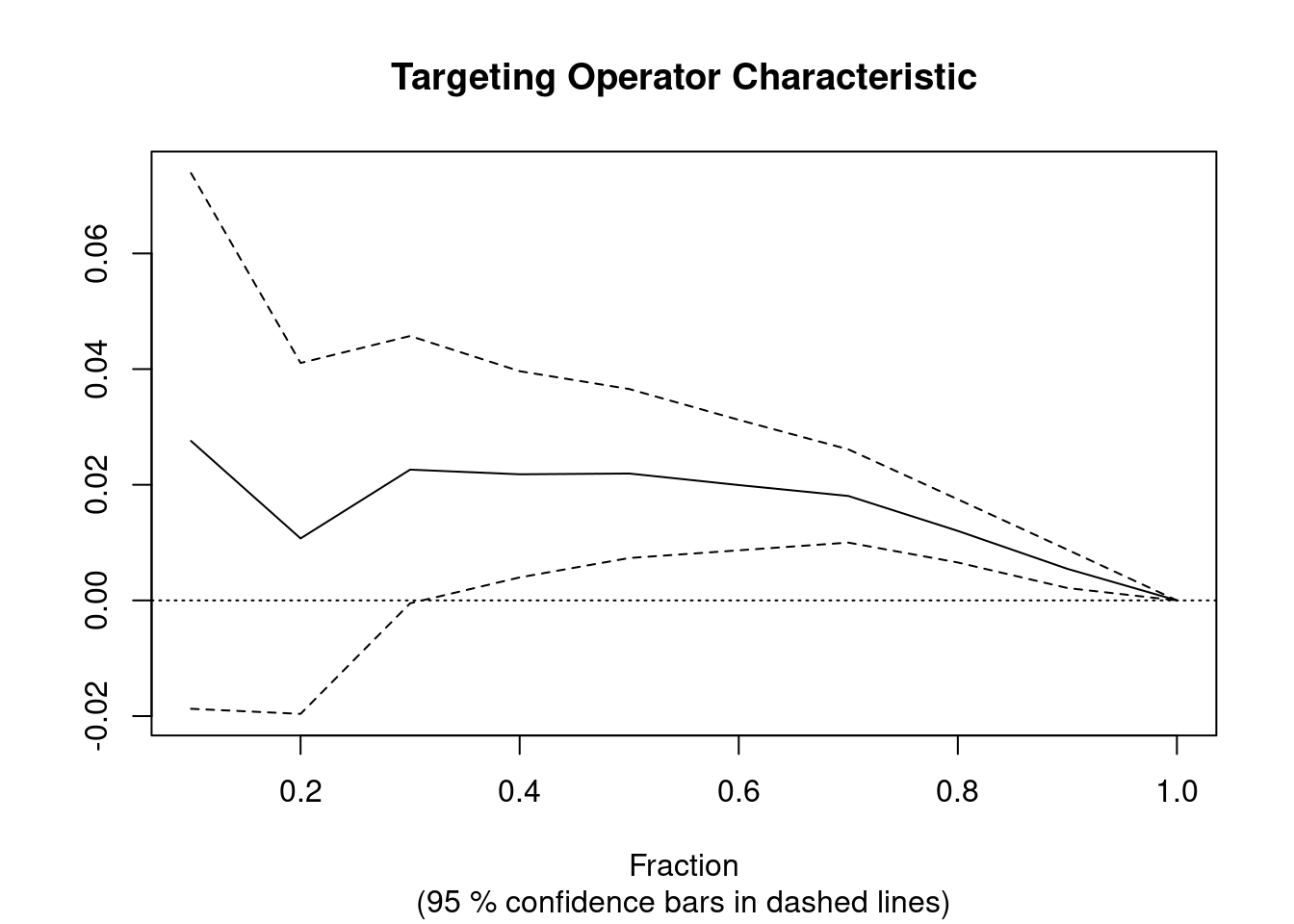}
          \caption{2018, late covariates}
      \end{subfigure}
     
      \caption{RATE estimates.}
      \label{fig:rate}
    \end{figure}

\FloatBarrier

\begin{table}[!htb]
    \centering
        \begin{tabular}[t]{llrrr}
        \toprule
        Year & Covariates & AUTOC & SE & $p$-value \\
        \midrule
        2017 & early & 0.01020 & 0.00588 & 0.04143 \\
         & late & 0.01351 & 0.00581 & 0.01001 \\
        \midrule
        2018 & early & 0.01107 & 0.00792 & 0.08103 \\
         & late & 0.01628 & 0.00768 & 0.01706 \\        
        \bottomrule
        \end{tabular}
    \caption{RATE-based omnibus test with one-sides $p$-value against the alternative of no benefit from targeting over random assignment.}
    \label{tbl:rate}
\end{table}

\FloatBarrier

\begin{figure}
  \centering
      \includegraphics[width=\textwidth]{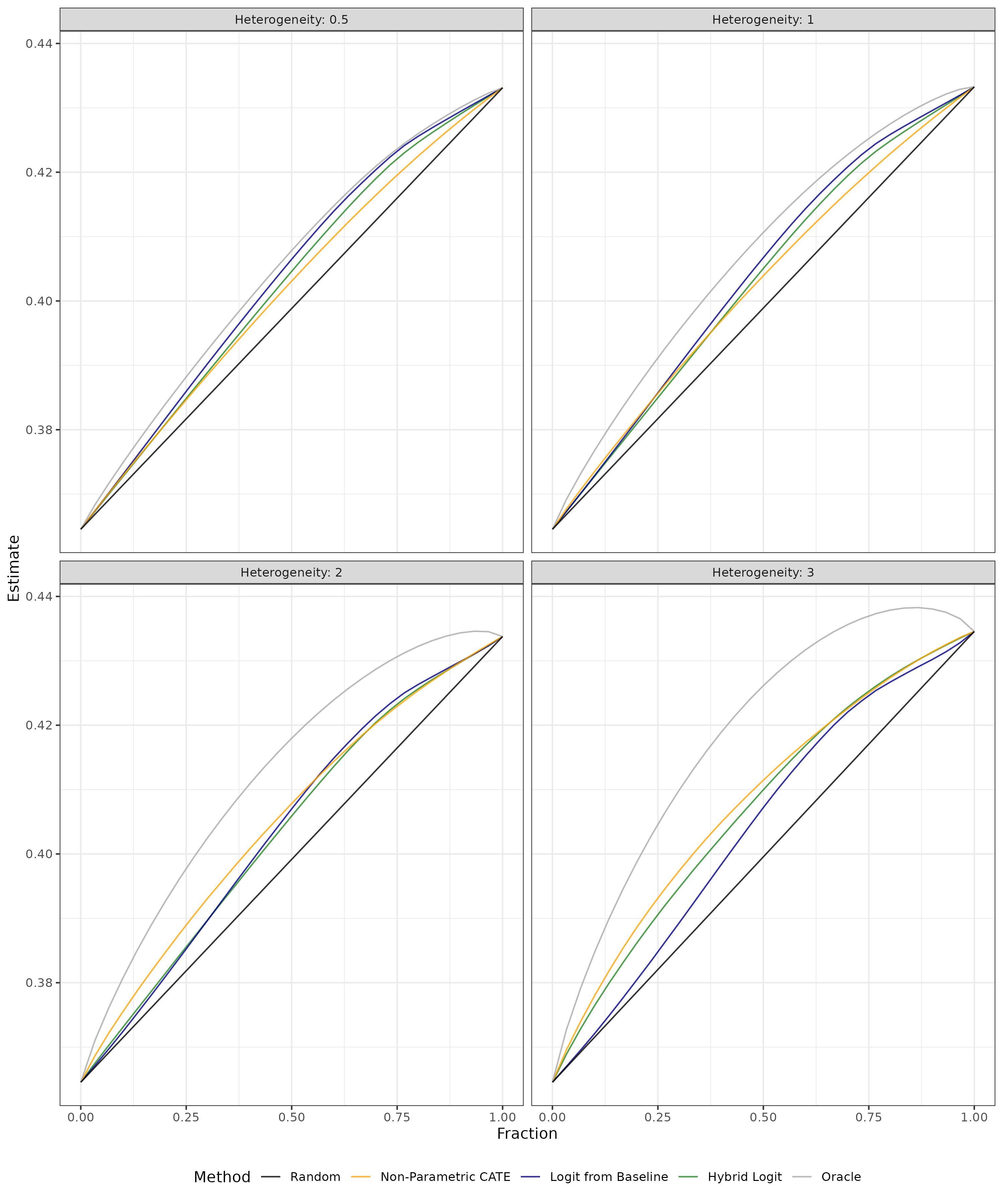}
  
  \caption{Performance of selected models in a simulation based on the 2017 data with early covariates and varying strengths of heterogeneity, as in \autoref{fig:simulation} but with only half the sample size for a total of 11,096 students.}
  
  \label{fig:simulation_halfsample}
\end{figure}

\FloatBarrier

  \begin{figure}[hbtp]
        \centering
      \centering
      \begin{subfigure}[t]{0.7\textwidth}
        \centering
        \includegraphics[width=\textwidth]{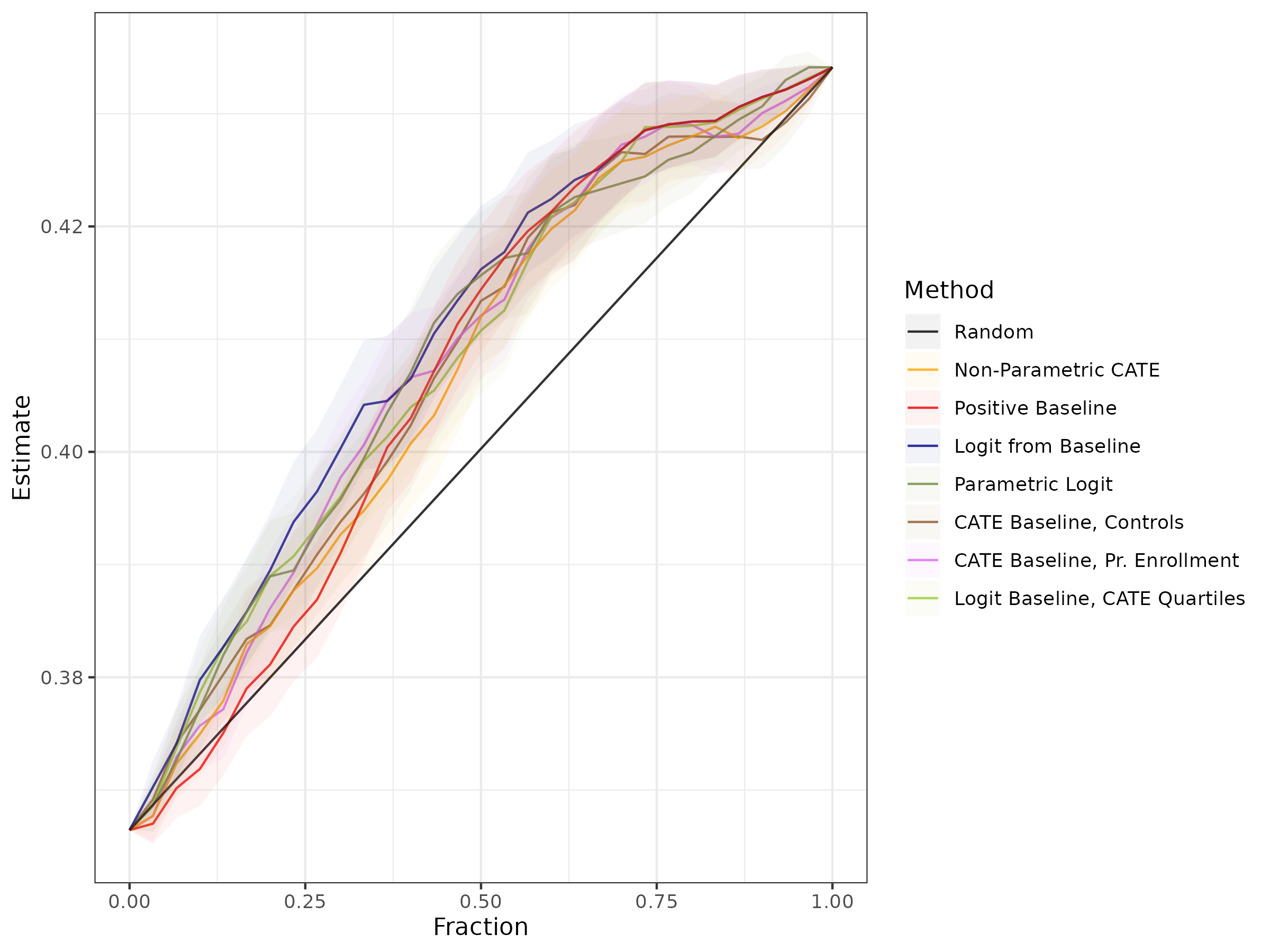}
        \caption{2017, early covariates}
      \end{subfigure}%

      \begin{subfigure}[t]{0.7\textwidth}
          \centering
          \includegraphics[width=\textwidth]{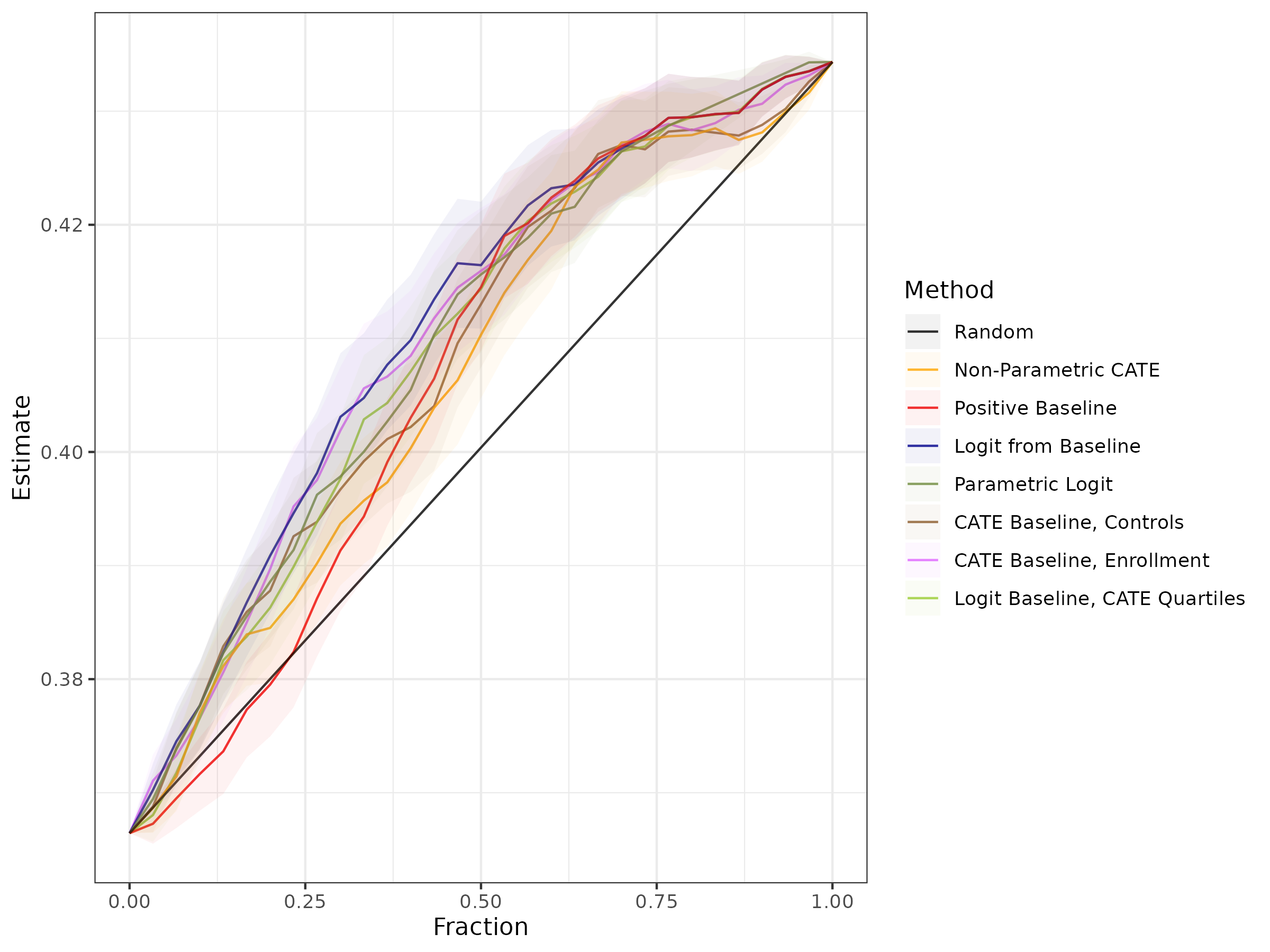}
          \caption{2017, late covariates}
      \end{subfigure}     
\caption{Total estimated FAFSA renewal rate ($y$-axis) by targeting a given fraction ($x$-axis) of students according to different cross-fitted predictions in the 2017 data as in \autoref{fig:policy_V1}, with additional targeting rules based on a causal-forest estimate from a prediction of the baseline and covariates (``CATE from Baseline, Controls'')
from a prediction of the baseline and predicted or actual enrollment (``CATE Baseline, Enrollment''),
as well as a logistic regression on baseline and treatment interacted with quartiles of estimated treatment effects (``Logit Baseline, CATE Quartiles'') and a treatment-interacted logistic regression on five variables chosen by the logit-LASSO (``Parametric Logit''). 
}
\label{fig:policy_V3}
\end{figure}

\FloatBarrier
    \newpage

    \begin{figure}[hbtp]
    \centering

    \begin{subfigure}[h]{\textwidth}
        \centering

        \begin{subfigure}[h]{.45\textwidth}
        \centering
        \includegraphics[width=\textwidth]{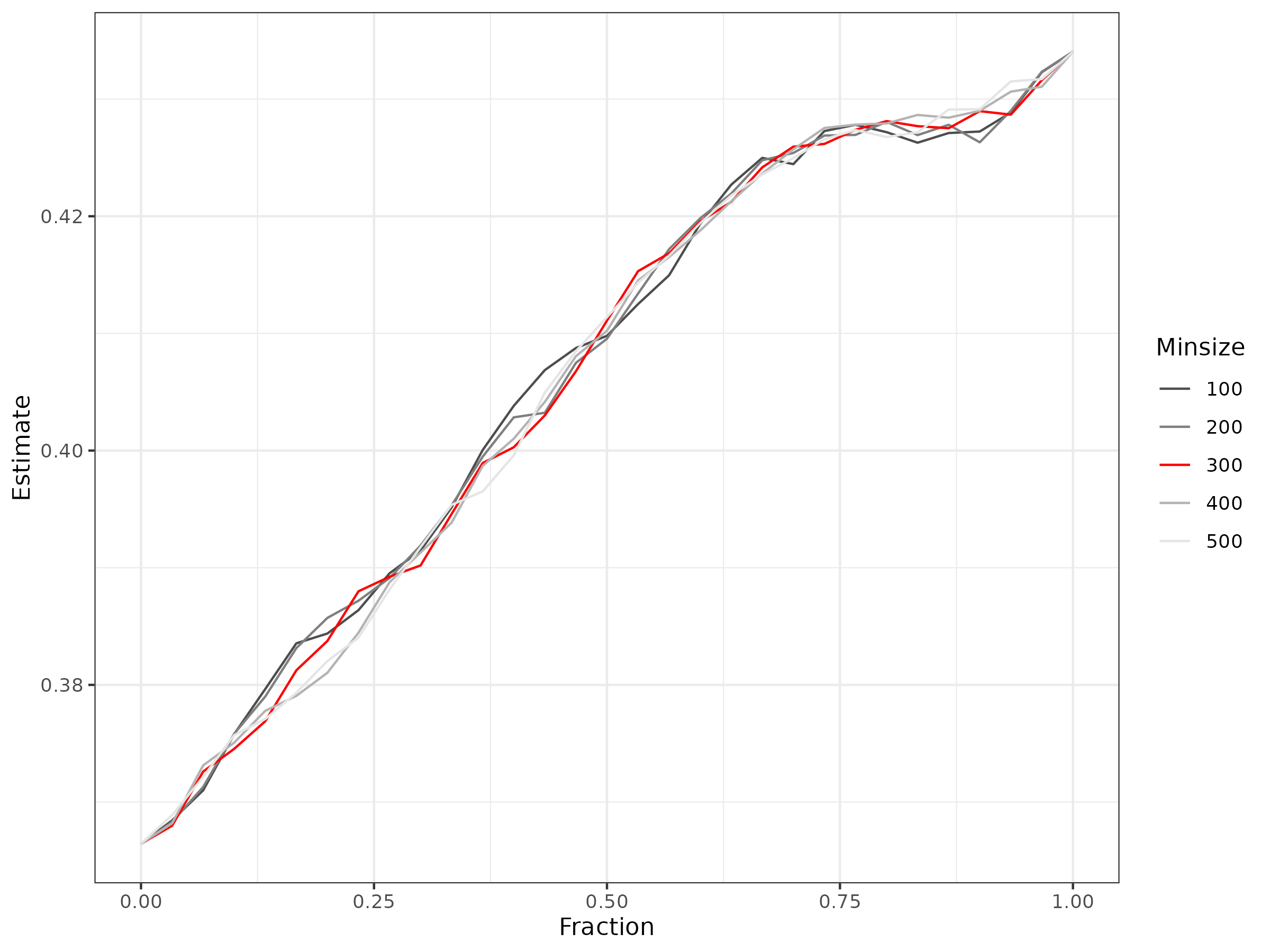}
        \caption{2017, early covariates}
       \end{subfigure}
        ~
        \begin{subfigure}[h]{.45\textwidth}
        \centering
        \includegraphics[width=\textwidth]{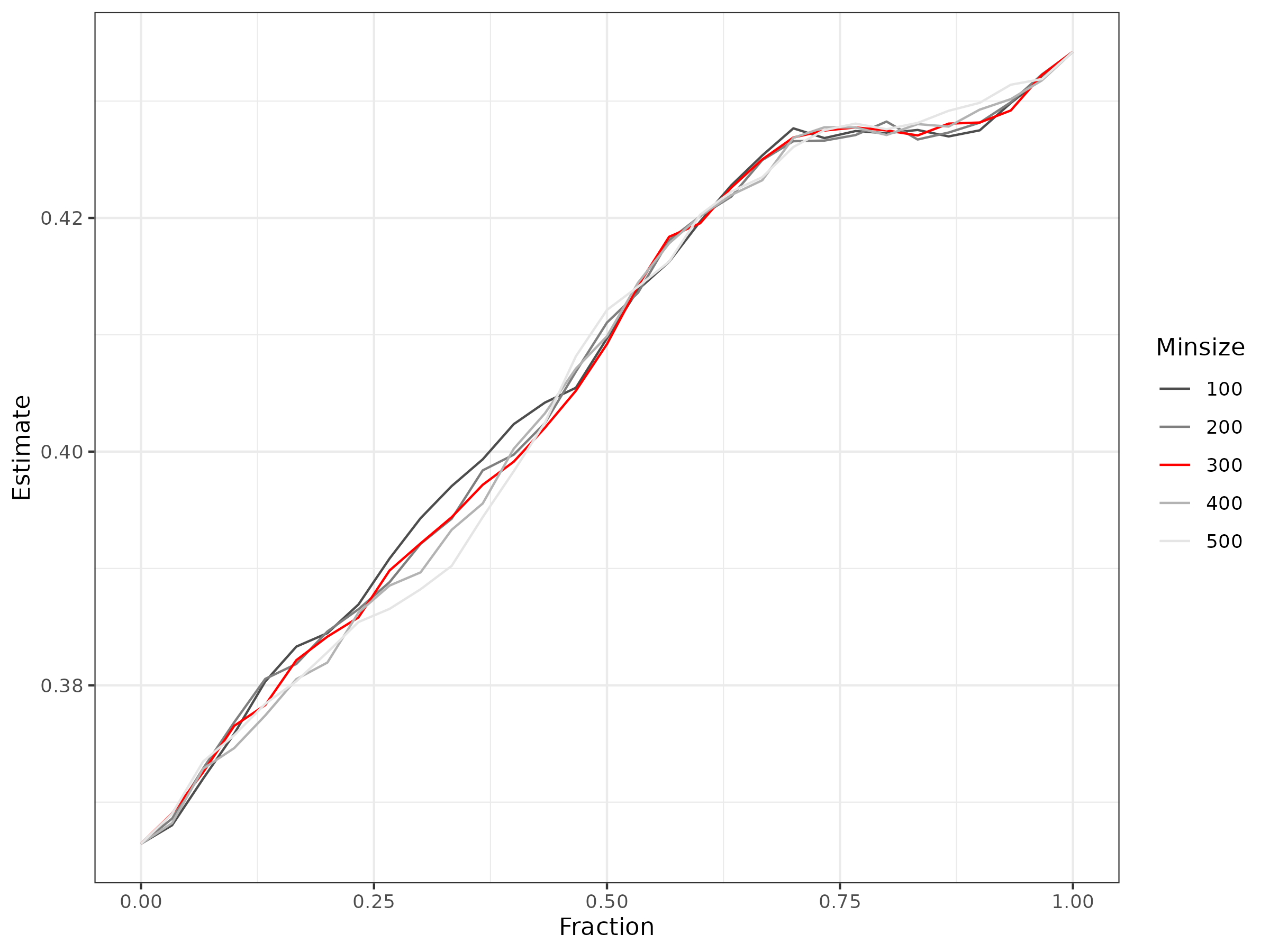}
        \caption{2017, late covariates}
       \end{subfigure}

       \caption*{(i) Minimum node size}

    \end{subfigure}

    \bigskip

        \begin{subfigure}[h]{\textwidth}
        \centering

        \begin{subfigure}[h]{.45\textwidth}
        \centering
        \includegraphics[width=\textwidth]{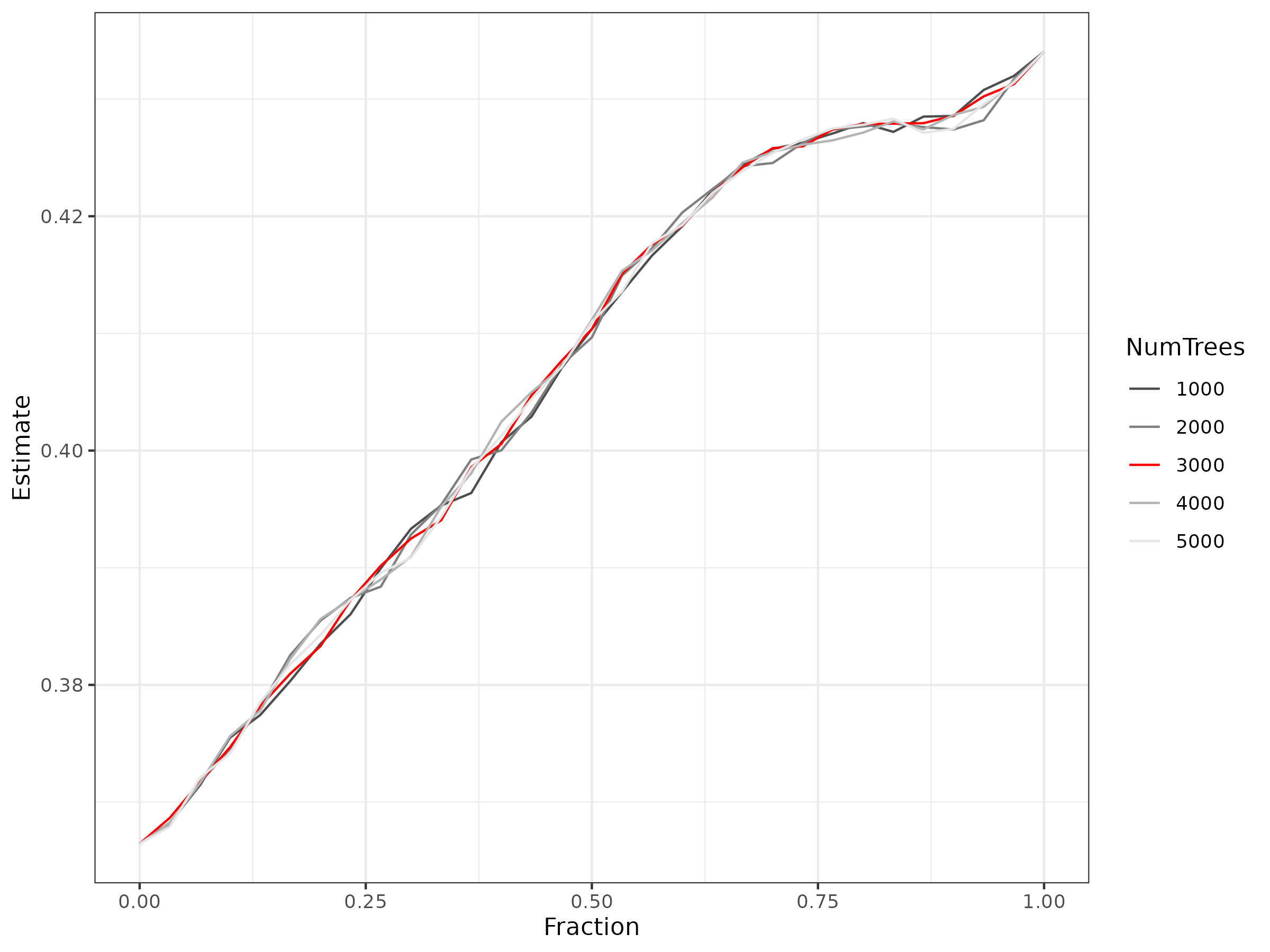}
        \caption{2017, early covariates}
       \end{subfigure}
        ~
        \begin{subfigure}[h]{.45\textwidth}
        \centering
        \includegraphics[width=\textwidth]{figures/fafsa2_early_full_tuning_numtrees.png}
        \caption{2017, late covariates}
       \end{subfigure}

       \caption*{(ii) Number of trees}

    \end{subfigure}
       
    \caption{
        The figures analyze the sensitivity of policies based on the non-parametric CATE estimate from \autoref{fig:policy_V1} on key tuning parameters of the causal forest, namely the (i) minimum node size and the (ii) number of trees.
        The red curve on each plot is the choice we use for targeting in \autoref{sec:targeting} and \autoref{sec:combined}.
    }
    \label{fig:robustness_tuning}
\end{figure}

    \FloatBarrier
    \newpage

    \begin{figure}[hbtp]
    \centering

    \begin{subfigure}[h]{\textwidth}
        \centering

        \begin{subfigure}[h]{.45\textwidth}
        \centering
        \includegraphics[width=\textwidth]{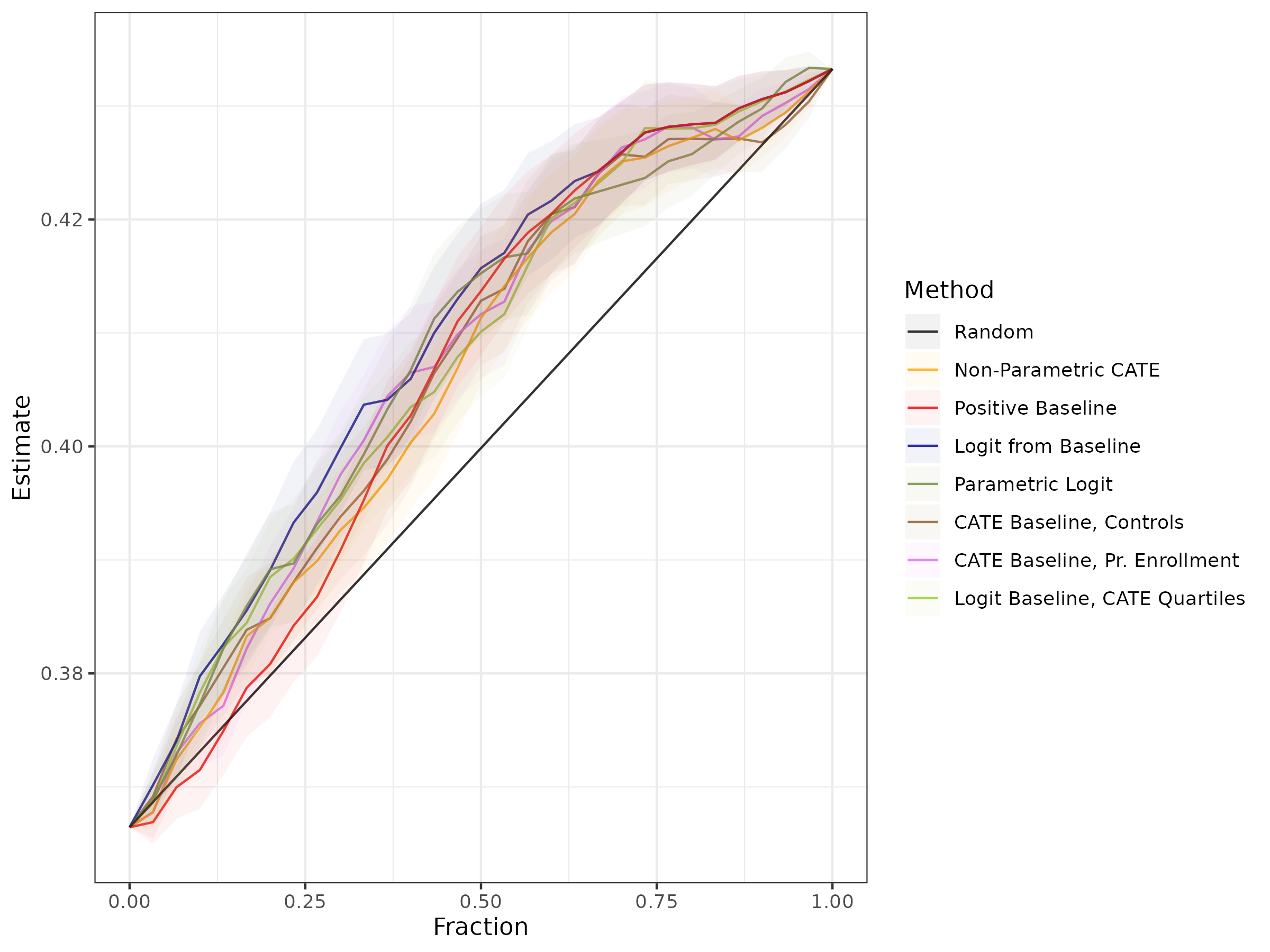}
        \caption{2017, early covariates}
       \end{subfigure}
        ~
        \begin{subfigure}[h]{.45\textwidth}
        \centering
        \includegraphics[width=\textwidth]{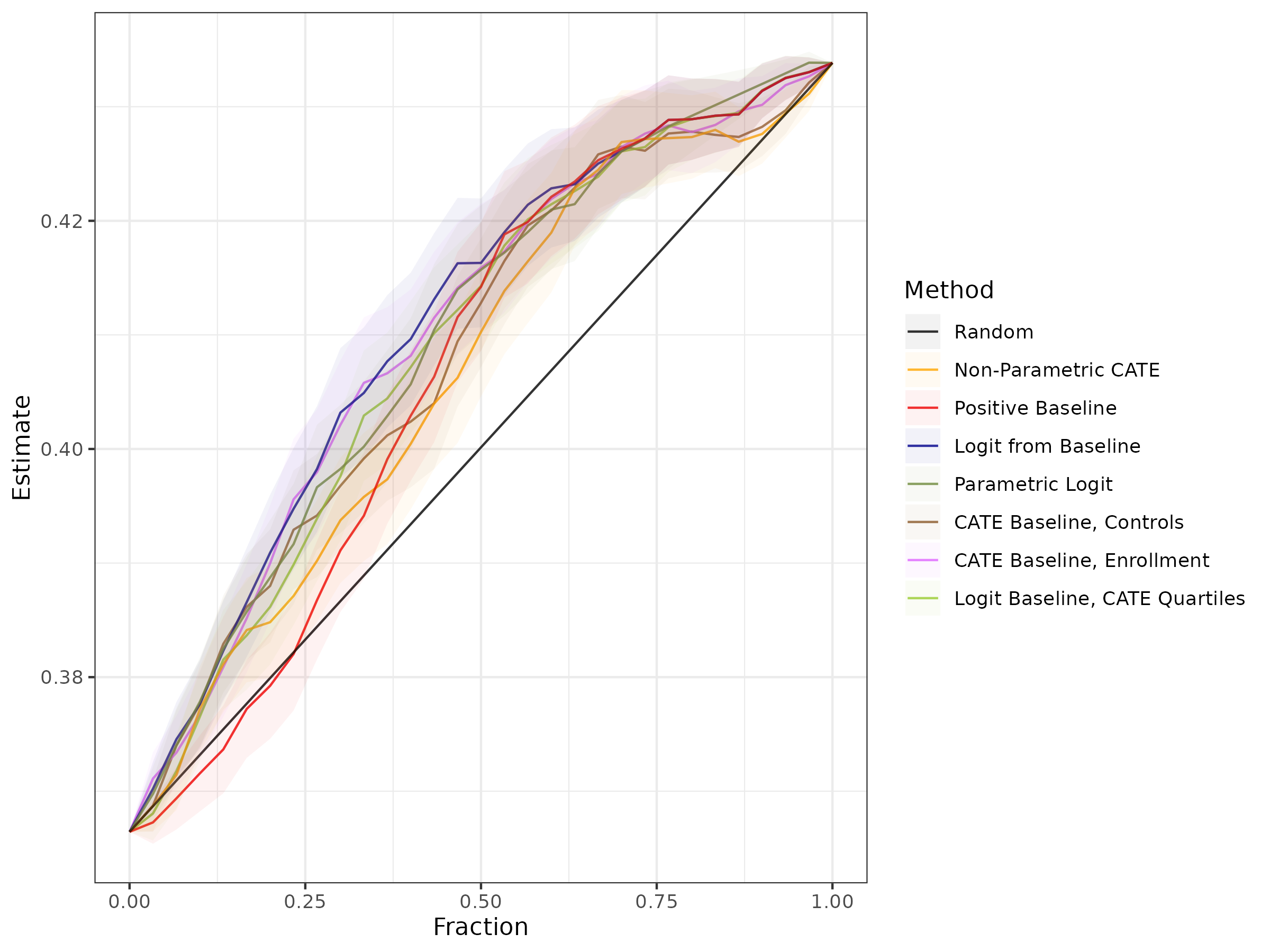}
        \caption{2017, late covariates}
       \end{subfigure}

       \caption*{(i) Same as \autoref{fig:policy_V3}, but cross-fold estimation of renewal rates are now performed using nuisance parameter estimates (of the CATE and the conditional outcome) that are re-estimated on the left-out fold using honest out-of-bag estimates in order to separate the evaluation from the estimation of policies on the training folds.}

    \end{subfigure}

    \bigskip

        \begin{subfigure}[h]{\textwidth}
        \centering

        \begin{subfigure}[h]{.45\textwidth}
        \centering
        \includegraphics[width=\textwidth]{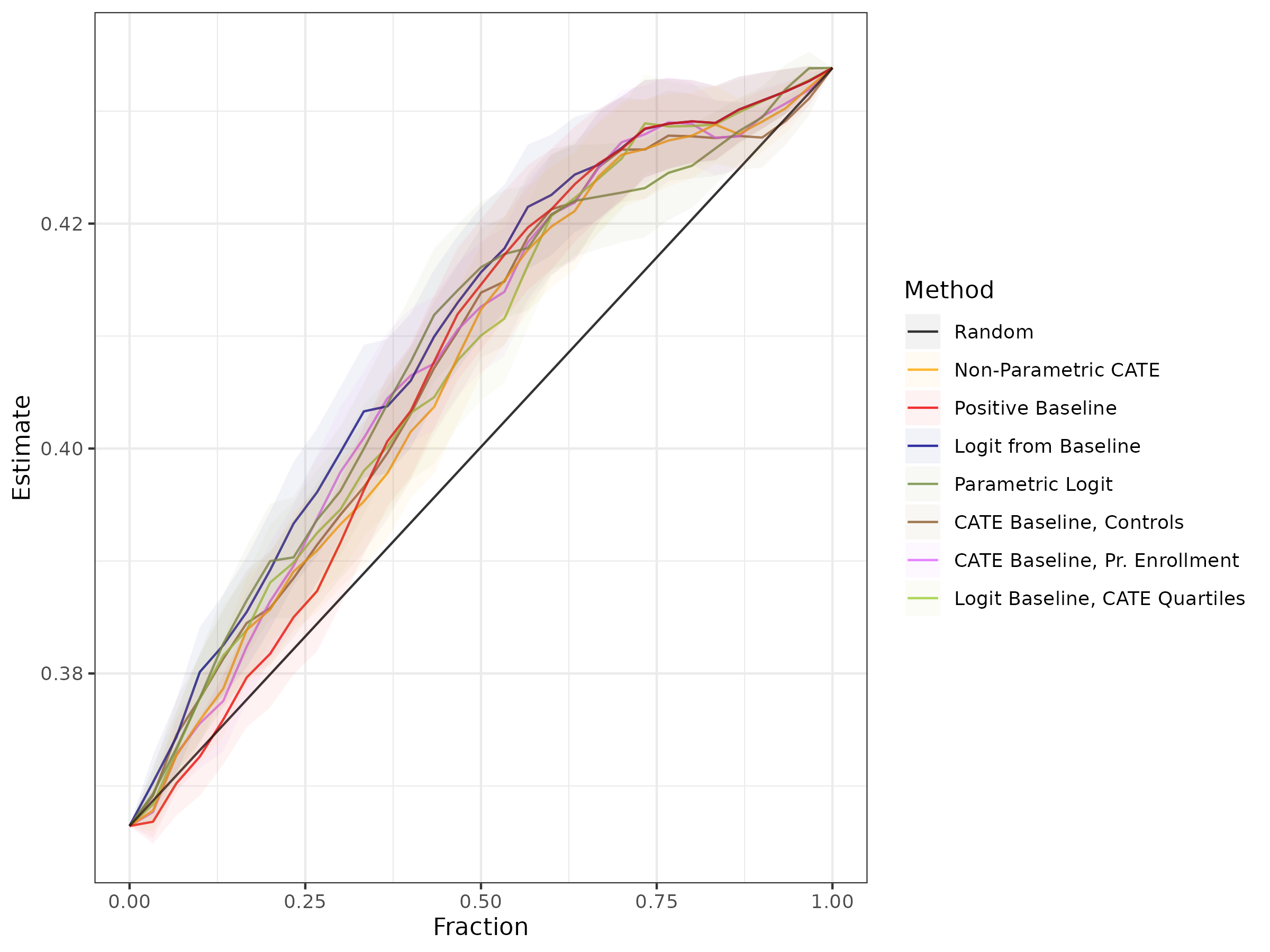}
        \caption{2017, early covariates}
       \end{subfigure}
        ~
        \begin{subfigure}[h]{.45\textwidth}
        \centering
        \includegraphics[width=\textwidth]{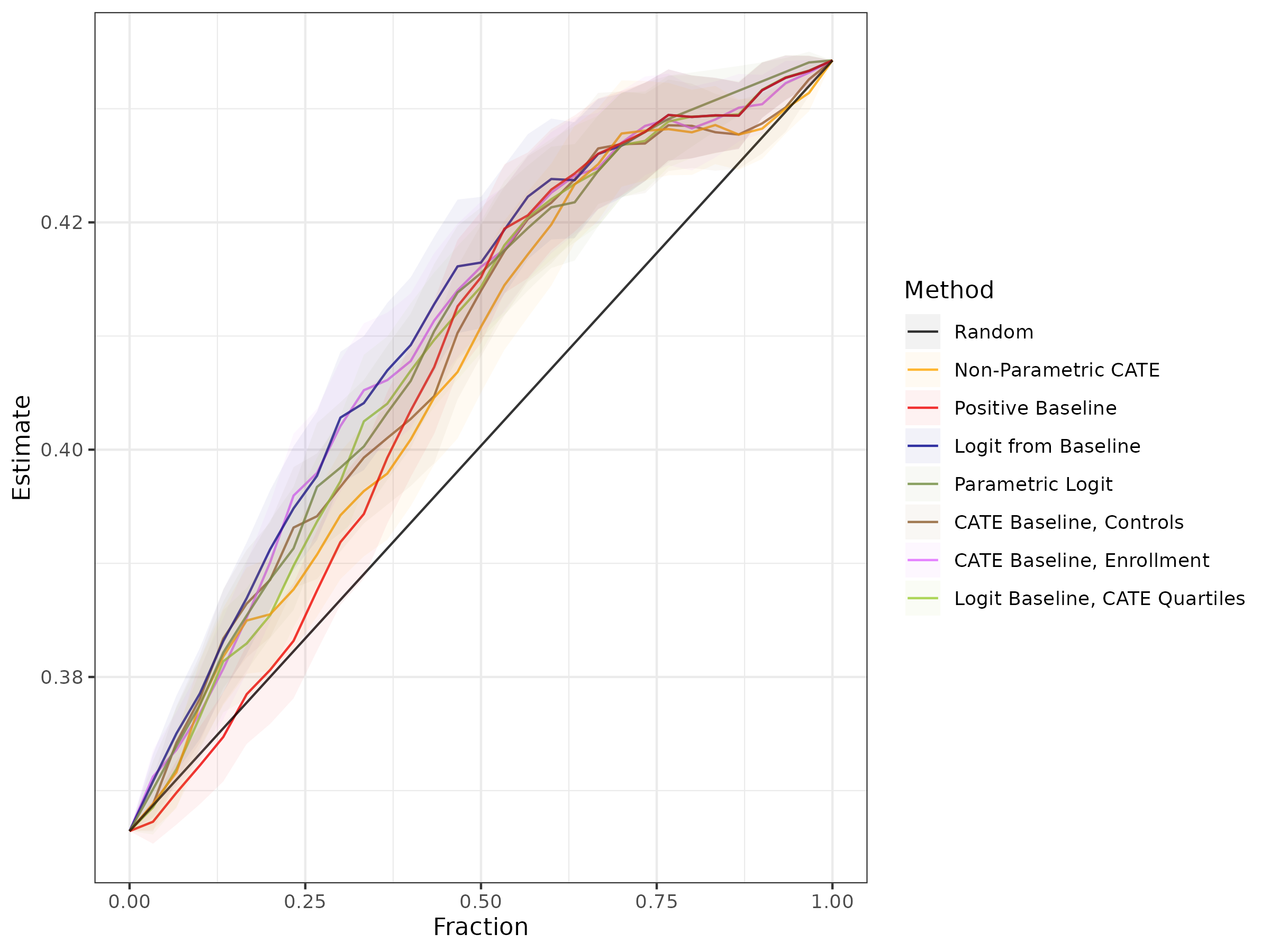}
        \caption{2017, late covariates}
       \end{subfigure}

       \caption*{(ii) Same as Panel~(i), but the propensity score is now also re-estimated on the left-out fold (rather than assumed constant) using a logistic regression.}

    \end{subfigure}
       
    \caption{
    The figure evaluates the same policies as \autoref{fig:policy_V3}, but makes changes to the estimation of renewal rates in the held-out folds that serve as additional robustness checks.
    Panel~(i) re-estimates nuisance parameters within fold and uses a fixed propensity score, and Panel~(ii) also re-estimates the propensity score.
    }

    \label{fig:robustness_reestimation}
\end{figure}

    \FloatBarrier
    \newpage

    \section{Evaluation of Assignment Policies}
    \label{apx:inference}

In the main article, we compare the precision of different treatment-effect estimates in terms of the average outcomes that can be achieved when we use them for targeting. 
In this section, we discuss estimation and inference on these average outcomes, which we use to obtain \autoref{fig:policy_V1}, as well as \autoref{fig:policy_V2} and \autoref{fig:policy_V3}.

Consider treatment assignment policies $\pi$ that map characteristics $X{=}x$ to probabilities $\pi(x) \in [0,1]$ of being treated. (This may include policies derived from treatment-effect estimates and random assignment, in which case $\pi(x) \equiv q$ with $q$ the probability of assignment.)
When treatment is assigned completely randomly (or randomly with known propensity score that only depends on $X$) in the existing data and $X$ is observed, then the average outcome $\E[\pi(X) Y(1) + (1-\pi(X)) Y(0)]$ under this policy, the total lift $\E[\pi(X) (Y(1) - Y(0))]$ relative to baseline, and the average treatment effect $\frac{\E[\pi(X) (Y(1) - Y(0))]}{\E[\pi(X)]}$ of those assigned to treatment are all identified, since $\E[Y(1)|X]=\E[Y|X,T{=}1], \E[Y(0)|X]=\E[Y|X,T{=}0]$ are.

Focusing on the case of average outcomes as in \autoref{fig:policy_V1}, we write $U(\pi) = \E[\pi(X) Y(1) + (1-\pi(X)) Y(0)$ for the expected outcome under this policy, which is identified by $U(\pi) = \E[\pi(X) Y |T{=}1] + \E[ (1-\pi(X)) Y|T{=}0]$ and could be estimated within fold $k$ by its sample analogue
\begin{align}
\label{eqn:simpledifference}
    \hat{U}_k(\pi) &= \frac{\sum_{i;k(i)=k} T_i \pi(X_i) Y_i}{\sum_{i=1}^n T_i} - \frac{\sum_{i=1}^n (1-T_i) (1-\pi(X_i)) Y_i}{\sum_{i=1}^n (1-T_i)}
    \\
    &=
    \frac{1}{n} \sum_{i;k(i)=k} \frac{T_i}{\hat{p}} \pi(X_i) Y_i  -  \frac{1 - T_i}{1-\hat{p}}(1-\pi(X_i)) Y_i \ 
    \text{  with  }  \ \hat{p} = \frac{\sum_{i;k(i)=k} T_i}{\sum_{i;k(i)=k} 1}
\end{align}
as in \citet{Hitsch2018-bw}, who consider the case of a known propensity score and non-stochastic assignment.

In our implementation for \autoref{fig:policy_V1}, we are specifically interested in making inference on differences $U(\pi) - U(\bar{\pi}_q)$ in outcomes of a policy $\pi$ that assigns students to treatment based on some rule (such as by ranking by estimated treatment effects) relative to the baseline policy $\bar{\pi}_q(x) \equiv q$ that assigns a \textit{random} fraction $q$ to treatment.
We note that for any two policies $\pi$ and $\bar{\pi}$, we have $U(\pi) - U(\bar{\pi}) = \E[(\pi(X) - \bar{\pi}(X)) Y(1) - Y(0)] = \E[(\pi(X) - \bar{\pi}(X)) \tau(X)]$, and for this specific baseline policy $\bar{\pi}_q$, we find $U(\bar{\pi}_q) = \E[Y(0)] + q \E[Y(1) - Y(0)] = \E[Y|T{=}0] + q \E[\tau(X)]$.%
\footnote{When propensity scores are non-constant, estimating $\E[Y(0)]$ will require additional care, and can be achieved by propensity-score weighting.}
Since $\E[Y|T{=}0]$ is readily estimated, we therefore now focus on efficient estimation and valid inference on weighted average treatment effects
$\tau_w = \E[w(X) \tau(X)]$,
where weights can be negative.
Once we have established estimation and inference for $\tau_w$,
we can estimate all quantities of interest via
\begin{align*}
    U(\pi) - U(\bar{\pi}_q) &= \tau_{\pi - \bar{\pi}_q},
    &
    U(\bar{\pi}_q) &= \E[Y|T{=}0] + \tau_{1},
    &
    U(\pi) &= U(\pi) + (U(\pi) - U(\bar{\pi}_q)).
\end{align*}

To improve efficiency and robustness of our estimate, as well as to ensure valid inference later on, we consider an augmented inverse propensity weighted (AIPW) estimator of $\tau_w = \E[w(X) \tau(X)]$.
Specifically, we assume that we have a consistent estimate $\hat{f}(x)$ of $\E[Y|X{=}x]$, a consistent estimate $\hat{\tau}(x)$ of $\tau(x)$, and a consistent propensity score estimate $\hat{p}(x)$ of $\E[T|X{=}x]$ available. The propensity score may be assumed to be constant when units are randomized unconditionally, in which case we may want to set $\hat{p}(x) \equiv \frac{\sum_{i=1}^n T_i}{n}$ analogously to above.
We assume that $\hat{f},\hat{\tau},\hat{p}$ are all fitted on separate data or using $k$-fold cross-fitting.
Writing $\hat{f}_1(x) = \hat{f}(x) + (1-\hat{p}(x)) \hat{\tau}(x), \hat{f}_0(x) = \hat{f}(x) - \hat{p}(x) \hat{\tau}(x)$,
the AIPW estimator is
\begin{align*}
    \hat{\tau}^\text{AIPW}_w
    &=
    \frac{1}{n} \sum_{i=1}^n w(X_i) \overbrace{\left( \hat{\tau}(X_i) + \frac{T_i - \hat{p}(X_i)}{\hat{p}(X_i) (1-\hat{p}(X_i))} (Y_i - \hat{f}_{T_i}(X_i)) \right)}^{=\hat{\tau}^\text{AIPW}(Y_i,T_i,X_i)}
    \\
    &=
    \frac{1}{n} \sum_{i=1}^n w(X_i)
      \frac{T_i - \hat{p}(X_i)}{\hat{p}(X_i) (1-\hat{p}(X_i))} (Y_i - \bar{f}(X_i))
\end{align*}
where $\bar{f}(x) = 
     (1- \hat{p}(x)) \hat{f}_1(x) 
     +\hat{p}(x) \hat{f}_0(x) = \hat{f}(x) + (1- 2 \hat{p}(x)) \hat{\tau}(x)
     = \hat{f}_0(x) + (1- \hat{p}(x)) \hat{\tau}(x)
     $.
This estimator is $\sqrt{n}$ consistent and asymptotically Normal under standard regularity conditions,
and it is exactly unbiased when the propensity score is known and remains consistent even when treatment-effect and outcome estimates are not.
We can consistently estimate its asymptotic variance by
\begin{align*}
    \hat{\sigma}^2_w
    =
    \frac{1}{n} \sum_{i=1}^m \left(w(X_i) \ \hat{\tau}^\text{AIPW}(Y_i,T_i,X_i) - \hat{\tau}^\text{AIPW}_w \right)^2
\end{align*}
to obtain standard error estimates $\frac{\hat{\sigma}_w}{\sqrt{n}}$ and a corresponding 95 \% confidence interval $\hat{\tau}^\text{AIPW}_w \pm 1.96 \cdot \frac{\hat{\sigma}_w}{\sqrt{n}}$.

Applying this estimator to the estimation for \autoref{fig:policy_V1},
we can estimate
\begin{align*}
    \hat{U}^\text{AIPW}(\bar{\pi}_q)
    &= \frac{\sum_{i=1}^n (1-T_i) Y_i}{\sum_{i=1}^n (1-T_i)} + q \  \frac{1}{n} \sum_{i=1}^n \hat{\tau}^\text{AIPW}(Y_i,T_i,X_i)
    \\
    \hat{U}^\text{AIPW}(\pi) &= \hat{U}^\text{AIPW}(\bar{\pi}_q) + \underbrace{\frac{1}{n} \sum_{i=1}^n (\pi(X_i) - q) \ \hat{\tau}^\text{AIPW}(Y_i,T_i,X_i)}_{=\hat{\Delta}(\pi)},
    \\
    \widehat{\text{SE}}\left( \hat{U}^\text{AIPW}(\pi) - \hat{U}^\text{AIPW}(\bar{\pi}_q) \right) &= \sqrt{\frac{1}{n^2} \sum_{i=1}^n \left((\pi(X_i) - q) \ \hat{\tau}^\text{AIPW}(Y_i,T_i,X_i) - \hat{\Delta}(\pi) \right)^2}.
\end{align*}
We note that, in particular, the estimation outcome of the random policy does not include any additional randomization, which would only add noise to the estimation.
Further, we obtain the estimator in \autoref{eqn:simpledifference} of $U(\pi)$ when we set $\hat{p}(x) \equiv \frac{\sum_{i=1}^n T_i}{n}, \hat{\tau}(x) \equiv 0, \hat{f}(x) \equiv 0$, which is generally inefficient.

So far, we have considered fixed policies $\pi$. However, in our application, policies are themselves estimated, such as the policy
\begin{align*}
    \hat{\pi}_{\hat{t}}(x) = \mathbbm{1}(\hat{\tau}(x) \geq \hat{t})
\end{align*}
where treatment effects $\hat{\tau}(x)$ and the cutoff $\hat{t}$ chosen to achieve a given proportion $q$ in treatment are all noisy.
We use cross-fitting to avoid biases from estimation in this case.
Specifically, we estimate the quantities of interest separately on each fold, using rankings estimated based on the other folds only, and then aggregate across all folds.
Under regularity conditions is the covariance between estimates from different folds of a lower order than the variance we estimate, allowing us to combine estimates and variance estimates across folds to obtain valid inference.

\end{document}